\documentclass[11pt]{article}

\usepackage{jheppub}
\usepackage[utf8]{inputenc}
\usepackage{booktabs}
\usepackage{graphicx}
\usepackage{bbold}
\usepackage{subcaption}
\usepackage{mathtools}
\usepackage{cancel}
\usepackage{datetime}
\usepackage{algorithm}
\usepackage{algorithmic}
\usepackage{xcolor}
\usepackage{xspace}
\usepackage[listings,skins]{tcolorbox}

\usepackage{xkeyval}

\usepackage{tikz,fp}
\usetikzlibrary{external,fixedpointarithmetic,decorations.pathmorphing,positioning,calc,fadings,decorations.markings,decorations.pathreplacing,patterns,arrows,arrows.meta}

\pgfdeclaremetadecoration{split graviton}{first wave}{
  \state{first wave}[width={.5*\pgfmetadecoratedpathlength-.5*min(3.68pt,.345*\pgfmetadecoratedpathlength)},next state=arrow]{
    \decoration{snake}
    \beforedecoration{\pgfpathmoveto{\pgfpointdecoratedpathfirst}}
  }
  \state{arrow}[width={min(3.68pt,.345*\pgfmetadecoratedpathlength)},next state=second wave]{
    \decoration{curveto}
    \afterdecoration{
      \pgfsetarrowsend{Stealth[length=\pgfdecoratedpathlength,width={.75*\pgfdecoratedpathlength}]}
      \pgfusepath{stroke}
      \pgfsetarrowsend{}
    }
  }
  \state{second wave}[width=\pgfmetadecoratedremainingdistance,next state=final]{
    \decoration{snake}
    \beforedecoration{\pgfpathmoveto{\pgfpointdecoratedpathfirst}}
    \afterdecoration{\pgfusepath{stroke}}
  }
  \state{final}{\decoration{moveto}}
}

\tikzset{
  label distance=-3pt,
  >=stealth,
  inner sep=1.5pt,
  boson/.style={
    decoration={snake, segment length=2mm, amplitude=0.5mm},
    decorate,
  },
  graviton/.style={
    line width=0.18pt,
    decoration={snake, segment length=1.3mm, amplitude=0.4mm},
    decorate
  },
  directed graviton/.style={
    line width=0.18pt,
    draw=none,
    postaction={graviton,decoration={split graviton}}
  },
  quark/.style={
    postaction={decorate},
    decoration={markings,mark=at position .5 with {\arrow[xshift=0.07cm,scale=1.15]{stealth}}}
  },
  aquark/.style={
    postaction={decorate},
    decoration={markings,mark=at position .5 with {\arrowreversed[xshift=-0.07cm,scale=1.15]{stealth}}}
  },
  gluonOld/.style={
    line width=0.7pt,
    decorate,
    decoration={coil,aspect=-0.5,amplitude=-2pt,segment length=2.2pt}
  },
  gluon/.style={
    line width=0.5pt,
    decorate,
    decoration={coil,aspect=-0.75,amplitude=-1.5pt,segment length=2pt}
  },
  agluon/.style={
    line width=0.5pt,
    decorate,
    decoration={coil,aspect=0.75,amplitude=1.5pt,segment length=2pt}
  },
  scalar/.style={
    dashed
  },
  dscalar/.style={
    dashed,
    postaction={decorate},
    decoration={markings,mark=at position 0.65 with {\draw[->] (-0.01,0) -- (0.07,0);}}
  },
  adscalar/.style={
    dashed,
    postaction={decorate},
    decoration={markings,mark=at position 0.65 with {\draw[->] (-0.01,0) -- (0.07,0);}}
  },
  line/.style={
  },
  doubleline/.style={
    double
  },
  dot/.style={
  	postaction={decorate},
    decoration={markings,mark=between positions 0.5 and 0.5 step 1 with {\draw[fill] (0,0) circle [radius=1.5pt];}}  
  },
  dots/.style={
  	postaction={decorate},
    decoration={markings,mark=between positions 0.33 and 0.66 step 0.33 with {\draw[fill] (0,0) circle [radius=1.5pt];}}  
  },
  dotss/.style={
  	postaction={decorate},
    decoration={markings,mark=between positions 0.25 and 0.75 step 0.25 with {\draw[fill] (0,0) circle [radius=1.5pt];}}  
  },
  blob/.style={
  	pattern=north west lines,
    draw=black
  },
  blobBlack/.style={
    draw=black,
    fill=black
  },
  blobWhite/.style={
    draw=black,
    fill=white
  },
  xSource/.style={
    circle, minimum width=5pt, draw, inner sep=0pt, path picture={\draw (path picture bounding box.south east) -- (path picture bounding box.north west) (path picture bounding box.south west) -- (path picture bounding box.north east);}
  },
  iter/.style={
    circle, minimum width=5pt, draw, inner sep=0pt
  }
}

\makeatletter
\def\gSetStdExtTypes#1{\presetkeys{gTypes}{eA=#1,eB=#1,eC=#1,eD=#1,eE=#1}{}}
\def\gSetStdIntTypes#1{\presetkeys{gTypes}{iA=#1,iB=#1,iC=#1,iD=#1,iE=#1,iF=#1,iG=#1,iH=#1,iI=#1,iJ=#1}{}}
\def\gSetStdTypes#1{\gSetStdExtTypes{#1}\gSetStdIntTypes{#1}}

\define@cmdkeys{general}{scale}
\define@cmdkeys{general}{rotate}
\define@cmdkeys{general}{font}
\define@cmdkeys{general}{yshift}
\presetkeys{general}{scale=1, rotate=0, font=\scriptsize, yshift=0pt}{}

\define@cmdkeys{gTypes}{eA,eB,eC,eD,eE,iA,iB,iC,iD,iE,iF,iG,iH,iI,iJ}
\gSetStdTypes{line}
\define@key{pre}{all}{\gSetStdTypes{#1}}
\define@key{pre}{allE}{\gSetStdExtTypes{#1}}
\define@key{pre}{allI}{\gSetStdIntTypes{#1}}

\define@cmdkeys{gLabels}{eLA,eLB,eLC,eLD,eLE,iLA,iLB,iLC,iLD,iLE,iLF,iLG,iLH,iLI,iLJ}
\presetkeys{gLabels}{eLA={},eLB={},eLC={},eLD={},eLE={},iLA={},iLB={},iLC={},iLD={},iLE={},iLF={},iLG={},iLH={},iLI={},iLJ={}}{}

\define@boolkeys{dots}{lDots,rDots}
\presetkeys{dots}{lDots=false, rDots=false}{}

\define@cmdkeys{blobs}{blobA,blobB,blobC}
\presetkeys{blobs}{blobA=blob,blobB=blob,blobC=blob}{}

\def\gTreeTri{\@ifnextchar[\@gTreeTri{\@gTreeTri[]}}
\def\@gTreeTri[#1]#2{%
  \setkeys*{pre}{#1}%
  \setrmkeys{general,gTypes,gLabels}%
  \begin{tikzpicture}
    [line width=1pt,
    baseline={([yshift=-0.5ex+\cmdKV@general@yshift]current bounding box.center)},
    scale=\cmdKV@general@scale,
    rotate=\cmdKV@general@rotate,
    font=\cmdKV@general@font]
    \draw[\cmdKV@gTypes@eA] (0.7,0.3) -- (0.4,0.3) node[label=(-180+\cmdKV@general@rotate):\cmdKV@gLabels@eLA] {};
    \draw[\cmdKV@gTypes@eB] (0.7,0.3) -- (0.9,0.6) node[label=(\cmdKV@general@rotate):\cmdKV@gLabels@eLB] {};
    \draw[\cmdKV@gTypes@eC] (0.7,0.3) -- (0.9,0) node[label=(\cmdKV@general@rotate):\cmdKV@gLabels@eLC] {};
    #2
  \end{tikzpicture}
  \gSetStdTypes{line}
}

\def\gTreeS{\@ifnextchar[\@gTreeS{\@gTreeS[]}}
\def\@gTreeS[#1]#2{%
  \setkeys*{pre}{#1}%
  \setrmkeys{general,gTypes,gLabels}%
  \begin{tikzpicture}
    [line width=1pt,
    baseline={([yshift=-0.5ex+\cmdKV@general@yshift]current bounding box.center)},
    scale=\cmdKV@general@scale,
    rotate=\cmdKV@general@rotate,
    font=\cmdKV@general@font]
    \draw[\cmdKV@gTypes@eA] (0.2,0.3) -- (0,0) node[label=(-180+\cmdKV@general@rotate):\cmdKV@gLabels@eLA] {};
    \draw[\cmdKV@gTypes@eB] (0.2,0.3) -- (0,0.6) node[label=(180+\cmdKV@general@rotate):\cmdKV@gLabels@eLB] {};
    \draw[\cmdKV@gTypes@eC] (0.7,0.3) -- (0.9,0.6) node[label=(\cmdKV@general@rotate):\cmdKV@gLabels@eLC] {};
    \draw[\cmdKV@gTypes@eD] (0.7,0.3) -- (0.9,0) node[label=(\cmdKV@general@rotate):\cmdKV@gLabels@eLD] {};    
    \draw[label distance=0,\cmdKV@gTypes@iA] (0.2,0.3) -- node[label=(90+\cmdKV@general@rotate):\cmdKV@gLabels@iLA] {} (0.7,0.3);
    #2
  \end{tikzpicture}
  \gSetStdTypes{line}
}

\def\gTreeT{\@ifnextchar[\@gTreeT{\@gTreeT[]}}
\def\@gTreeT[#1]#2{%
  \setkeys*{pre}{#1}%
  \setrmkeys{general,gTypes,gLabels}%
  \begin{tikzpicture}
    [line width=1pt,
    baseline={([yshift=-0.5ex+\cmdKV@general@yshift]current bounding box.center)},
    scale=\cmdKV@general@scale,
    rotate=\cmdKV@general@rotate,
    font=\cmdKV@general@font]
    \draw[\cmdKV@gTypes@eA] (0.45,0.1) -- (0,0) node[label=(-180+\cmdKV@general@rotate):\cmdKV@gLabels@eLA] {};
    \draw[\cmdKV@gTypes@eB] (0.45,0.5) -- (0,0.6) node[label=(180+\cmdKV@general@rotate):\cmdKV@gLabels@eLB] {};
    \draw[\cmdKV@gTypes@eC] (0.45,0.5) -- (0.9,0.6) node[label=(\cmdKV@general@rotate):\cmdKV@gLabels@eLC] {};
    \draw[\cmdKV@gTypes@eD] (0.45,0.1) -- (0.9,0) node[label=(\cmdKV@general@rotate):\cmdKV@gLabels@eLD] {};
    \draw[label distance=0,\cmdKV@gTypes@iA] (0.45,0.5) -- node[label=(\cmdKV@general@rotate):\cmdKV@gLabels@iLA] {} (0.45,0.1);
    #2
  \end{tikzpicture}
  \gSetStdTypes{line}
}

\def\gTreeU{\@ifnextchar[\@gTreeU{\@gTreeU[]}}
\def\@gTreeU[#1]#2{%
  \setkeys*{pre}{#1}%
  \setrmkeys{general,gTypes,gLabels}%
  \begin{tikzpicture}
    [line width=1pt,
    baseline={([yshift=-0.5ex+\cmdKV@general@yshift]current bounding box.center)},
    scale=\cmdKV@general@scale,
    rotate=\cmdKV@general@rotate,
    font=\cmdKV@general@font]
    \draw[\cmdKV@gTypes@eA] (0.45,0.5) -- (0,0) node[label=(-180+\cmdKV@general@rotate):\cmdKV@gLabels@eLA] {};
    \foreach \x in {5,...,14} \fill[opacity=0.12,white] (0.27,0.3) circle (\x/100);
    \draw[\cmdKV@gTypes@eB] (0.45,0.1) -- (0,0.6) node[label=(180+\cmdKV@general@rotate):\cmdKV@gLabels@eLB] {};
    \draw[\cmdKV@gTypes@eC] (0.45,0.5) -- (0.9,0.6) node[label=(\cmdKV@general@rotate):\cmdKV@gLabels@eLC] {};
    \draw[\cmdKV@gTypes@eD] (0.45,0.1) -- (0.9,0) node[label=(\cmdKV@general@rotate):\cmdKV@gLabels@eLD] {};
    \draw[label distance=0,\cmdKV@gTypes@iA] (0.45,0.5) -- node[label=(\cmdKV@general@rotate):\cmdKV@gLabels@iLA] {} (0.45,0.1);
    #2
  \end{tikzpicture}
  \gSetStdTypes{line}
}

\def\gTreeFourPt{\@ifnextchar[\@gTreeFourPt{\@gTreeFourPt[]}}
\def\@gTreeFourPt[#1]#2{%
  \setkeys*{pre}{#1}%
  \setrmkeys{general,gTypes,gLabels}%
  \begin{tikzpicture}
    [line width=1pt,
    baseline={([yshift=-0.5ex+\cmdKV@general@yshift]current bounding box.center)},
    scale=\cmdKV@general@scale,
    rotate=\cmdKV@general@rotate,
    font=\cmdKV@general@font]
    \draw[\cmdKV@gTypes@eA] (0,0) -- (-0.4,-0.4) node[label=(-180+\cmdKV@general@rotate):\cmdKV@gLabels@eLA] {};
    \draw[\cmdKV@gTypes@eB] (0,0) -- (-0.4,0.4) node[label=(180+\cmdKV@general@rotate):\cmdKV@gLabels@eLB] {};
    \draw[\cmdKV@gTypes@eC] (0,0) -- (0.4,0.4) node[label=(\cmdKV@general@rotate):\cmdKV@gLabels@eLC] {};
    \draw[\cmdKV@gTypes@eD] (0,0) -- (0.4,-0.4) node[label=(\cmdKV@general@rotate):\cmdKV@gLabels@eLD] {};
    #2
  \end{tikzpicture}
  \gSetStdTypes{line}
}

\def\gTadA{\@ifnextchar[\@gTadA{\@gTadA[]}}
\def\@gTadA[#1]#2{%
  \setkeys*{pre}{#1}%
  \setrmkeys{general,gTypes,gLabels}%
  \begin{tikzpicture}
    [line width=1pt,
    baseline={([yshift=-0.5ex+\cmdKV@general@yshift]current bounding box.center)},
    scale=\cmdKV@general@scale,
    rotate=\cmdKV@general@rotate,
    font=\cmdKV@general@font]
    \draw[\cmdKV@gTypes@eA] (-0.7,0.3) -- (-0.9,0) node[label=(180+\cmdKV@general@rotate):\cmdKV@gLabels@eLA] {};
    \draw[\cmdKV@gTypes@eB] (-0.7,0.3) -- (-0.9,0.6) node[label=(180+\cmdKV@general@rotate):\cmdKV@gLabels@eLB] {};
    \draw[\cmdKV@gTypes@eC] (0.15,0.45) -- (0.5,0.6) node[label=(\cmdKV@general@rotate):\cmdKV@gLabels@eLC] {};
    \draw[\cmdKV@gTypes@eD] (-0.3,0.3) -- (-0.2,0) node[label=(\cmdKV@general@rotate):\cmdKV@gLabels@eLD] {};
    \draw[label distance=0,\cmdKV@gTypes@iA] (-0.7,0.3) -- node[label=(90+\cmdKV@general@rotate):\cmdKV@gLabels@iLA] {} (-0.3,0.3);
    \draw[label distance=0,\cmdKV@gTypes@iB] (-0.3,0.3) -- node[label=(-80+\cmdKV@general@rotate):\cmdKV@gLabels@iLB] {} (0.15,0.45);
    \draw[label distance=0,\cmdKV@gTypes@iC] (0.15,0.45) -- node[label=(45+\cmdKV@general@rotate):\cmdKV@gLabels@iLC] {} (-0.18,0.69);
    \draw[label distance=-0.5,\cmdKV@gTypes@iD] (-0.18,0.69) .. controls ($(-0.18,0.69) + (-0.12,-0.12)$) and ($(-0.35,0.87) + (-0.12,-0.12)$) .. (-0.35,0.87) node[label=(180+\cmdKV@general@rotate):\cmdKV@gLabels@iLD] {} .. controls ($(-0.35,0.87) + (0.12,0.12)$) and ($(-0.18,0.69) + (0.12,0.12)$) .. (-0.18,0.69);
    #2
  \end{tikzpicture}
  \gSetStdTypes{line}
}

\def\gTadB{\@ifnextchar[\@gTadB{\@gTadB[]}}
\def\@gTadB[#1]#2{%
  \setkeys*{pre}{#1}%
  \setrmkeys{general,gTypes,gLabels}%
  \begin{tikzpicture}
    [line width=1pt,
    baseline={([yshift=-0.5ex+\cmdKV@general@yshift]current bounding box.center)},
    scale=\cmdKV@general@scale,
    rotate=\cmdKV@general@rotate,
    font=\cmdKV@general@font]
    \draw[\cmdKV@gTypes@eA] (0,0.3) -- (-0.2,0) node[label=(-180+\cmdKV@general@rotate):\cmdKV@gLabels@eLA] {};
    \draw[\cmdKV@gTypes@eB] (0,0.3) -- (-0.2,0.6) node[label=(180+\cmdKV@general@rotate):\cmdKV@gLabels@eLB] {};
    \draw[\cmdKV@gTypes@eC] (0.7,0.3) -- (0.9,0.6) node[label=(\cmdKV@general@rotate):\cmdKV@gLabels@eLC] {};
    \draw[\cmdKV@gTypes@eD] (0.7,0.3) -- (0.9,0) node[label=(\cmdKV@general@rotate):\cmdKV@gLabels@eLD] {};
    \draw[label distance=0,\cmdKV@gTypes@iA] (0,0.3) -- node[label=(-90+\cmdKV@general@rotate):\cmdKV@gLabels@iLA] {} (0.35,0.3);
    \draw[label distance=0,\cmdKV@gTypes@iB] (0.35,0.3) -- node[label=(\cmdKV@general@rotate):\cmdKV@gLabels@iLB] {} (0.35,0.6);
    \draw[label distance=-0.5,\cmdKV@gTypes@iC] (0.35,0.6) .. controls (0.18,0.6) and (0.18,0.85) .. (0.35,0.85) .. controls (0.52,0.85) and (0.52,0.6) .. node[label=(\cmdKV@general@rotate):\cmdKV@gLabels@iLC] {} (0.35,0.6);
    \draw[label distance=0,\cmdKV@gTypes@iD] (0.35,0.3) -- node[label=(-90+\cmdKV@general@rotate):\cmdKV@gLabels@iLD] {} (0.7,0.3);
    #2
  \end{tikzpicture}
  \gSetStdTypes{line}
}
\def\gBubA{\@ifnextchar[\@gBubA{\@gBubA[]}}
\def\@gBubA[#1]#2{%
  \setkeys*{pre}{#1}%
  \setrmkeys{general,gTypes,gLabels}%
  \begin{tikzpicture}
    [line width=1pt,
    baseline={([yshift=-0.5ex+\cmdKV@general@yshift]current bounding box.center)},
    scale=\cmdKV@general@scale,
    rotate=\cmdKV@general@rotate,
    font=\cmdKV@general@font]
    \draw[\cmdKV@gTypes@eA] (0.5,0.7) -- (0.2,0.7) node[label=(180+\cmdKV@general@rotate):\cmdKV@gLabels@eLA] {};
    \draw[\cmdKV@gTypes@eB] (0.9,0.7) -- (1.2,0.7) node[label=(\cmdKV@general@rotate):\cmdKV@gLabels@eLB] {};
    \draw[label distance=0,\cmdKV@gTypes@iA] (0.5,0.7) .. controls (0.5,0.95) and (0.9,0.95) .. node[label=(90+\cmdKV@general@rotate):\cmdKV@gLabels@iLA] {} (0.9,0.7);
    \draw[label distance=0,\cmdKV@gTypes@iB] (0.9,0.7) .. controls (0.9,0.45) and (0.5,0.45) .. node[label=(-90+\cmdKV@general@rotate):\cmdKV@gLabels@iLB] {} (0.5,0.7);
    #2
  \end{tikzpicture}
  \gSetStdTypes{line}
}
\def\gBubB{\@ifnextchar[\@gBubB{\@gBubB[]}}
\def\@gBubB[#1]#2{%
  \setkeys*{pre}{#1}%
  \setrmkeys{general,gTypes,gLabels}%
  \begin{tikzpicture}
    [line width=1pt,
    baseline={([yshift=-0.5ex+\cmdKV@general@yshift]current bounding box.center)},
    scale=\cmdKV@general@scale,
    rotate=\cmdKV@general@rotate,
    font=\cmdKV@general@font]
    \draw[\cmdKV@gTypes@eA] (0.2,0.7) -- (0,0.4) node[label=(-180+\cmdKV@general@rotate):\cmdKV@gLabels@eLA] {};
    \draw[\cmdKV@gTypes@eB] (0.2,0.7) -- (0,1) node[label=(180+\cmdKV@general@rotate):\cmdKV@gLabels@eLB] {};
    \draw[\cmdKV@gTypes@eC] (1.2,0.7) -- (1.4,1) node[label=(\cmdKV@general@rotate):\cmdKV@gLabels@eLC] {};
    \draw[\cmdKV@gTypes@eD] (1.2,0.7) -- (1.4,0.4) node[label=(\cmdKV@general@rotate):\cmdKV@gLabels@eLD] {};
    \draw[label distance=0,\cmdKV@gTypes@iA] (0.2,0.7) -- node[label=(90+\cmdKV@general@rotate):\cmdKV@gLabels@iLA] {} (0.5,0.7);
    \draw[label distance=0,\cmdKV@gTypes@iB] (0.5,0.7) .. controls (0.5,0.95) and (0.9,0.95) .. node[label=(90+\cmdKV@general@rotate):\cmdKV@gLabels@iLB] {} (0.9,0.7);
    \draw[label distance=0,\cmdKV@gTypes@iC] (0.9,0.7) .. controls (0.9,0.45) and (0.5,0.45) .. node[label=(-90+\cmdKV@general@rotate):\cmdKV@gLabels@iLC] {} (0.5,0.7);
    \draw[label distance=0,\cmdKV@gTypes@iD] (0.9,0.7) -- node[label=(90+\cmdKV@general@rotate):\cmdKV@gLabels@iLD] {} (1.2,0.7);
    #2
  \end{tikzpicture}
  \gSetStdTypes{line}
}

\def\gBubC{\@ifnextchar[\@gBubC{\@gBubC[]}}
\def\@gBubC[#1]#2{%
  \setkeys*{pre}{#1}%
  \setrmkeys{general,gTypes,gLabels}%
  \begin{tikzpicture}
    [line width=1pt,
    baseline={([yshift=-0.5ex+\cmdKV@general@yshift]current bounding box.center)},
    scale=\cmdKV@general@scale,
    rotate=\cmdKV@general@rotate,
    font=\cmdKV@general@font]
    \draw[\cmdKV@gTypes@eA] (-0.7,0.3) -- (-0.9,0) node[label=(180+\cmdKV@general@rotate):\cmdKV@gLabels@eLA] {};
    \draw[\cmdKV@gTypes@eB] (-0.7,0.3) -- (-0.9,0.6) node[label=(180+\cmdKV@general@rotate):\cmdKV@gLabels@eLB] {};
    \draw[\cmdKV@gTypes@eC] (0.3,0.5) -- (0.5,0.6) node[label=(\cmdKV@general@rotate):\cmdKV@gLabels@eLC] {};
    \draw[\cmdKV@gTypes@eD] (-0.3,0.3) -- (-0.2,0) node[label=(\cmdKV@general@rotate):\cmdKV@gLabels@eLD] {};
    \draw[label distance=0,\cmdKV@gTypes@iA] (-0.7,0.3) -- node[label=(90+\cmdKV@general@rotate):\cmdKV@gLabels@iLA] {} (-0.3,0.3);
    \draw[label distance=0,\cmdKV@gTypes@iB] (-0.3,0.3) -- node[label=(90+\cmdKV@general@rotate):\cmdKV@gLabels@iLB] {} (-0,0.4);
    \draw[label distance=0,\cmdKV@gTypes@iC] (0,0.4) .. controls (-0.03,0.6) and (0.22,0.65) .. node[label=(90+\cmdKV@general@rotate):\cmdKV@gLabels@iLC] {} (0.3,0.5);
    \draw[label distance=0,\cmdKV@gTypes@iD] (0.3,0.5) .. controls (0.38,0.3) and (0.08,0.21) .. node[label=(-90+\cmdKV@general@rotate):\cmdKV@gLabels@iLD] {} (0,0.4);
    #2
  \end{tikzpicture}
  \gSetStdTypes{line}
}
\def\gTriA{\@ifnextchar[\@gTriA{\@gTriA[]}}
\def\@gTriA[#1]#2{%
  \setkeys*{pre}{#1}%
  \setrmkeys{general,gTypes,gLabels}%
  \begin{tikzpicture}
    [line width=1pt,
    baseline={([yshift=-0.5ex+\cmdKV@general@yshift]current bounding box.center)},
    scale=\cmdKV@general@scale,
    rotate=\cmdKV@general@rotate,
    font=\cmdKV@general@font]
    \draw[\cmdKV@gTypes@eC] (0:0.3) -- (0:0.55) node[label=(\cmdKV@general@rotate):\cmdKV@gLabels@eLC] {};
    \draw[\cmdKV@gTypes@eA] (-120:0.3) -- (-120:0.55) node[label=(-180+\cmdKV@general@rotate):\cmdKV@gLabels@eLA] {};
    \draw[\cmdKV@gTypes@eB] (120:0.3) -- (120:0.55) node[label=(180+\cmdKV@general@rotate):\cmdKV@gLabels@eLB] {};    
    \draw[label distance=0,\cmdKV@gTypes@iA] (-120:0.3) -- node[label=(180+\cmdKV@general@rotate):\cmdKV@gLabels@iLA] {} (120:0.3);
    \draw[label distance=0,\cmdKV@gTypes@iB] (120:0.3) -- node[label=(60+\cmdKV@general@rotate):\cmdKV@gLabels@iLB] {} (0:0.3);
    \draw[label distance=0,\cmdKV@gTypes@iC] (0:0.3) -- node[label=(-60+\cmdKV@general@rotate):\cmdKV@gLabels@iLC] {} (-120:0.3);
    #2
  \end{tikzpicture}
  \gSetStdTypes{line}
}
\def\gTriB{\@ifnextchar[\@gTriB{\@gTriB[]}}
\def\@gTriB[#1]#2{%
  \setkeys*{pre}{#1}%
  \setrmkeys{general,gTypes,gLabels}%
  \begin{tikzpicture}
    [line width=1pt,
    baseline={([yshift=-0.5ex+\cmdKV@general@yshift]current bounding box.center)},
    scale=\cmdKV@general@scale,
    rotate=\cmdKV@general@rotate,
    font=\cmdKV@general@font]
    \draw[\cmdKV@gTypes@eA] (0.3,0.45) -- (0,0.3) node[label=(-180+\cmdKV@general@rotate):\cmdKV@gLabels@eLA] {};
    \draw[\cmdKV@gTypes@eB] (0.3,0.95) -- (0,1.1) node[label=(180+\cmdKV@general@rotate):\cmdKV@gLabels@eLB] {};
    \draw[\cmdKV@gTypes@eC] (1,0.7) -- (1.3,1.1) node[label=(\cmdKV@general@rotate):\cmdKV@gLabels@eLC] {};
    \draw[\cmdKV@gTypes@eD] (1,0.7) -- (1.3,0.3) node[label=(\cmdKV@general@rotate):\cmdKV@gLabels@eLD] {};
    \draw[label distance=0,\cmdKV@gTypes@iA] (0.3,0.45) -- node[label=(180+\cmdKV@general@rotate):\cmdKV@gLabels@iLA] {} (0.3,0.95);
    \draw[label distance=0,\cmdKV@gTypes@iB] (0.3,0.95) -- node[label=(70+\cmdKV@general@rotate):\cmdKV@gLabels@iLB] {} (0.7,0.7);
    \draw[label distance=0,\cmdKV@gTypes@iC] (0.7,0.7) -- node[label=(-70+\cmdKV@general@rotate):\cmdKV@gLabels@iLC] {} (0.3,0.45);
    \draw[label distance=0,\cmdKV@gTypes@iD] (1,0.7) -- node[label=(90+\cmdKV@general@rotate):\cmdKV@gLabels@iLD] {} (0.7,0.7);
    #2
  \end{tikzpicture}
  \gSetStdTypes{line}
}
\def\gBox{\@ifnextchar[\@gBox{\@gBox[]}}
\def\@gBox[#1]#2{%
  \setkeys*{pre}{#1}%
  \setrmkeys{general,gTypes,gLabels}%
  \begin{tikzpicture}
    [line width=1pt,
    baseline={([yshift=-0.5ex+\cmdKV@general@yshift]current bounding box.center)},
    scale=\cmdKV@general@scale,
    rotate=\cmdKV@general@rotate,
    font=\cmdKV@general@font]
    \draw[\cmdKV@gTypes@eA] (0.25,0.25) -- (0,0) node[label=(-180+\cmdKV@general@rotate):\cmdKV@gLabels@eLA] {};
    \draw[\cmdKV@gTypes@eB] (0.25,0.75) -- (0,1) node[label=(180+\cmdKV@general@rotate):\cmdKV@gLabels@eLB] {};
    \draw[\cmdKV@gTypes@eC] (0.75,0.75) -- (1,1) node[label=(\cmdKV@general@rotate):\cmdKV@gLabels@eLC] {};
    \draw[\cmdKV@gTypes@eD] (0.75,0.25) -- (1,0) node[label=(\cmdKV@general@rotate):\cmdKV@gLabels@eLD] {};
    \draw[label distance=0,\cmdKV@gTypes@iA] (0.25,0.25) -- node[label=(180+\cmdKV@general@rotate):\cmdKV@gLabels@iLA] {} (0.25,0.75);
    \draw[label distance=0,\cmdKV@gTypes@iB] (0.25,0.75) -- node[label=(90+\cmdKV@general@rotate):\cmdKV@gLabels@iLB] {} (0.75,0.75);
    \draw[label distance=0,\cmdKV@gTypes@iC] (0.75,0.75) -- node[label=(\cmdKV@general@rotate):\cmdKV@gLabels@iLC] {} (0.75,0.25);
    \draw[label distance=0,\cmdKV@gTypes@iD] (0.75,0.25) -- node[label=(-90+\cmdKV@general@rotate):\cmdKV@gLabels@iLD] {} (0.25,0.25);
    #2
  \end{tikzpicture}
  \gSetStdTypes{line}
}
\def\gPenta{\@ifnextchar[\@gPenta{\@gPenta[]}}
\def\@gPenta[#1]#2{%
  \setkeys*{pre}{#1}%
  \setrmkeys{general,gTypes,gLabels}%
  \begin{tikzpicture}
    [line width=1pt,
    baseline={([yshift=-0.5ex+\cmdKV@general@yshift]current bounding box.center)},
    scale=\cmdKV@general@scale,
    rotate=\cmdKV@general@rotate,
    font=\cmdKV@general@font]
    \draw[\cmdKV@gTypes@eA] (-144:0.35) -- (-144:0.75) node[label=(-144+\cmdKV@general@rotate):\cmdKV@gLabels@eLA] {};
    \draw[\cmdKV@gTypes@eB] (144:0.35) -- (144:0.75) node[label=(144+\cmdKV@general@rotate):\cmdKV@gLabels@eLB] {};
    \draw[\cmdKV@gTypes@eC] (72:0.35) -- (72:0.75) node[label=(72+\cmdKV@general@rotate):\cmdKV@gLabels@eLC] {};
    \draw[\cmdKV@gTypes@eD] (0:0.35) -- (0:0.75) node[label=(\cmdKV@general@rotate):\cmdKV@gLabels@eLD] {};
    \draw[\cmdKV@gTypes@eE] (-72:0.35) -- (-72:0.75) node[label=(-72+\cmdKV@general@rotate):\cmdKV@gLabels@eLE] {};
    \draw[label distance=0,\cmdKV@gTypes@iA] (-144:0.35) -- node[label=(180+\cmdKV@general@rotate):\cmdKV@gLabels@iLA] {} (144:0.35);
    \draw[label distance=0,\cmdKV@gTypes@iB] (144:0.35) -- node[label=(108+\cmdKV@general@rotate):\cmdKV@gLabels@iLB] {} (72:0.35);
    \draw[label distance=0,\cmdKV@gTypes@iC] (72:0.35) -- node[label=(36+\cmdKV@general@rotate):\cmdKV@gLabels@iLC] {} (0:0.35);
    \draw[label distance=0,\cmdKV@gTypes@iD] (0:0.35) -- node[label=(-36+\cmdKV@general@rotate):\cmdKV@gLabels@iLD] {} (-72:0.35);
    \draw[label distance=0,\cmdKV@gTypes@iE] (-72:0.35) -- node[label=(-108+\cmdKV@general@rotate):\cmdKV@gLabels@iLE] {} (-144:0.35);
    #2
  \end{tikzpicture}
  \gSetStdTypes{line}
}

\def\gBoxBox{\@ifnextchar[\@gBoxBox{\@gBoxBox[]}}
\def\@gBoxBox[#1]#2{%
  \setkeys*{pre}{#1}%
  \setrmkeys{general,gTypes,gLabels}%
  \begin{tikzpicture}
    [line width=1pt,
    baseline={([yshift=-0.5ex+\cmdKV@general@yshift]current bounding box.center)},
    scale=\cmdKV@general@scale,
    rotate=\cmdKV@general@rotate,
    font=\cmdKV@general@font]
    \draw[\cmdKV@gTypes@eA] (0.3,0.3) -- (0,0) node[label=(-180+\cmdKV@general@rotate):\cmdKV@gLabels@eLA] {};
    \draw[\cmdKV@gTypes@eB] (0.3,0.9) -- (0,1.2) node[label=(180+\cmdKV@general@rotate):\cmdKV@gLabels@eLB] {};
    \draw[\cmdKV@gTypes@eC] (1.5,0.9) -- (1.8,1.2) node[label=(\cmdKV@general@rotate):\cmdKV@gLabels@eLC] {};
    \draw[\cmdKV@gTypes@eD] (1.5,0.3) -- (1.8,0) node[label=(\cmdKV@general@rotate):\cmdKV@gLabels@eLD] {};
    \draw[label distance=0,\cmdKV@gTypes@iA] (0.3,0.3) -- node[label=(180+\cmdKV@general@rotate):\cmdKV@gLabels@iLA] {} (0.3,0.9);
    \draw[label distance=0,\cmdKV@gTypes@iB] (0.3,0.9) -- node[label=(90+\cmdKV@general@rotate):\cmdKV@gLabels@iLB] {} (0.9,0.9);
    \draw[label distance=0,\cmdKV@gTypes@iC] (0.9,0.9) -- node[label=(90+\cmdKV@general@rotate):\cmdKV@gLabels@iLC] {} (1.5,0.9);
    \draw[label distance=0,\cmdKV@gTypes@iD] (1.5,0.9) -- node[label=(\cmdKV@general@rotate):\cmdKV@gLabels@iLD] {} (1.5,0.3);
    \draw[label distance=0,\cmdKV@gTypes@iE] (1.5,0.3) -- node[label=(-90+\cmdKV@general@rotate):\cmdKV@gLabels@iLE] {} (0.9,0.3);
    \draw[label distance=0,\cmdKV@gTypes@iF] (0.9,0.3) -- node[label=(-90+\cmdKV@general@rotate):\cmdKV@gLabels@iLF] {} (0.3,0.3);
    \draw[label distance=0,\cmdKV@gTypes@iG] (0.9,0.9) -- node[label=(\cmdKV@general@rotate):\cmdKV@gLabels@iLG] {} (0.9,0.3);
    #2
  \end{tikzpicture}
  \gSetStdTypes{line}
}
\def\gBoxBoxNP{\@ifnextchar[\@gBoxBoxNP{\@gBoxBoxNP[]}}
\def\@gBoxBoxNP[#1]#2{%
  \setkeys*{pre}{#1}%
  \setrmkeys{general,gTypes,gLabels}%
  \begin{tikzpicture}
    [line width=1pt,
    baseline={([yshift=-0.5ex+\cmdKV@general@yshift]current bounding box.center)},
    scale=\cmdKV@general@scale,
    rotate=\cmdKV@general@rotate,
    font=\cmdKV@general@font]
    \draw[\cmdKV@gTypes@eA] (-1.9,0.3) -- (-2.2,0) node[label=(180+\cmdKV@general@rotate):\cmdKV@gLabels@eLA] {};
    \draw[\cmdKV@gTypes@eB] (-1.9,1.1) -- (-2.2,1.4) node[label=(180+\cmdKV@general@rotate):\cmdKV@gLabels@eLB] {};
    \draw[\cmdKV@gTypes@eC] (-0.3,0.7) -- (0,0.7) node[label=(\cmdKV@general@rotate):\cmdKV@gLabels@eLC] {};
    \draw[\cmdKV@gTypes@eD] (-1.1,0.7) -- (-1.4,0.7) node[label=(180+\cmdKV@general@rotate):\cmdKV@gLabels@eLD] {};
    \draw[label distance=0,\cmdKV@gTypes@iA] (-1.9,0.3) -- node[label=(180+\cmdKV@general@rotate):\cmdKV@gLabels@iLA] {} (-1.9,1.1);
    \draw[label distance=0,\cmdKV@gTypes@iB] (-1.9,1.1) -- node[label=(90+\cmdKV@general@rotate):\cmdKV@gLabels@iLB] {} (-0.7,1.1);
    \draw[label distance=0,\cmdKV@gTypes@iC] (-0.7,1.1) -- node[label=(45+\cmdKV@general@rotate):\cmdKV@gLabels@iLC] {} (-0.3,0.7);
    \draw[label distance=0,\cmdKV@gTypes@iD] (-0.3,0.7) -- node[label=(-45+\cmdKV@general@rotate):\cmdKV@gLabels@iLD] {} (-0.7,0.3);
    \draw[label distance=0,\cmdKV@gTypes@iE] (-0.7,0.3) -- node[label=(-90+\cmdKV@general@rotate):\cmdKV@gLabels@iLE] {} (-1.9,0.3);
    \draw[label distance=0,\cmdKV@gTypes@iF] (-0.7,1.1) -- node[label=(180+\cmdKV@general@rotate):\cmdKV@gLabels@iLF] {} (-1.1,0.7);
    \draw[label distance=0,\cmdKV@gTypes@iG] (-1.1,0.7) -- node[label=(-180+\cmdKV@general@rotate):\cmdKV@gLabels@iLG] {} (-0.7,0.3);
    #2
  \end{tikzpicture}
  \gSetStdTypes{line}
}

\def\gBoxBoxNPB{\@ifnextchar[\@gBoxBoxNPB{\@gBoxBoxNPB[]}}
\def\@gBoxBoxNPB[#1]#2{%
  \setkeys*{pre}{#1}%
  \setrmkeys{general,gTypes,gLabels}%
  \begin{tikzpicture}
    [line width=1pt,
    baseline={([yshift=-0.5ex+\cmdKV@general@yshift]current bounding box.center)},
    scale=\cmdKV@general@scale,
    rotate=\cmdKV@general@rotate,
    font=\cmdKV@general@font]
    \draw[\cmdKV@gTypes@eA] (0.3,0.3) -- (0,0) node[label=(-180+\cmdKV@general@rotate):\cmdKV@gLabels@eLA] {};
    \draw[\cmdKV@gTypes@eB] (0.3,0.9) -- (0,1.2) node[label=(180+\cmdKV@general@rotate):\cmdKV@gLabels@eLB] {};
    \draw[\cmdKV@gTypes@eC] (1.5,0.9) -- (1.8,1.2) node[label=(\cmdKV@general@rotate):\cmdKV@gLabels@eLC] {};
    \draw[\cmdKV@gTypes@eD] (1.5,0.3) -- (1.8,0) node[label=(\cmdKV@general@rotate):\cmdKV@gLabels@eLD] {};
    \draw[label distance=0,\cmdKV@gTypes@iA] (0.3,0.3) -- node[label=(180+\cmdKV@general@rotate):\cmdKV@gLabels@iLA] {} (0.3,0.9);
    \draw[label distance=0,\cmdKV@gTypes@iB] (0.3,0.9) -- node[label=(90+\cmdKV@general@rotate):\cmdKV@gLabels@iLB] {} (0.9,0.9);
    \draw[label distance=0,\cmdKV@gTypes@iC] (0.9,0.9) -- node[label=(90+\cmdKV@general@rotate):\cmdKV@gLabels@iLC] {} (1.5,0.3);
    \foreach \x in {5,...,14} \fill[opacity=0.12,white] (1.2,0.6) circle (\x/100);
    \draw[label distance=0,\cmdKV@gTypes@iD] (1.5,0.3) -- node[label=(\cmdKV@general@rotate):\cmdKV@gLabels@iLD] {} (1.5,0.9);
    \draw[label distance=0,\cmdKV@gTypes@iE] (1.5,0.9) -- node[label=(-90+\cmdKV@general@rotate):\cmdKV@gLabels@iLE] {} (0.9,0.3);
    \draw[label distance=0,\cmdKV@gTypes@iF] (0.9,0.3) -- node[label=(-90+\cmdKV@general@rotate):\cmdKV@gLabels@iLF] {} (0.3,0.3);
    \draw[label distance=-1,\cmdKV@gTypes@iG] (0.9,0.9) -- node[label=(180+\cmdKV@general@rotate):\cmdKV@gLabels@iLG] {} (0.9,0.3);
    #2
  \end{tikzpicture}
  \gSetStdTypes{line}
}

\def\gBoxBoxNPC{\@ifnextchar[\@gBoxBoxNPC{\@gBoxBoxNPC[]}}
\def\@gBoxBoxNPC[#1]#2{%
  \setkeys*{pre}{#1}%
  \setrmkeys{general,gTypes,gLabels}%
  \begin{tikzpicture}
    [line width=1pt,
    baseline={([yshift=-0.5ex+\cmdKV@general@yshift]current bounding box.center)},
    scale=\cmdKV@general@scale,
    rotate=\cmdKV@general@rotate,
    font=\cmdKV@general@font]
    \draw[\cmdKV@gTypes@eA] (0.3,0.3) -- (0,0) node[label=(-180+\cmdKV@general@rotate):\cmdKV@gLabels@eLA] {};
    \draw[\cmdKV@gTypes@eB] (0.3,0.9) -- (0,1.2) node[label=(180+\cmdKV@general@rotate):\cmdKV@gLabels@eLB] {};
    \draw[\cmdKV@gTypes@eC] (1.5,0.9) -- (1.8,1.2) node[label=(\cmdKV@general@rotate):\cmdKV@gLabels@eLC] {};
    \draw[\cmdKV@gTypes@eD] (1.5,0.3) -- (1.8,0) node[label=(\cmdKV@general@rotate):\cmdKV@gLabels@eLD] {};
    \draw[label distance=0,\cmdKV@gTypes@iA] (0.3,0.3) -- node[label=(180+\cmdKV@general@rotate):\cmdKV@gLabels@iLA] {} (0.3,0.9);
    \draw[label distance=0,\cmdKV@gTypes@iB] (0.3,0.9) -- node[label=(90+\cmdKV@general@rotate):\cmdKV@gLabels@iLB] {} (0.9,0.9);
    \draw[label distance=0,\cmdKV@gTypes@iC] (0.9,0.9) -- node[label=(90+\cmdKV@general@rotate):\cmdKV@gLabels@iLC] {} (1.5,0.9);
    \draw[label distance=0,\cmdKV@gTypes@iD] (1.5,0.9) -- node[label=(\cmdKV@general@rotate):\cmdKV@gLabels@iLD] {} (0.9,0.3);
    \foreach \x in {5,...,14} \fill[opacity=0.12,white] (1.2,0.6) circle (\x/100);
    \draw[label distance=0,\cmdKV@gTypes@iE] (1.5,0.3) -- node[label=(-90+\cmdKV@general@rotate):\cmdKV@gLabels@iLE] {} (0.9,0.3);
    \draw[label distance=0,\cmdKV@gTypes@iF] (0.9,0.3) -- node[label=(-90+\cmdKV@general@rotate):\cmdKV@gLabels@iLF] {} (0.3,0.3);
    \draw[label distance=-1,\cmdKV@gTypes@iG] (0.9,0.9) -- node[label=(180+\cmdKV@general@rotate):\cmdKV@gLabels@iLG] {} (1.5,0.3);
    #2
  \end{tikzpicture}
  \gSetStdTypes{line}
}

\def\gBoxTriNPA{\@ifnextchar[\@gBoxTriNPA{\@gBoxTriNPA[]}}
\def\@gBoxTriNPA[#1]#2{%
  \setkeys*{pre}{#1}%
  \setrmkeys{general,gTypes,gLabels}%
  \begin{tikzpicture}
    [line width=1pt,
    baseline={([yshift=-0.5ex+\cmdKV@general@yshift]current bounding box.center)},
    scale=\cmdKV@general@scale,
    rotate=\cmdKV@general@rotate,
    font=\cmdKV@general@font]
    \draw[\cmdKV@gTypes@eA] (0.3,0.6) -- (0,0) node[label=(-180+\cmdKV@general@rotate):\cmdKV@gLabels@eLA] {};
    \draw[\cmdKV@gTypes@eB] (0.3,0.6) -- (0,1.2) node[label=(180+\cmdKV@general@rotate):\cmdKV@gLabels@eLB] {};
    \draw[\cmdKV@gTypes@eC] (1.5,0.9) -- (1.8,1.2) node[label=(\cmdKV@general@rotate):\cmdKV@gLabels@eLC] {};
    \draw[\cmdKV@gTypes@eD] (1.5,0.3) -- (1.8,0) node[label=(\cmdKV@general@rotate):\cmdKV@gLabels@eLD] {};
    \draw[label distance=0,\cmdKV@gTypes@iA] (0.3,0.6) -- node[label=(90+\cmdKV@general@rotate):\cmdKV@gLabels@iLA] {} (0.9,0.9);
    \draw[label distance=0,\cmdKV@gTypes@iB] (0.9,0.9) -- node[label=(90+\cmdKV@general@rotate):\cmdKV@gLabels@iLB] {} (1.5,0.9);
    \draw[label distance=-1,\cmdKV@gTypes@iC] (0.9,0.9) -- node[label=(180+\cmdKV@general@rotate):\cmdKV@gLabels@iLC] {} (1.5,0.3);
    \foreach \x in {5,...,14} \fill[opacity=0.12,white] (1.2,0.6) circle (\x/100);
    \draw[label distance=0,\cmdKV@gTypes@iD] (1.5,0.3) -- node[label=(-90+\cmdKV@general@rotate):\cmdKV@gLabels@iLD] {} (0.9,0.3);
    \draw[label distance=0,\cmdKV@gTypes@iE] (0.9,0.3) -- node[label=(-90+\cmdKV@general@rotate):\cmdKV@gLabels@iLE] {} (0.3,0.6);
    \draw[label distance=-1,\cmdKV@gTypes@iF] (1.5,0.9) -- node[label=(180+\cmdKV@general@rotate):\cmdKV@gLabels@iLF] {} (0.9,0.3);
    #2
  \end{tikzpicture}
  \gSetStdTypes{line}
}

\def\gTriPenta{\@ifnextchar[\@gTriPenta{\@gTriPenta[]}}
\def\@gTriPenta[#1]#2{%
  \setkeys*{pre}{#1}%
  \setrmkeys{general,gTypes,gLabels}%
  \begin{tikzpicture}
    [line width=1pt,
    baseline={([yshift=-0.5ex+\cmdKV@general@yshift]current bounding box.center)},
    scale=\cmdKV@general@scale,
    rotate=\cmdKV@general@rotate,
    font=\cmdKV@general@font]
    \draw[\cmdKV@gTypes@eA] (-108:0.5) -- (-135:1) node[label=(145+\cmdKV@general@rotate):\cmdKV@gLabels@eLA] {};
    \draw[\cmdKV@gTypes@eB] (180:0.5) -- (180:1) node[label=(180+\cmdKV@general@rotate):\cmdKV@gLabels@eLB] {};
    \draw[\cmdKV@gTypes@eC] (108:0.5) -- (135:1) node[label=(-145+\cmdKV@general@rotate):\cmdKV@gLabels@eLC] {};
    \draw[\cmdKV@gTypes@eD] (0:1.2) -- (0:1.5) node[label=(\cmdKV@general@rotate):\cmdKV@gLabels@eLD] {};
    \draw[label distance=0,\cmdKV@gTypes@iA] (-108:0.5) -- node[label=(-135+\cmdKV@general@rotate):\cmdKV@gLabels@iLA] {} (-180:0.5);
    \draw[label distance=0,\cmdKV@gTypes@iB] (180:0.5) -- node[label=(135+\cmdKV@general@rotate):\cmdKV@gLabels@iLB] {} (108:0.5);
    \draw[label distance=0,\cmdKV@gTypes@iC] (108:0.5) -- node[label=(72+\cmdKV@general@rotate):\cmdKV@gLabels@iLC] {} (36:0.5);
    \draw[label distance=0,\cmdKV@gTypes@iD] (36:0.5) -- node[label=(72+\cmdKV@general@rotate):\cmdKV@gLabels@iLD] {} (0:1.2);
    \draw[label distance=2,\cmdKV@gTypes@iE] (0:1.2) -- node[label=(-90+\cmdKV@general@rotate):\cmdKV@gLabels@iLE] {} (-36:0.5);
    \draw[label distance=2,\cmdKV@gTypes@iF] (-36:0.5) -- node[label=(-90+\cmdKV@general@rotate):\cmdKV@gLabels@iLF] {} (-108:0.5);
    \draw[label distance=0,\cmdKV@gTypes@iG] (36:0.5) -- node[label=(180+\cmdKV@general@rotate):\cmdKV@gLabels@iLG] {} (-36:0.5);
    #2
  \end{tikzpicture}
  \gSetStdTypes{line}
}
\def\gTriTri{\@ifnextchar[\@gTriTri{\@gTriTri[]}}
\def\@gTriTri[#1]#2{%
  \setkeys*{pre}{#1}%
  \setrmkeys{general,gTypes,gLabels}%
  \begin{tikzpicture}
    [line width=1pt,
    baseline={([yshift=-0.5ex+\cmdKV@general@yshift]current bounding box.center)},
    scale=\cmdKV@general@scale,
    rotate=\cmdKV@general@rotate,
    font=\cmdKV@general@font]
    \draw[\cmdKV@gTypes@eA] (0.3,0.3) -- (0,0) node[label=(-180+\cmdKV@general@rotate):\cmdKV@gLabels@eLA] {};
    \draw[\cmdKV@gTypes@eB] (0.3,1.1) -- (0,1.4) node[label=(180+\cmdKV@general@rotate):\cmdKV@gLabels@eLB] {};
    \draw[\cmdKV@gTypes@eC] (1.9,1.1) -- (2.2,1.4) node[label=(\cmdKV@general@rotate):\cmdKV@gLabels@eLC] {};
    \draw[\cmdKV@gTypes@eD] (1.9,0.3) -- (2.2,0) node[label=(\cmdKV@general@rotate):\cmdKV@gLabels@eLD] {};
    \draw[label distance=0,\cmdKV@gTypes@iA] (0.3,0.3) -- node[label=(180+\cmdKV@general@rotate):\cmdKV@gLabels@iLA] {} (0.3,1.1);
    \draw[label distance=0,\cmdKV@gTypes@iB] (0.3,1.1) -- node[label=(45+\cmdKV@general@rotate):\cmdKV@gLabels@iLB] {} (0.9,0.7);
    \draw[label distance=0,\cmdKV@gTypes@iC] (0.9,0.7) -- node[label=(-45+\cmdKV@general@rotate):\cmdKV@gLabels@iLC] {} (0.3,0.3);
    \draw[label distance=0,\cmdKV@gTypes@iD] (0.9,0.7) -- node[label=(90+\cmdKV@general@rotate):\cmdKV@gLabels@iLD] {} (1.3,0.7);
    \draw[label distance=0,\cmdKV@gTypes@iE] (1.3,0.7) -- node[label=(135+\cmdKV@general@rotate):\cmdKV@gLabels@iLE] {} (1.9,1.1);
    \draw[label distance=0,\cmdKV@gTypes@iF] (1.9,1.1) -- node[label=(\cmdKV@general@rotate):\cmdKV@gLabels@iLF] {} (1.9,0.3);
    \draw[label distance=0,\cmdKV@gTypes@iG] (1.9,0.3) -- node[label=(-135+\cmdKV@general@rotate):\cmdKV@gLabels@iLG] {} (1.3,0.7);
    #2
  \end{tikzpicture}
  \gSetStdTypes{line}
}

\def\gBoxBubA{\@ifnextchar[\@gBoxBubA{\@gBoxBubA[]}}
\def\@gBoxBubA[#1]#2{%
  \setkeys*{pre}{#1}%
  \setrmkeys{general,gTypes,gLabels}%
  \begin{tikzpicture}
    [line width=1pt,
    baseline={([yshift=-0.5ex+\cmdKV@general@yshift]current bounding box.center)},
    scale=\cmdKV@general@scale,
    rotate=\cmdKV@general@rotate,
    font=\cmdKV@general@font]
    \draw[\cmdKV@gTypes@eA] (0.1,0.1) -- (-0.1,-0.1) node[label=(-180+\cmdKV@general@rotate):\cmdKV@gLabels@eLA] {};
    \draw[\cmdKV@gTypes@eB] (0.1,0.9) -- (-0.1,1.1) node[label=(180+\cmdKV@general@rotate):\cmdKV@gLabels@eLB] {};
    \draw[\cmdKV@gTypes@eC] (0.9,0.9) -- (1.1,1.1) node[label=(\cmdKV@general@rotate):\cmdKV@gLabels@eLC] {};
    \draw[\cmdKV@gTypes@eD] (0.9,0.1) -- (1.1,-0.1) node[label=(\cmdKV@general@rotate):\cmdKV@gLabels@eLD] {};
    \draw[label distance=0,\cmdKV@gTypes@iA] (0.1,0.1) -- node[label=(180+\cmdKV@general@rotate):\cmdKV@gLabels@iLA] {} (0.1,0.9);
    \draw[label distance=0,\cmdKV@gTypes@iB] (0.1,0.9) -- node[label=(90+\cmdKV@general@rotate):\cmdKV@gLabels@iLB] {} (0.32,0.9);
    \draw[label distance=0,\cmdKV@gTypes@iC] (0.32,0.9) .. controls (0.32,1.15) and (0.68,1.15) .. node[label=(90+\cmdKV@general@rotate):\cmdKV@gLabels@iLC] {} (0.68,0.9);
    \draw[label distance=0,\cmdKV@gTypes@iD] (0.68,0.9) .. controls (0.68,0.65) and (0.32,0.65) .. node[label=(-90+\cmdKV@general@rotate):\cmdKV@gLabels@iLD] {} (0.32,0.9);
    \draw[label distance=0,\cmdKV@gTypes@iE] (0.68,0.9) -- node[label=(90+\cmdKV@general@rotate):\cmdKV@gLabels@iLE] {} (0.9,0.9);
    \draw[label distance=0,\cmdKV@gTypes@iF] (0.9,0.9) -- node[label=(\cmdKV@general@rotate):\cmdKV@gLabels@iLF] {} (0.9,0.1);
    \draw[label distance=0,\cmdKV@gTypes@iG] (0.9,0.1) -- node[label=(-90+\cmdKV@general@rotate):\cmdKV@gLabels@iLG] {} (0.1,0.1);
    #2
  \end{tikzpicture}
  \gSetStdTypes{line}
}

\def\gBoxBubB{\@ifnextchar[\@gBoxBubB{\@gBoxBubB[]}}
\def\@gBoxBubB[#1]#2{%
  \setkeys*{pre}{#1}%
  \setrmkeys{general,gTypes,gLabels}%
  \begin{tikzpicture}
    [line width=1pt,
    baseline={([yshift=-0.5ex+\cmdKV@general@yshift]current bounding box.center)},
    scale=\cmdKV@general@scale,
    rotate=\cmdKV@general@rotate,
    font=\cmdKV@general@font]
    \draw[label distance=-1,\cmdKV@gTypes@eA] (0,-0.4) -- (0,-0.7) node[label=(180+\cmdKV@general@rotate):\cmdKV@gLabels@eLA] {};
    \draw[\cmdKV@gTypes@eB] (-0.4,0) -- (-0.7,0) node[label=(180+\cmdKV@general@rotate):\cmdKV@gLabels@eLB] {};
    \draw[label distance=-1,\cmdKV@gTypes@eC] (0,0.4) -- (0,0.7) node[label=(180+\cmdKV@general@rotate):\cmdKV@gLabels@eLC] {};
    \draw[\cmdKV@gTypes@eD] (1.1,0) -- (1.4,0) node[label=(\cmdKV@general@rotate):\cmdKV@gLabels@eLD] {};
    \draw[label distance=0,\cmdKV@gTypes@iA] (0,-0.4) -- node[label=(-135+\cmdKV@general@rotate):\cmdKV@gLabels@iLA] {} (-0.4,0);
    \draw[label distance=0,\cmdKV@gTypes@iB] (-0.4,0) -- node[label=(135+\cmdKV@general@rotate):\cmdKV@gLabels@iLB] {} (0,0.4);
    \draw[label distance=0,\cmdKV@gTypes@iC] (0,0.4) -- node[label=(45+\cmdKV@general@rotate):\cmdKV@gLabels@iLC] {} (0.4,0);
    \draw[label distance=0,\cmdKV@gTypes@iD] (0.4,0) -- node[label=(-45+\cmdKV@general@rotate):\cmdKV@gLabels@iLD] {} (0,-0.4);
    \draw[label distance=0,\cmdKV@gTypes@iE] (0.4,0) --  node[label=(90+\cmdKV@general@rotate):\cmdKV@gLabels@iLE] {} (0.7,0);
    \draw[label distance=0,\cmdKV@gTypes@iF] (0.7,0) .. controls (0.7,0.26) and (1.1,0.26) .. node[label=(90+\cmdKV@general@rotate):\cmdKV@gLabels@iLF] {} (1.1,0);
    \draw[label distance=0,\cmdKV@gTypes@iG] (1.1,0) .. controls (1.1,-0.26) and (0.7,-0.26) ..  node[label=(-90+\cmdKV@general@rotate):\cmdKV@gLabels@iLG] {} (0.7,0);
    #2
  \end{tikzpicture}
  \gSetStdTypes{line}
}

\def\gPentaTad{\@ifnextchar[\@gPentaTad{\@gPentaTad[]}}
\def\@gPentaTad[#1]#2{%
  \setkeys*{pre}{#1}%
  \setrmkeys{general,gTypes,gLabels}%
  \begin{tikzpicture}
    [line width=1pt,
    baseline={([yshift=-0.5ex+\cmdKV@general@yshift]current bounding box.center)},
    scale=\cmdKV@general@scale,
    rotate=\cmdKV@general@rotate,
    font=\cmdKV@general@font]
    \draw[label distance=-0.5,\cmdKV@gTypes@eA] (108:0.35) -- (108:0.75) node[label=(180+\cmdKV@general@rotate):\cmdKV@gLabels@eLA] {};
    \draw[\cmdKV@gTypes@eB] (36:0.35) -- (36:0.75) node[label=(36+\cmdKV@general@rotate):\cmdKV@gLabels@eLB] {};
    \draw[\cmdKV@gTypes@eC] (-36:0.35) -- (-36:0.75) node[label=(-36+\cmdKV@general@rotate):\cmdKV@gLabels@eLC] {};
    \draw[label distance=-0.5,\cmdKV@gTypes@eD] (-108:0.35) -- (-108:0.75) node[label=(-180+\cmdKV@general@rotate):\cmdKV@gLabels@eLD] {};
    \draw[label distance=0,\cmdKV@gTypes@iA] (108:0.35) -- node[label=(72+\cmdKV@general@rotate):\cmdKV@gLabels@iLA] {} (36:0.35);
    \draw[label distance=0,\cmdKV@gTypes@iB] (36:0.35) -- node[label=(\cmdKV@general@rotate):\cmdKV@gLabels@iLB] {} (-36:0.35);
    \draw[label distance=0,\cmdKV@gTypes@iC] (-36:0.35) -- node[label=(-72+\cmdKV@general@rotate):\cmdKV@gLabels@iLC] {} (-108:0.35);
    \draw[label distance=0,\cmdKV@gTypes@iD] (-108:0.35) -- node[label=(-144+\cmdKV@general@rotate):\cmdKV@gLabels@iLD] {} (-180:0.35);
    \draw[label distance=0,\cmdKV@gTypes@iE] (180:0.35) -- node[label=(144+\cmdKV@general@rotate):\cmdKV@gLabels@iLE] {} (108:0.35);
    \draw[label distance=0,\cmdKV@gTypes@iF] (180:0.35) -- node[label=(90+\cmdKV@general@rotate):\cmdKV@gLabels@iLF] {} (180:0.75);
    \draw[label distance=-0.5,\cmdKV@gTypes@iG] (180:0.75) .. controls ($(180:0.75)+(0,-0.18)$) and ($(180:1)+(0,-0.18)$) .. (180:1) node[label=(180+\cmdKV@general@rotate):\cmdKV@gLabels@iLG] {} .. controls ($(180:1)+(0,0.18)$) and ($(180:0.75)+(0,0.18)$) .. (180:0.75);
    #2
  \end{tikzpicture}
  \gSetStdTypes{line}
}

\def\gBoxTadA{\@ifnextchar[\@gBoxTadA{\@gBoxTadA[]}}
\def\@gBoxTadA[#1]#2{%
  \setkeys*{pre}{#1}%
  \setrmkeys{general,gTypes,gLabels}%
  \begin{tikzpicture}
    [line width=1pt,
    baseline={([yshift=-0.5ex+\cmdKV@general@yshift]current bounding box.center)},
    scale=\cmdKV@general@scale,
    rotate=\cmdKV@general@rotate,
    font=\cmdKV@general@font]
    \draw[label distance=-0.5,\cmdKV@gTypes@eA] (0,0.4) -- (0,0.7) node[label=(\cmdKV@general@rotate):\cmdKV@gLabels@eLA] {};
    \draw[\cmdKV@gTypes@eB] (0.4,0) -- (0.7,0) node[label=(\cmdKV@general@rotate):\cmdKV@gLabels@eLB] {};
    \draw[label distance=-0.5,\cmdKV@gTypes@eC] (0,-0.4) -- (0,-0.7) node[label=(\cmdKV@general@rotate):\cmdKV@gLabels@eLC] {};
    \draw[\cmdKV@gTypes@eD] (-0.7,0) -- (-0.7,-0.3) node[label=(-90+\cmdKV@general@rotate):\cmdKV@gLabels@eLD] {};
    \draw[label distance=0,\cmdKV@gTypes@iA] (0,0.4) -- node[label=(45+\cmdKV@general@rotate):\cmdKV@gLabels@iLA] {} (0.4,0);
    \draw[label distance=0,\cmdKV@gTypes@iB] (0.4,0) -- node[label=(-45+\cmdKV@general@rotate):\cmdKV@gLabels@iLB] {} (0,-0.4);
    \draw[label distance=0,\cmdKV@gTypes@iC] (0,-0.4) -- node[label=(-135+\cmdKV@general@rotate):\cmdKV@gLabels@iLC] {} (-0.4,0);
    \draw[label distance=0,\cmdKV@gTypes@iD] (-0.4,0) -- node[label=(135+\cmdKV@general@rotate):\cmdKV@gLabels@iLD] {} (0,0.4);
    \draw[label distance=0,\cmdKV@gTypes@iE] (-0.4,0) --  node[label=(90+\cmdKV@general@rotate):\cmdKV@gLabels@iLE] {} (-0.7,0);
    \draw[label distance=0,\cmdKV@gTypes@iF] (-0.7,0) -- node[label=(90+\cmdKV@general@rotate):\cmdKV@gLabels@iLF] {} (-1.1,0);
    \draw[label distance=-0.5,\cmdKV@gTypes@iG] (-1.1,0) .. controls (-1.1,-0.18) and (-1.35,-0.18) .. (-1.35,0) node[label=(180+\cmdKV@general@rotate):\cmdKV@gLabels@iLG] {} .. controls (-1.35,0.18) and (-1.1,0.18) .. (-1.1,0);
    #2
  \end{tikzpicture}
  \gSetStdTypes{line}
}

\def\gSunsetA{\@ifnextchar[\@gSunsetA{\@gSunsetA[]}}
\def\@gSunsetA[#1]#2{%
  \setkeys*{pre}{#1}%
  \setrmkeys{general,gTypes,gLabels}%
  \begin{tikzpicture}
    [line width=1pt,
    baseline={([yshift=-0.5ex+\cmdKV@general@yshift]current bounding box.center)},
    scale=\cmdKV@general@scale,
    rotate=\cmdKV@general@rotate,
    font=\cmdKV@general@font]
    \draw[\cmdKV@gTypes@eA] (0.3,0.6) -- (0,0) node[label=(-180+\cmdKV@general@rotate):\cmdKV@gLabels@eLA] {};
    \draw[\cmdKV@gTypes@eB] (0.3,0.6) -- (0,1.2) node[label=(180+\cmdKV@general@rotate):\cmdKV@gLabels@eLB] {};
    \draw[\cmdKV@gTypes@eC] (1.5,0.6) -- (1.8,1.2) node[label=(\cmdKV@general@rotate):\cmdKV@gLabels@eLC] {};
    \draw[\cmdKV@gTypes@eD] (1.5,0.6) -- (1.8,0) node[label=(\cmdKV@general@rotate):\cmdKV@gLabels@eLD] {};
    \draw[label distance=0,\cmdKV@gTypes@iA] (0.3,0.6) to[bend left=90] node[label=(90+\cmdKV@general@rotate):\cmdKV@gLabels@iLA] {} (1.5,0.6);
    \draw[label distance=0,\cmdKV@gTypes@iB] (1.5,0.6) to[bend left=90] node[label=(-90+\cmdKV@general@rotate):\cmdKV@gLabels@iLB] {} (0.3,0.6);
    \draw[label distance=-2,\cmdKV@gTypes@iC] (0.3,0.6) -- node[label=(90+\cmdKV@general@rotate):\cmdKV@gLabels@iLC] {} (1.5,0.6);
    #2
  \end{tikzpicture}
  \gSetStdTypes{line}
}

\def\gTriBubA{\@ifnextchar[\@gTriBubA{\@gTriBubA[]}}
\def\@gTriBubA[#1]#2{%
  \setkeys*{pre}{#1}%
  \setrmkeys{general,gTypes,gLabels}%
  \begin{tikzpicture}
    [line width=1pt,
    baseline={([yshift=-0.5ex+\cmdKV@general@yshift]current bounding box.center)},
    scale=\cmdKV@general@scale,
    rotate=\cmdKV@general@rotate,
    font=\cmdKV@general@font]
    \draw[\cmdKV@gTypes@eA] (0.4,0.3) -- (0,0) node[label=(-180+\cmdKV@general@rotate):\cmdKV@gLabels@eLA] {};
    \draw[\cmdKV@gTypes@eB] (0.4,0.9) -- (0,1.2) node[label=(180+\cmdKV@general@rotate):\cmdKV@gLabels@eLB] {};
    \draw[\cmdKV@gTypes@eC] (1.4,0.6) -- (1.8,1.2) node[label=(\cmdKV@general@rotate):\cmdKV@gLabels@eLC] {};
    \draw[\cmdKV@gTypes@eD] (1.4,0.6) -- (1.8,0) node[label=(\cmdKV@general@rotate):\cmdKV@gLabels@eLD] {};
    \draw[label distance=0,\cmdKV@gTypes@iA] (0.4,0.3) to[bend left=40] node[label=(180+\cmdKV@general@rotate):\cmdKV@gLabels@iLA] {} (0.4,0.9);
    \draw[label distance=0,\cmdKV@gTypes@iB] (0.4,0.9) -- node[label=(90+\cmdKV@general@rotate):\cmdKV@gLabels@iLB] {} (1.4,0.6);
    \draw[label distance=0,\cmdKV@gTypes@iC] (1.4,0.6) -- node[label=(-90+\cmdKV@general@rotate):\cmdKV@gLabels@iLC] {} (0.4,0.3);
    \draw[label distance=-1,\cmdKV@gTypes@iD] (0.4,0.9) to[bend left=40] node[label=(\cmdKV@general@rotate):\cmdKV@gLabels@iLD] {} (0.4,0.3);
    #2
  \end{tikzpicture}
  \gSetStdTypes{line}
}

\def\gBubBubA{\@ifnextchar[\@gBubBubA{\@gBubBubA[]}}
\def\@gBubBubA[#1]#2{%
  \setkeys*{pre}{#1}%
  \setrmkeys{general,gTypes,gLabels}%
  \begin{tikzpicture}
    [line width=1pt,
    baseline={([yshift=-0.5ex+\cmdKV@general@yshift]current bounding box.center)},
    scale=\cmdKV@general@scale,
    rotate=\cmdKV@general@rotate,
    font=\cmdKV@general@font]
    \draw[\cmdKV@gTypes@eA] (0.3,0.6) -- (0,0) node[label=(-180+\cmdKV@general@rotate):\cmdKV@gLabels@eLA] {};
    \draw[\cmdKV@gTypes@eB] (0.3,0.6) -- (0,1.2) node[label=(180+\cmdKV@general@rotate):\cmdKV@gLabels@eLB] {};
    \draw[\cmdKV@gTypes@eC] (1.5,0.6) -- (1.8,1.2) node[label=(\cmdKV@general@rotate):\cmdKV@gLabels@eLC] {};
    \draw[\cmdKV@gTypes@eD] (1.5,0.6) -- (1.8,0) node[label=(\cmdKV@general@rotate):\cmdKV@gLabels@eLD] {};
    \draw[label distance=0,\cmdKV@gTypes@iA] (0.3,0.6) to[bend left=90] node[label=(90+\cmdKV@general@rotate):\cmdKV@gLabels@iLA] {} (0.9,0.6);
    \draw[label distance=0,\cmdKV@gTypes@iB] (0.9,0.6) to[bend left=90] node[label=(90+\cmdKV@general@rotate):\cmdKV@gLabels@iLB] {} (1.5,0.6);
    \draw[label distance=0,\cmdKV@gTypes@iC] (1.5,0.6) to[bend left=90] node[label=(-90+\cmdKV@general@rotate):\cmdKV@gLabels@iLC] {} (0.9,0.6);
    \draw[label distance=0,\cmdKV@gTypes@iD] (0.9,0.6) to[bend left=90] node[label=(-90+\cmdKV@general@rotate):\cmdKV@gLabels@iLD] {} (0.3,0.6);
    #2
  \end{tikzpicture}
  \gSetStdTypes{line}
}

\def\gBoxBubC{\@ifnextchar[\@gBoxBubC{\@gBoxBubC[]}}
\def\@gBoxBubC[#1]#2{%
  \setkeys*{pre}{#1}%
  \setrmkeys{general,gTypes,gLabels}%
  \begin{tikzpicture}
    [line width=1pt,
    baseline={([yshift=-0.5ex+\cmdKV@general@yshift]current bounding box.center)},
    scale=\cmdKV@general@scale,
    rotate=\cmdKV@general@rotate,
    font=\cmdKV@general@font]
    \draw[\cmdKV@gTypes@eA] (0.1,0.1) -- (-0.1,-0.1) node[label=(-180+\cmdKV@general@rotate):\cmdKV@gLabels@eLA] {};
    \draw[\cmdKV@gTypes@eB] (0.1,0.9) -- (-0.1,1.1) node[label=(180+\cmdKV@general@rotate):\cmdKV@gLabels@eLB] {};
    \draw[\cmdKV@gTypes@eC] (0.9,0.9) -- (1.1,1.1) node[label=(\cmdKV@general@rotate):\cmdKV@gLabels@eLC] {};
    \draw[\cmdKV@gTypes@eD] (0.9,0.1) -- (1.1,-0.1) node[label=(\cmdKV@general@rotate):\cmdKV@gLabels@eLD] {};
    \draw[label distance=0,\cmdKV@gTypes@iA] (0.1,0.1) -- node[label=(180+\cmdKV@general@rotate):\cmdKV@gLabels@iLA] {} (0.1,0.9);
    \draw[label distance=0,\cmdKV@gTypes@iB] (0.1,0.9) to[bend left] node[label=(90+\cmdKV@general@rotate):\cmdKV@gLabels@iLB] {} (0.9,0.9);
    \draw[label distance=0,\cmdKV@gTypes@iC] (0.9,0.9) to[bend left] node[label=(-90+\cmdKV@general@rotate):\cmdKV@gLabels@iLC] {} (0.1,0.9);
    \draw[label distance=0,\cmdKV@gTypes@iD] (0.9,0.9) -- node[label=(\cmdKV@general@rotate):\cmdKV@gLabels@iLD] {} (0.9,0.1);
    \draw[label distance=0,\cmdKV@gTypes@iE] (0.9,0.1) -- node[label=(-90+\cmdKV@general@rotate):\cmdKV@gLabels@iLE] {} (0.1,0.1);
    #2
  \end{tikzpicture}
  \gSetStdTypes{line}
}

\def\gBoxZ{\@ifnextchar[\@gBoxZ{\@gBoxZ[]}}
\def\@gBoxZ[#1]#2{%
  \setkeys*{pre}{#1}%
  \setrmkeys{general,gTypes,gLabels}%
  \begin{tikzpicture}
    [line width=1pt,
    baseline={([yshift=-0.5ex+\cmdKV@general@yshift]current bounding box.center)},
    scale=\cmdKV@general@scale,
    rotate=\cmdKV@general@rotate,
    font=\cmdKV@general@font]
    \draw[\cmdKV@gTypes@eA] (0.15,0.15) -- (-0.1,-0.1) node[label=(-180+\cmdKV@general@rotate):\cmdKV@gLabels@eLA] {};
    \draw[\cmdKV@gTypes@eB] (0.15,0.85) -- (-0.1,1.1) node[label=(180+\cmdKV@general@rotate):\cmdKV@gLabels@eLB] {};
    \draw[\cmdKV@gTypes@eC] (0.85,0.85) -- (1.1,1.1) node[label=(\cmdKV@general@rotate):\cmdKV@gLabels@eLC] {};
    \draw[\cmdKV@gTypes@eD] (0.85,0.15) -- (1.1,-0.1) node[label=(\cmdKV@general@rotate):\cmdKV@gLabels@eLD] {};
    \draw[label distance=0,\cmdKV@gTypes@iA] (0.15,0.15) -- node[label=(180+\cmdKV@general@rotate):\cmdKV@gLabels@iLA] {} (0.15,0.85);
    \draw[label distance=0,\cmdKV@gTypes@iB] (0.15,0.85) -- node[label=(90+\cmdKV@general@rotate):\cmdKV@gLabels@iLB] {} (0.85,0.85);
    \draw[label distance=0,\cmdKV@gTypes@iC] (0.85,0.85) -- node[label=(\cmdKV@general@rotate):\cmdKV@gLabels@iLC] {} (0.85,0.15);
    \draw[label distance=0,\cmdKV@gTypes@iD] (0.85,0.15) -- node[label=(-90+\cmdKV@general@rotate):\cmdKV@gLabels@iLD] {} (0.15,0.15);
    \draw[label distance=-4,\cmdKV@gTypes@iE] (0.15,0.15) -- node[label=(135+\cmdKV@general@rotate):\cmdKV@gLabels@iLE] {} (0.85,0.85);
    #2
  \end{tikzpicture}
  \gSetStdTypes{line}
}

\def\gBoxBoxBoxA{\@ifnextchar[\@gBoxBoxBoxA{\@gBoxBoxBoxA[]}}
\def\@gBoxBoxBoxA[#1]#2{%
  \setkeys*{pre}{#1}%
  \setrmkeys{general,gTypes,gLabels}%
  \begin{tikzpicture}
    [line width=1pt,
    baseline={([yshift=-0.5ex+\cmdKV@general@yshift]current bounding box.center)},
    scale=\cmdKV@general@scale,
    rotate=\cmdKV@general@rotate,
    font=\cmdKV@general@font]
    \draw[\cmdKV@gTypes@eA] (-0.3,0.3) -- (-0.6,0) node[label=(-180+\cmdKV@general@rotate):\cmdKV@gLabels@eLA] {};
    \draw[\cmdKV@gTypes@eB] (-0.3,0.9) -- (-0.6,1.2) node[label=(180+\cmdKV@general@rotate):\cmdKV@gLabels@eLB] {};
    \draw[\cmdKV@gTypes@eC] (1.5,0.9) -- (1.8,1.2) node[label=(\cmdKV@general@rotate):\cmdKV@gLabels@eLC] {};
    \draw[\cmdKV@gTypes@eD] (1.5,0.3) -- (1.8,0) node[label=(\cmdKV@general@rotate):\cmdKV@gLabels@eLD] {};
    \draw[label distance=0,\cmdKV@gTypes@iA] (-0.3,0.3) -- node[label=(180+\cmdKV@general@rotate):\cmdKV@gLabels@iLA] {} (-0.3,0.9);
    \draw[label distance=0,\cmdKV@gTypes@iB] (-0.3,0.9) -- node[label=(90+\cmdKV@general@rotate):\cmdKV@gLabels@iLB] {} (0.3,0.9);
    \draw[label distance=0,\cmdKV@gTypes@iC] (0.3,0.9) -- node[label=(90+\cmdKV@general@rotate):\cmdKV@gLabels@iLC] {} (0.9,0.9);
    \draw[label distance=0,\cmdKV@gTypes@iD] (0.9,0.9) -- node[label=(90+\cmdKV@general@rotate):\cmdKV@gLabels@iLD] {} (1.5,0.9);
    \draw[label distance=0,\cmdKV@gTypes@iE] (1.5,0.9) -- node[label=(\cmdKV@general@rotate):\cmdKV@gLabels@iLE] {} (1.5,0.3);
    \draw[label distance=0,\cmdKV@gTypes@iF] (1.5,0.3) -- node[label=(-90+\cmdKV@general@rotate):\cmdKV@gLabels@iLF] {} (0.9,0.3);
    \draw[label distance=0,\cmdKV@gTypes@iG] (0.9,0.3) -- node[label=(-90+\cmdKV@general@rotate):\cmdKV@gLabels@iLG] {} (0.3,0.3);
    \draw[label distance=0,\cmdKV@gTypes@iH] (0.3,0.3) -- node[label=(-90+\cmdKV@general@rotate):\cmdKV@gLabels@iLH] {} (-0.3,0.3);
    \draw[label distance=0,\cmdKV@gTypes@iI] (0.3,0.9) -- node[label=(\cmdKV@general@rotate):\cmdKV@gLabels@iLI] {} (0.3,0.3);
    \draw[label distance=0,\cmdKV@gTypes@iJ] (0.9,0.9) -- node[label=(\cmdKV@general@rotate):\cmdKV@gLabels@iLJ] {} (0.9,0.3);
    #2
  \end{tikzpicture}
  \gSetStdTypes{line}
}

\def\gBoxBoxBoxB{\@ifnextchar[\@gBoxBoxBoxB{\@gBoxBoxBoxB[]}}
\def\@gBoxBoxBoxB[#1]#2{%
  \setkeys*{pre}{#1}%
  \setrmkeys{general,gTypes,gLabels}%
  \begin{tikzpicture}
    [line width=1pt,
    baseline={([yshift=-0.5ex+\cmdKV@general@yshift]current bounding box.center)},
    scale=\cmdKV@general@scale,
    rotate=\cmdKV@general@rotate,
    font=\cmdKV@general@font]
    \draw[\cmdKV@gTypes@eA] (0.3,0.3) -- (0,0) node[label=(-180+\cmdKV@general@rotate):\cmdKV@gLabels@eLA] {};
    \draw[\cmdKV@gTypes@eB] (0.3,1.5) -- (0,1.8) node[label=(180+\cmdKV@general@rotate):\cmdKV@gLabels@eLB] {};
    \draw[\cmdKV@gTypes@eC] (1.5,1.5) -- (1.8,1.8) node[label=(\cmdKV@general@rotate):\cmdKV@gLabels@eLC] {};
    \draw[\cmdKV@gTypes@eD] (1.5,0.3) -- (1.8,0) node[label=(\cmdKV@general@rotate):\cmdKV@gLabels@eLD] {};
    \draw[label distance=0,\cmdKV@gTypes@iA] (0.3,0.3) -- node[label=(180+\cmdKV@general@rotate):\cmdKV@gLabels@iLA] {} (0.3,0.9);
    \draw[label distance=0,\cmdKV@gTypes@iB] (0.3,0.9) -- node[label=(180+\cmdKV@general@rotate):\cmdKV@gLabels@iLB] {} (0.3,1.5);
    \draw[label distance=0,\cmdKV@gTypes@iC] (0.3,1.5) -- node[label=(90+\cmdKV@general@rotate):\cmdKV@gLabels@iLC] {} (0.9,1.5);
    \draw[label distance=0,\cmdKV@gTypes@iD] (0.9,1.5) -- node[label=(90+\cmdKV@general@rotate):\cmdKV@gLabels@iLD] {} (1.5,1.5);
    \draw[label distance=0,\cmdKV@gTypes@iE] (1.5,1.5) -- node[label=(\cmdKV@general@rotate):\cmdKV@gLabels@iLE] {} (1.5,0.3);
    \draw[label distance=0,\cmdKV@gTypes@iF] (1.5,0.3) -- node[label=(-90+\cmdKV@general@rotate):\cmdKV@gLabels@iLF] {} (0.9,0.3);
    \draw[label distance=0,\cmdKV@gTypes@iG] (0.9,0.3) -- node[label=(-90+\cmdKV@general@rotate):\cmdKV@gLabels@iLG] {} (0.3,0.3);
    \draw[label distance=0,\cmdKV@gTypes@iH] (0.9,1.5) -- node[label=(\cmdKV@general@rotate):\cmdKV@gLabels@iLH] {} (0.9,0.9);
    \draw[label distance=0,\cmdKV@gTypes@iI] (0.9,0.9) -- node[label=(\cmdKV@general@rotate):\cmdKV@gLabels@iLI] {} (0.9,0.3);
    \draw[label distance=0,\cmdKV@gTypes@iJ] (0.9,0.9) -- node[label=(90+\cmdKV@general@rotate):\cmdKV@gLabels@iLJ] {} (0.3,0.9);
    #2
  \end{tikzpicture}
  \gSetStdTypes{line}
}

\def\gEFTTree{\@ifnextchar[\@gEFTTree{\@gEFTTree[]}}
\def\@gEFTTree[#1]#2{%
  \setkeys*{pre}{#1}%
  \setrmkeys{general,gTypes,gLabels}%
  \begin{tikzpicture}
    [line width=1pt,
    baseline={([yshift=-0.5ex+\cmdKV@general@yshift]current bounding box.center)},
    scale=\cmdKV@general@scale,
    rotate=\cmdKV@general@rotate,
    font=\cmdKV@general@font]
    \draw[\cmdKV@gTypes@iA] (0,0.5) node[label=(90+\cmdKV@general@rotate):\cmdKV@gLabels@eLA] {} -- node[label=(\cmdKV@general@rotate):\cmdKV@gLabels@iLA] {} (0,-0.5) node[label=(-90+\cmdKV@general@rotate):\cmdKV@gLabels@eLB] {};
    \draw[fill] (0,0.5) circle (0.02);
    \draw[fill] (0,-0.5) circle (0.02);
    #2
  \end{tikzpicture}
  \gSetStdTypes{line}
}

\def\gEFTV{\@ifnextchar[\@gEFTV{\@gEFTV[]}}
\def\@gEFTV[#1]#2{%
  \setkeys*{pre}{#1}%
  \setrmkeys{general,gTypes,gLabels}%
  \begin{tikzpicture}
    [line width=1pt,
    baseline={([yshift=-0.5ex+\cmdKV@general@yshift]current bounding box.center)},
    scale=\cmdKV@general@scale,
    rotate=\cmdKV@general@rotate,
    font=\cmdKV@general@font]
    \draw[\cmdKV@gTypes@iA] (0,0.5) node[label=(90+\cmdKV@general@rotate):\cmdKV@gLabels@eLA] {} -- node[label=(200+\cmdKV@general@rotate):\cmdKV@gLabels@iLA] {} (0.3,-0.5) node[label=(-90+\cmdKV@general@rotate):\cmdKV@gLabels@eLC] {};
    \draw[\cmdKV@gTypes@iB] (0.6,0.5) node[label=(90+\cmdKV@general@rotate):\cmdKV@gLabels@eLB] {} -- node[label=(-20+\cmdKV@general@rotate):\cmdKV@gLabels@iLB] {} (0.3,-0.5);
    \draw[fill] (0,0.5) circle (0.02);
    \draw[fill] (0.3,-0.5) circle (0.02);
    \draw[fill] (0.6,0.5) circle (0.02);
    #2
  \end{tikzpicture}
  \gSetStdTypes{line}
}

\def\gEFTY{\@ifnextchar[\@gEFTY{\@gEFTY[]}}
\def\@gEFTY[#1]#2{%
  \setkeys*{pre}{#1}%
  \setrmkeys{general,gTypes,gLabels}%
  \begin{tikzpicture}
    [line width=1pt,
    baseline={([yshift=-0.5ex+\cmdKV@general@yshift]current bounding box.center)},
    scale=\cmdKV@general@scale,
    rotate=\cmdKV@general@rotate,
    font=\cmdKV@general@font]
    \draw[\cmdKV@gTypes@iA] (0,0.5) node[label=(90+\cmdKV@general@rotate):\cmdKV@gLabels@eLA] {} -- node[label=(225+\cmdKV@general@rotate):\cmdKV@gLabels@iLA] {} (0.3,0);
    \draw[\cmdKV@gTypes@iB] (0.6,0.5) node[label=(90+\cmdKV@general@rotate):\cmdKV@gLabels@eLB] {} -- node[label=(-45+\cmdKV@general@rotate):\cmdKV@gLabels@iLB] {} (0.3,0);
    \draw[\cmdKV@gTypes@iC] (0.3,0) -- node[label=(\cmdKV@general@rotate):\cmdKV@gLabels@iLC] {} (0.3,-0.5) node[label=(-90+\cmdKV@general@rotate):\cmdKV@gLabels@eLC] {};
    \draw[fill] (0,0.5) circle (0.02);
    \draw[fill] (0.6,0.5) circle (0.02);
    \draw[fill] (0.3,-0.5) circle (0.02);
    #2
  \end{tikzpicture}
  \gSetStdTypes{line}
}

\def\gEFTPsi{\@ifnextchar[\@gEFTPsi{\@gEFTPsi[]}}
\def\@gEFTPsi[#1]#2{%
  \setkeys*{pre}{#1}%
  \setrmkeys{general,gTypes,gLabels}%
  \begin{tikzpicture}
    [line width=1pt,
    baseline={([yshift=-0.5ex+\cmdKV@general@yshift]current bounding box.center)},
    scale=\cmdKV@general@scale,
    rotate=\cmdKV@general@rotate,
    font=\cmdKV@general@font]
    \draw[\cmdKV@gTypes@iA] (0,0.5) node[label=(90+\cmdKV@general@rotate):\cmdKV@gLabels@eLA] {} -- node[label=(225+\cmdKV@general@rotate):\cmdKV@gLabels@iLA] {} (0.5,0);
    \draw[\cmdKV@gTypes@iB] (0.5,0.5) node[label=(90+\cmdKV@general@rotate):\cmdKV@gLabels@eLB] {} -- node[label=(\cmdKV@general@rotate):\cmdKV@gLabels@iLB] {} (0.5,0);
    \draw[\cmdKV@gTypes@iC] (1,0.5) node[label=(90+\cmdKV@general@rotate):\cmdKV@gLabels@eLC] {} -- node[label=(-45+\cmdKV@general@rotate):\cmdKV@gLabels@iLC] {} (0.5,0);
    \draw[\cmdKV@gTypes@iD] (0.5,0) -- node[label=(\cmdKV@general@rotate):\cmdKV@gLabels@iLD] {} (0.5,-0.5) node[label=(-90+\cmdKV@general@rotate):\cmdKV@gLabels@eLD] {};
    \draw[fill] (0,0.5) circle (0.02);
    \draw[fill] (0.5,0.5) circle (0.02);
    \draw[fill] (1,0.5) circle (0.02);
    \draw[fill] (0.5,-0.5) circle (0.02);
    #2
  \end{tikzpicture}
  \gSetStdTypes{line}
}

\def\gEFTVinY{\@ifnextchar[\@gEFTVinY{\@gEFTVinY[]}}
\def\@gEFTVinY[#1]#2{%
  \setkeys*{pre}{#1}%
  \setrmkeys{general,gTypes,gLabels}%
  \begin{tikzpicture}
    [line width=1pt,
    baseline={([yshift=-0.5ex+\cmdKV@general@yshift]current bounding box.center)},
    scale=\cmdKV@general@scale,
    rotate=\cmdKV@general@rotate,
    font=\cmdKV@general@font]
    \draw[\cmdKV@gTypes@iA] (0,0.5) node[label=(90+\cmdKV@general@rotate):\cmdKV@gLabels@eLA] {} -- node[label=(225+\cmdKV@general@rotate):\cmdKV@gLabels@iLA] {} (0.25,0.25);
    \draw[\cmdKV@gTypes@iE] (0.25,0.25) -- node[label=(225+\cmdKV@general@rotate):\cmdKV@gLabels@iLE] {} (0.5,0);
    \draw[\cmdKV@gTypes@iB] (0.5,0.5) node[label=(90+\cmdKV@general@rotate):\cmdKV@gLabels@eLB] {} -- node[label=(\cmdKV@general@rotate):\cmdKV@gLabels@iLB] {} (0.25,0.25);
    \draw[\cmdKV@gTypes@iC] (1,0.5) node[label=(90+\cmdKV@general@rotate):\cmdKV@gLabels@eLC] {} -- node[label=(-45+\cmdKV@general@rotate):\cmdKV@gLabels@iLC] {} (0.5,0);
    \draw[\cmdKV@gTypes@iD] (0.5,0) -- node[label=(\cmdKV@general@rotate):\cmdKV@gLabels@iLD] {} (0.5,-0.5) node[label=(-90+\cmdKV@general@rotate):\cmdKV@gLabels@eLD] {};
    \draw[fill] (0,0.5) circle (0.02);
    \draw[fill] (0.5,0.5) circle (0.02);
    \draw[fill] (1,0.5) circle (0.02);
    \draw[fill] (0.5,-0.5) circle (0.02);
    #2
  \end{tikzpicture}
  \gSetStdTypes{line}
}

\def\gEFTYinvY{\@ifnextchar[\@gEFTYinvY{\@gEFTYinvY[]}}
\def\@gEFTYinvY[#1]#2{%
  \setkeys*{pre}{#1}%
  \setrmkeys{general,gTypes,gLabels}%
  \begin{tikzpicture}
    [line width=1pt,
    baseline={([yshift=-0.5ex+\cmdKV@general@yshift]current bounding box.center)},
    scale=\cmdKV@general@scale,
    rotate=\cmdKV@general@rotate,
    font=\cmdKV@general@font]
    \draw[\cmdKV@gTypes@iA] (0,0.5) node[label=(90+\cmdKV@general@rotate):\cmdKV@gLabels@eLA] {} -- node[label=(225+\cmdKV@general@rotate):\cmdKV@gLabels@iLA] {} (0.3,0.166);
    \draw[\cmdKV@gTypes@iB] (0.6,0.5) node[label=(90+\cmdKV@general@rotate):\cmdKV@gLabels@eLB] {} -- node[label=(-45+\cmdKV@general@rotate):\cmdKV@gLabels@iLB] {} (0.3,0.166);
    \draw[\cmdKV@gTypes@iC] (0.3,-0.166) -- node[label=(135+\cmdKV@general@rotate):\cmdKV@gLabels@iLC] {} (0,-0.5) node[label=(-90+\cmdKV@general@rotate):\cmdKV@gLabels@eLC] {}; 
    \draw[\cmdKV@gTypes@iD] (0.3,-0.166) -- node[label=(45+\cmdKV@general@rotate):\cmdKV@gLabels@iLD] {} (0.6,-0.5) node[label=(-90+\cmdKV@general@rotate):\cmdKV@gLabels@eLD] {};
    \draw[\cmdKV@gTypes@iE] (0.3,0.166) -- node[label=(\cmdKV@general@rotate):\cmdKV@gLabels@iLE] {} (0.3,-0.166);
    \draw[fill] (0,0.5) circle (0.02);
    \draw[fill] (0.6,0.5) circle (0.02);
    \draw[fill] (0,-0.5) circle (0.02);
    \draw[fill] (0.6,-0.5) circle (0.02);
    #2
  \end{tikzpicture}
  \gSetStdTypes{line}
}

\def\gEFTX{\@ifnextchar[\@gEFTX{\@gEFTX[]}}
\def\@gEFTX[#1]#2{%
  \setkeys*{pre}{#1}%
  \setrmkeys{general,gTypes,gLabels}%
  \begin{tikzpicture}
    [line width=1pt,
    baseline={([yshift=-0.5ex+\cmdKV@general@yshift]current bounding box.center)},
    scale=\cmdKV@general@scale,
    rotate=\cmdKV@general@rotate,
    font=\cmdKV@general@font]
    \draw[\cmdKV@gTypes@iA] (0,0.5) node[label=(90+\cmdKV@general@rotate):\cmdKV@gLabels@eLA] {} -- node[label=(225+\cmdKV@general@rotate):\cmdKV@gLabels@iLA] {} (0.3,0);
    \draw[\cmdKV@gTypes@iB] (0.6,0.5) node[label=(90+\cmdKV@general@rotate):\cmdKV@gLabels@eLB] {} -- node[label=(-45+\cmdKV@general@rotate):\cmdKV@gLabels@iLB] {} (0.3,0);
    \draw[\cmdKV@gTypes@iC] (0.3,0) -- node[label=(135+\cmdKV@general@rotate):\cmdKV@gLabels@iLC] {} (0,-0.5) node[label=(-90+\cmdKV@general@rotate):\cmdKV@gLabels@eLC] {}; 
    \draw[\cmdKV@gTypes@iD] (0.3,0) -- node[label=(45+\cmdKV@general@rotate):\cmdKV@gLabels@iLD] {} (0.6,-0.5) node[label=(-90+\cmdKV@general@rotate):\cmdKV@gLabels@eLD] {};
    \draw[fill] (0,0.5) circle (0.02);
    \draw[fill] (0.6,0.5) circle (0.02);
    \draw[fill] (0,-0.5) circle (0.02);
    \draw[fill] (0.6,-0.5) circle (0.02);
    #2
  \end{tikzpicture}
  \gSetStdTypes{line}
}

\def\gEFTH{\@ifnextchar[\@gEFTH{\@gEFTH[]}}
\def\@gEFTH[#1]#2{%
  \setkeys*{pre}{#1}%
  \setrmkeys{general,gTypes,gLabels}%
  \begin{tikzpicture}
    [line width=1pt,
    baseline={([yshift=-0.5ex+\cmdKV@general@yshift]current bounding box.center)},
    scale=\cmdKV@general@scale,
    rotate=\cmdKV@general@rotate,
    font=\cmdKV@general@font]
    \draw[\cmdKV@gTypes@iA] (0,0.5) node[label=(90+\cmdKV@general@rotate):\cmdKV@gLabels@eLA] {} -- node[label=(180+\cmdKV@general@rotate):\cmdKV@gLabels@iLA] {} (0,0);
    \draw[\cmdKV@gTypes@iB] (0.6,0.5) node[label=(90+\cmdKV@general@rotate):\cmdKV@gLabels@eLB] {} -- node[label=(\cmdKV@general@rotate):\cmdKV@gLabels@iLB] {} (0.6,0);
    \draw[\cmdKV@gTypes@iC] (0.0,0) -- node[label=(180+\cmdKV@general@rotate):\cmdKV@gLabels@iLC] {} (0,-0.5) node[label=(-90+\cmdKV@general@rotate):\cmdKV@gLabels@eLC] {}; 
    \draw[\cmdKV@gTypes@iD] (0.6,0) -- node[label=(\cmdKV@general@rotate):\cmdKV@gLabels@iLD] {} (0.6,-0.5) node[label=(-90+\cmdKV@general@rotate):\cmdKV@gLabels@eLD] {};
    \draw[\cmdKV@gTypes@iE] (0,0) -- node[label=(90+\cmdKV@general@rotate):\cmdKV@gLabels@iLE] {} (0.6,0);
    \draw[fill] (0,0.5) circle (0.02);
    \draw[fill] (0.6,0.5) circle (0.02);
    \draw[fill] (0,-0.5) circle (0.02);
    \draw[fill] (0.6,-0.5) circle (0.02);
    #2
  \end{tikzpicture}
  \gSetStdTypes{line}
}

\def\cTree{\@ifnextchar[\@cTree{\@cTree[]}}
\def\@cTree[#1]#2{%
  \setkeys*{pre}{#1}%
  \setrmkeys{general,gTypes,gLabels,blobs}%
  \begin{tikzpicture}
    [line width=1pt,
    baseline={([yshift=-0.5ex+\cmdKV@general@yshift]current bounding box.center)},
    scale=\cmdKV@general@scale,
    rotate=\cmdKV@general@rotate,
    font=\cmdKV@general@font]
    \fill[\cmdKV@blobs@blobA] (0,0) ellipse (0.25 and 0.25);
    \draw[\cmdKV@gTypes@eA] ($ (0,0) + (-135:0.25) $) -- ($ (0,0) + (-135:0.6) $) node[label=(-180+\cmdKV@general@rotate):\cmdKV@gLabels@eLA] {};
    \draw[\cmdKV@gTypes@eB] ($ (0,0) + (135:0.25) $) -- ($ (0,0) + (135:0.6) $) node[label=(180+\cmdKV@general@rotate):\cmdKV@gLabels@eLB] {};
    \draw[\cmdKV@gTypes@eC] ($ (0,0) + (45:0.25) $) -- ($ (0,0) + (45:0.6) $) node[label=(\cmdKV@general@rotate):\cmdKV@gLabels@eLC] {};
    \draw[\cmdKV@gTypes@eD] ($ (0,0) + (-45:0.25) $) -- ($ (0,0) + (-45:0.6) $) node[label=(\cmdKV@general@rotate):\cmdKV@gLabels@eLD] {};
    \node[label=(90+\cmdKV@general@rotate):\cmdKV@gLabels@iLA] at (0,0.3) {};
    \node[label=(\cmdKV@general@rotate):\cmdKV@gLabels@iLB] at (0.3,0) {};
    #2
  \end{tikzpicture}
  \gSetStdTypes{line}
}

\def\cBoxA{\@ifnextchar[\@cBoxA{\@cBoxA[]}}
\def\@cBoxA[#1]#2{%
  \setkeys*{pre}{#1}%
  \setrmkeys{general,gTypes,gLabels,dots,blobs}%
  \begin{tikzpicture}
    [line width=1pt,
    baseline={([yshift=-0.5ex+\cmdKV@general@yshift]current bounding box.center)},
    scale=\cmdKV@general@scale,
    rotate=\cmdKV@general@rotate,
    font=\cmdKV@general@font]
    \fill[\cmdKV@blobs@blobA] (-0.4,0) ellipse (0.2 and 0.4);
    \fill[\cmdKV@blobs@blobB] (0.4,0) ellipse (0.2 and 0.4);
    \draw[\cmdKV@gTypes@eA] ($ (-0.4,0) + (-135:0.26) $) -- ($ (-0.4,0) + (-135:0.7) $) node[label=(-180+\cmdKV@general@rotate):\cmdKV@gLabels@eLA] {};
    \draw[\cmdKV@gTypes@eB] ($ (-0.4,0) + (135:0.26) $) -- ($ (-0.4,0) + (135:0.7) $) node[label=(180+\cmdKV@general@rotate):\cmdKV@gLabels@eLB] {};
    \draw[\cmdKV@gTypes@eC] ($ (0.4,0) + (45:0.26) $) -- ($ (0.4,0) + (45:0.7) $) node[label=(\cmdKV@general@rotate):\cmdKV@gLabels@eLC] {};
    \draw[\cmdKV@gTypes@eD] ($ (0.4,0) + (-45:0.26) $) -- ($ (0.4,0) + (-45:0.7) $) node[label=(\cmdKV@general@rotate):\cmdKV@gLabels@eLD] {};
    \draw[label distance=1,\cmdKV@gTypes@iA] ($ (-0.4,0) + (45:0.26) $) -- node[label=(90+\cmdKV@general@rotate):\cmdKV@gLabels@iLA] {}($ (0.4,0) + (135:0.26) $);
    \draw[label distance=1,\cmdKV@gTypes@iB] ($ (0.4,0) + (-135:0.26) $) -- node[label=(-90+\cmdKV@general@rotate):\cmdKV@gLabels@iLB] {}($ (-0.4,0) + (-45:0.26) $);
    \node[label=(180+\cmdKV@general@rotate):\cmdKV@gLabels@iLC] at (-0.6,0) {};
    \node[label=(\cmdKV@general@rotate):\cmdKV@gLabels@iLD] at (0.6,0) {};
    \ifKV@dots@lDots
      \draw[dotted] ($(-0.4,0) + (-150:0.7)$) to[bend left] ($(-0.4,0) + (150:0.7)$);%
    \fi
    \ifKV@dots@rDots
      \draw[dotted] ($(0.4,0) + (30:0.7)$) to[bend left] ($(0.4,0) + (-30:0.7)$);%
    \fi
    #2
  \end{tikzpicture}
  \gSetStdTypes{line}
}

\def\cBoxAA{\@ifnextchar[\@cBoxAA{\@cBoxAA[]}}
\def\@cBoxAA[#1]#2{%
  \setkeys*{pre}{#1}%
  \setrmkeys{general,gTypes,gLabels,dots,blobs}%
  \begin{tikzpicture}
    [line width=1pt,
    baseline={([yshift=-0.5ex+\cmdKV@general@yshift]current bounding box.center)},
    scale=\cmdKV@general@scale,
    rotate=\cmdKV@general@rotate,
    font=\cmdKV@general@font]
    \fill[\cmdKV@blobs@blobA] (-0.4,0) ellipse (0.2 and 0.4);
    \fill[\cmdKV@blobs@blobB] (0.4,0) ellipse (0.2 and 0.4);
    \draw[\cmdKV@gTypes@eA] ($ (-0.4,0) + (-135:0.26) $) -- ($ (-0.4,0) + (-135:0.7) $) node[label=(-180+\cmdKV@general@rotate):\cmdKV@gLabels@eLA] {};
    \draw[\cmdKV@gTypes@eB] ($ (-0.4,0) + (135:0.26) $) -- ($ (-0.4,0) + (135:0.7) $) node[label=(180+\cmdKV@general@rotate):\cmdKV@gLabels@eLB] {};
    \draw[\cmdKV@gTypes@eC] ($ (0.4,0) + (45:0.26) $) -- ($ (0.4,0) + (45:0.7) $) node[label=(\cmdKV@general@rotate):\cmdKV@gLabels@eLC] {};
    \draw[\cmdKV@gTypes@eD] ($ (0.4,0) + (-45:0.26) $) -- ($ (0.4,0) + (-45:0.7) $) node[label=(\cmdKV@general@rotate):\cmdKV@gLabels@eLD] {};
    \draw[label distance=1,\cmdKV@gTypes@iA] ($ (-0.4,0) + (60:0.32) $) to[bend left] node[label=(90+\cmdKV@general@rotate):\cmdKV@gLabels@iLA] {}($ (0.4,0) + (120:0.32) $);
    \draw[label distance=1,\cmdKV@gTypes@iB] ($ (0.4,0) + (-120:0.32) $) to[bend left]  node[label=(-90+\cmdKV@general@rotate):\cmdKV@gLabels@iLB] {}($ (-0.4,0) + (-60:0.32) $);
    \draw[dotted] (0,0.25) -- (0,-0.25);
    \node[label=(180+\cmdKV@general@rotate):\cmdKV@gLabels@iLC] at (-0.6,0) {};
    \node[label=(\cmdKV@general@rotate):\cmdKV@gLabels@iLD] at (0.6,0) {};
    \ifKV@dots@lDots
      \draw[dotted] ($(-0.4,0) + (-150:0.7)$) to[bend left] ($(-0.4,0) + (150:0.7)$);%
    \fi
    \ifKV@dots@rDots
      \draw[dotted] ($(0.4,0) + (30:0.7)$) to[bend left] ($(0.4,0) + (-30:0.7)$);%
    \fi
    #2
  \end{tikzpicture}
  \gSetStdTypes{line}
}

\def\cBoxB{\@ifnextchar[\@cBoxB{\@cBoxB[]}}
\def\@cBoxB[#1]#2{%
  \setkeys*{pre}{#1}%
  \setrmkeys{general,gTypes,gLabels,blobs}%
  \begin{tikzpicture}
    [line width=1pt,
    baseline={([yshift=-0.5ex+\cmdKV@general@yshift]current bounding box.center)},
    scale=\cmdKV@general@scale,
    rotate=\cmdKV@general@rotate,
    font=\cmdKV@general@font]
    \fill[\cmdKV@blobs@blobA] (0,-0.3) ellipse (0.4 and 0.2);
    \fill[\cmdKV@blobs@blobB] (0,0.3) ellipse (0.4 and 0.2);
    \draw[\cmdKV@gTypes@eA] ($ (0,-0.3) + (180:0.4) $) -- ($ (0,-0.3) + (-170:0.7) $) node[label=(180+\cmdKV@general@rotate):\cmdKV@gLabels@eLA] {};
    \draw[\cmdKV@gTypes@eB] ($ (0,0.3) + (180:0.4) $) -- ($ (0,0.3) + (170:0.7) $) node[label=(180+\cmdKV@general@rotate):\cmdKV@gLabels@eLB] {};
    \draw[\cmdKV@gTypes@eC] ($ (0,0.3) + (0:0.4) $) -- ($ (0,0.3) + (10:0.7) $) node[label=(\cmdKV@general@rotate):\cmdKV@gLabels@eLC] {};
    \draw[\cmdKV@gTypes@eD] ($ (0,-0.3) + (0:0.4) $) -- ($ (0,-0.3) + (-10:0.7) $) node[label=(\cmdKV@general@rotate):\cmdKV@gLabels@eLD] {};
    \draw[label distance=1,\cmdKV@gTypes@iA] ($ (0,-0.3) + (135:0.26) $) -- node[label=(-180+\cmdKV@general@rotate):\cmdKV@gLabels@iLA] {}($ (0,0.3) + (-135:0.26) $);
    \draw[label distance=1,\cmdKV@gTypes@iB] ($ (0,0.3) + (-45:0.26) $) -- node[label=(\cmdKV@general@rotate):\cmdKV@gLabels@iLB] {}($ (0,-0.3) + (45:0.26) $);
    \node[label=(90+\cmdKV@general@rotate):\cmdKV@gLabels@iLC] at (0,0.5) {};
    \node[label=(-90+\cmdKV@general@rotate):\cmdKV@gLabels@iLD] at (0,-0.5) {};
    #2
  \end{tikzpicture}
  \gSetStdTypes{line}
}

\def\cBoxBoxA{\@ifnextchar[\@cBoxBoxA{\@cBoxBoxA[]}}
\def\@cBoxBoxA[#1]#2{%
  \setkeys*{pre}{#1}%
  \setrmkeys{general,gTypes,gLabels,blobs}%
  \begin{tikzpicture}
    [line width=1pt,
    baseline={([yshift=-0.5ex+\cmdKV@general@yshift]current bounding box.center)},
    scale=\cmdKV@general@scale,
    rotate=\cmdKV@general@rotate,
    font=\cmdKV@general@font]
    \fill[\cmdKV@blobs@blobA] (-0.4,0) ellipse (0.2 and 0.4);
    \fill[\cmdKV@blobs@blobB] (0.4,0) ellipse (0.2 and 0.4);
    \fill[\cmdKV@blobs@blobC] (1.2,0) ellipse (0.2 and 0.4);
    \draw[\cmdKV@gTypes@eA] ($ (-0.4,0) + (-135:0.26) $) -- ($ (-0.4,0) + (-135:0.7) $) node[label=(-180+\cmdKV@general@rotate):\cmdKV@gLabels@eLA] {};
    \draw[\cmdKV@gTypes@eB] ($ (-0.4,0) + (135:0.26) $) -- ($ (-0.4,0) + (135:0.7) $) node[label=(180+\cmdKV@general@rotate):\cmdKV@gLabels@eLB] {};
    \draw[\cmdKV@gTypes@eC] ($ (1.2,0) + (45:0.26) $) -- ($ (1.2,0) + (45:0.7) $) node[label=(\cmdKV@general@rotate):\cmdKV@gLabels@eLC] {};
    \draw[\cmdKV@gTypes@eD] ($ (1.2,0) + (-45:0.26) $) -- ($ (1.2,0) + (-45:0.7) $) node[label=(\cmdKV@general@rotate):\cmdKV@gLabels@eLD] {};
    \draw[label distance=1,\cmdKV@gTypes@iA] ($ (-0.4,0) + (45:0.26) $) -- node[label=(90+\cmdKV@general@rotate):\cmdKV@gLabels@iLA] {}($ (0.4,0) + (135:0.26) $);
    \draw[label distance=1,\cmdKV@gTypes@iB] ($ (0.4,0) + (45:0.26) $) -- node[label=(90+\cmdKV@general@rotate):\cmdKV@gLabels@iLB] {}($ (1.2,0) + (135:0.26) $);
    \draw[label distance=1,\cmdKV@gTypes@iC] ($ (1.2,0) + (-135:0.26) $) -- node[label=(-90+\cmdKV@general@rotate):\cmdKV@gLabels@iLC] {}($ (0.4,0) + (-45:0.26) $);
    \draw[label distance=1,\cmdKV@gTypes@iD] ($ (0.4,0) + (-135:0.26) $) -- node[label=(-90+\cmdKV@general@rotate):\cmdKV@gLabels@iLD] {}($ (-0.4,0) + (-45:0.26) $);
    \node[label=(180+\cmdKV@general@rotate):\cmdKV@gLabels@iLE] at (-0.6,0) {};
    \node[label=(\cmdKV@general@rotate):\cmdKV@gLabels@iLF] at (0.6,0) {};
    \node[label=(\cmdKV@general@rotate):\cmdKV@gLabels@iLG] at (1.4,0) {};
    #2
  \end{tikzpicture}
  \gSetStdTypes{line}
}

\def\cBoxBoxB{\@ifnextchar[\@cBoxBoxB{\@cBoxBoxB[]}}
\def\@cBoxBoxB[#1]#2{%
  \setkeys*{pre}{#1}%
  \setrmkeys{general,gTypes,gLabels}%
  \begin{tikzpicture}
    [line width=1pt,
    baseline={([yshift=-0.5ex+\cmdKV@general@yshift]current bounding box.center)},
    scale=\cmdKV@general@scale,
    rotate=\cmdKV@general@rotate,
    font=\cmdKV@general@font]
    \fill[\cmdKV@blobs@blobA] (0,0.3) ellipse (0.4 and 0.2);
    \fill[\cmdKV@blobs@blobB] (0,-0.3) ellipse (0.4 and 0.2);
    \fill[\cmdKV@blobs@blobC] (1,0) ellipse (0.2 and 0.4);
    \draw[\cmdKV@gTypes@eA] ($ (0,-0.3) + (180:0.4) $) -- ($ (0,-0.3) + (-170:0.7) $) node[label=(180+\cmdKV@general@rotate):\cmdKV@gLabels@eLA] {};
    \draw[\cmdKV@gTypes@eB] ($ (0,0.3) + (180:0.4) $) -- ($ (0,0.3) + (170:0.7) $) node[label=(180+\cmdKV@general@rotate):\cmdKV@gLabels@eLB] {};
    \draw[\cmdKV@gTypes@eC] ($ (1,0) + (45:0.26) $) -- ($ (1,0) + (45:0.7) $) node[label=(\cmdKV@general@rotate):\cmdKV@gLabels@eLC] {};
    \draw[\cmdKV@gTypes@eD] ($ (1,0) + (-45:0.26) $) -- ($ (1,0) + (-45:0.7) $) node[label=(\cmdKV@general@rotate):\cmdKV@gLabels@eLD] {};
    \draw[label distance=1,\cmdKV@gTypes@iA] ($ (0,-0.3) + (135:0.26) $) -- node[label=(-180+\cmdKV@general@rotate):\cmdKV@gLabels@iLA] {}($ (0,0.3) + (-135:0.26) $);
    \draw[label distance=1,\cmdKV@gTypes@iB] ($ (0,0.3) + (0:0.4) $) -- node[label=(90+\cmdKV@general@rotate):\cmdKV@gLabels@iLB] {} ($ (1,0) + (135:0.26) $);
    \draw[label distance=1,\cmdKV@gTypes@iC] ($ (1,0) + (-135:0.26) $) -- node[label=(-90+\cmdKV@general@rotate):\cmdKV@gLabels@iLC] {} ($ (0,-0.3) + (0:0.4) $);
    \draw[label distance=1,\cmdKV@gTypes@iD] ($ (0,0.3) + (-45:0.26) $) -- node[label=(\cmdKV@general@rotate):\cmdKV@gLabels@iLD] {}($ (0,-0.3) + (45:0.26) $);
    \node[label=(90+\cmdKV@general@rotate):\cmdKV@gLabels@iLE] at (0,0.5) {};
    \node[label=(\cmdKV@general@rotate):\cmdKV@gLabels@iLF] at (1.2,0) {};
    \node[label=(-90+\cmdKV@general@rotate):\cmdKV@gLabels@iLG] at (0,-0.5) {};
    #2
  \end{tikzpicture}
  \gSetStdTypes{line}
}

\def\cNPBox{\@ifnextchar[\@cNPBox{\@cNPBox[]}}
\def\@cNPBox[#1]#2{%
  \setkeys*{pre}{#1}%
  \setrmkeys{general,gTypes,gLabels,blob}%
  \begin{tikzpicture}
    [line width=1pt,
    baseline={([yshift=-0.5ex+\cmdKV@general@yshift]current bounding box.center)},
    scale=\cmdKV@general@scale,
    rotate=\cmdKV@general@rotate,
    font=\cmdKV@general@font]
    \fill[\cmdKV@blobs@blobA] (-0.6,0) ellipse (0.2 and 0.4);
    \fill[\cmdKV@blobs@blobB] (0.4,0) ellipse (0.2 and 0.4);
    \draw[\cmdKV@gTypes@eA] ($ (-0.6,0) + (-180:0.2) $) -- ($ (-0.6,0) + (-180:0.6) $) node[label=(-180+\cmdKV@general@rotate):\cmdKV@gLabels@eLA] {};
    \draw[\cmdKV@gTypes@eB] ($ (-0.6,0) + (0:0.2) $) -- (-0.18,0) node[label=(\cmdKV@general@rotate):\cmdKV@gLabels@eLB] {};
    \draw[\cmdKV@gTypes@eC] ($ (0.4,0) + (45:0.26) $) -- ($ (0.4,0) + (45:0.7) $) node[label=(\cmdKV@general@rotate):\cmdKV@gLabels@eLC] {};
    \draw[\cmdKV@gTypes@eD] ($ (0.4,0) + (-45:0.26) $) -- ($ (0.4,0) + (-45:0.7) $) node[label=(\cmdKV@general@rotate):\cmdKV@gLabels@eLD] {};
    \draw[label distance=1,\cmdKV@gTypes@iA] ($ (-0.6,0) + (60:0.3) $) -- node[label=(90+\cmdKV@general@rotate):\cmdKV@gLabels@iLA] {}($ (0.4,0) + (135:0.26) $);
    \draw[label distance=1,\cmdKV@gTypes@iB] ($ (0.4,0) + (-135:0.26) $) -- node[label=(-90+\cmdKV@general@rotate):\cmdKV@gLabels@iLB] {}($ (-0.6,0) + (-60:0.3) $);
    \node[label=(\cmdKV@general@rotate):\cmdKV@gLabels@iLD] at (0.6,0) {};
    #2
  \end{tikzpicture}
  \gSetStdTypes{line}
}

\def\cFourPtRec{\@ifnextchar[\@cFourPtRec{\@cFourPtRec[]}}
\def\@cFourPtRec[#1]#2{%
  \setkeys*{pre}{#1}%
  \setrmkeys{general,gTypes,gLabels,blobs}%
  \begin{tikzpicture}
    [line width=1pt,
    baseline={([yshift=-0.5ex+\cmdKV@general@yshift]current bounding box.center)},
    scale=\cmdKV@general@scale,
    rotate=\cmdKV@general@rotate,
    font=\cmdKV@general@font]
    \fill[\cmdKV@blobs@blobA] (-0.5,0) ellipse (0.2 and 0.3);
    \fill[\cmdKV@blobs@blobB] (0.5,0) ellipse (0.3 and 0.4);
    \filldraw[fill=white,draw=black] (0.55,0.15) circle (0.1);
    \filldraw[fill=white,draw=black] (0.45,-0.1) circle (0.1);
    \draw[\cmdKV@gTypes@eA] ($(-0.5,0) + (135:0.23)$) -- ($(-0.5,0) + (135:0.5)$) node[label=(180+\cmdKV@general@rotate):\cmdKV@gLabels@eLA] {};
    \draw[\cmdKV@gTypes@eB] ($(-0.5,0) + (-135:0.23)$) -- ($(-0.5,0) + (-135:0.5)$) node[label=(180+\cmdKV@general@rotate):\cmdKV@gLabels@eLB] {};
    \draw[\cmdKV@gTypes@eC] ($(0.5,0) + (45:0.35)$) -- ($(0.5,0) + (45:0.7)$) node[label=(\cmdKV@general@rotate):\cmdKV@gLabels@eLC] {};
    \draw[\cmdKV@gTypes@eD] ($(0.5,0) + (-45:0.35)$) -- ($(0.5,0) + (-45:0.7)$) node[label=(\cmdKV@general@rotate):\cmdKV@gLabels@eLD] {};
    \draw[label distance=1,\cmdKV@gTypes@iA] ($(-0.5,0) + (30:0.23)$) -- node[label=(90+\cmdKV@general@rotate):\cmdKV@gLabels@iLA] {} ($(0.5,0) + (157:0.3)$);
    \draw[label distance=1,\cmdKV@gTypes@iB] ($(0.5,0) + (-157:0.3)$) -- node[label=(-90+\cmdKV@general@rotate):\cmdKV@gLabels@iLB] {} ($(-0.5,0) + (-30:0.23)$);
    #2
 \end{tikzpicture}
  \gSetStdTypes{line}
}

\makeatother

\usetikzlibrary{fit}
\usetikzlibrary{decorations.markings}
\usetikzlibrary{calc,arrows.meta}
\numberwithin{equation}{section}

\algsetup{indent=2em}

\definecolor{celllabel}{RGB}{85,105,135}
\newcommand{\codefont}{\small\fontencoding{T1}\fontfamily{lmtt}\selectfont}
\lstdefinestyle{codeexample}{
  basicstyle=\codefont,
  columns=fullflexible, keepspaces=true, showstringspaces=false,
  upquote=true, breaklines=true, breakatwhitespace=false,
  literate={-}{{\char45}}1,
  aboveskip=0pt, belowskip=0pt
}
\newtcblisting{terminal}{
  enhanced, listing only, listing options={style=codeexample,language=bash},
  colback=black!5, colframe=black!20, boxrule=0.4pt, sharp corners,
  boxsep=0pt, left=8pt, right=60pt, top=6pt, bottom=6pt,
  before skip=6pt, after skip=6pt,
  overlay={\node[anchor=east, inner sep=0pt,
    font=\footnotesize\sffamily\bfseries, text=black!65]
    at ([xshift=-8pt]frame.east) {Terminal};}
}
\newtcblisting[auto counter]{mathematicacell}{
  enhanced, listing only,
  listing options={style=codeexample,language=Mathematica,
    basicstyle=\codefont\bfseries,
    keywordstyle=\color{black}, stringstyle=\color{black!65},
    commentstyle=\color{black!50}},
  frame hidden, interior hidden, boxrule=0pt, boxsep=0pt,
  left=52pt, right=8pt, top=5pt, bottom=5pt,
  before skip=6pt, after skip=6pt,
  overlay={
    \node[anchor=north east, inner sep=0pt,
      font=\scriptsize\ttfamily, text=celllabel]
      at ([xshift=44pt,yshift=-6pt]frame.north west)
      {In[\thetcbcounter]:=};
    \draw[celllabel, line width=0.5pt]
      ([xshift=-3pt]frame.north east) -- (frame.north east) --
      (frame.south east) -- ([xshift=-3pt]frame.south east);
  }
}

\newcommand\dd{{\textnormal d}}

\newcommand{\balpha}{{\boldsymbol \alpha}}

\newcommand{\bp}{{\boldsymbol p}}

\newcommand{\ba}{{\boldsymbol a}}

\newcommand{\bg}{{\boldsymbol g}}

\newcommand\cD{\mathcal{D}}
\newcommand\cO{\mathcal{O}}

\newcommand\cS{\mathcal{S}}

\newcommand\cL{\mathcal{L}}

\newcommand\cF{\mathcal{F}}

\makeatletter
\newcommand{\Biggg}{\bBigg@{3.5}}
\makeatother

\author[a]{Gregor Kälin}
\author[b,c]{Jungwon Lim}
\affiliation[a]{Institute for Theoretical Physics, ETH Zürich, 8093 Zürich, Switzerland}
\affiliation[b]{Max-Planck-Institut f\"ur Physik, Werner-Heisenberg-Institut, Boltzmannstr. 8, 85748 Garching, Germany}
\affiliation[c]{Deutsches Elektronen-Synchrotron DESY, Notkestr. 85, 22607 Hamburg, Germany}
\emailAdd{kaelingre@gmail.com}
\emailAdd{wonlim@desy.de}

\preprint{
  \begin{minipage}[t]{8cm}
    \begin{flushright}
      MPP-2026-191 \\ DESY-26-143
    \end{flushright}
\end{minipage}}

\title{Graph-Based Algorithms for Worldline EFT Integrands}

\abstract{
We present novel graph-centered algorithms for the construction and representation of worldline effective field theory (WEFT) integrands that separate the topological and algebraic structure of loop integrands into independent \emph{graph--numerator pairs}.
In this representation, the graph encodes all propagators, worldline source and sink vertices, momentum routing, $i0$ prescriptions, and proper-time dependence, while the numerator carries only the minimal algebraic, tensorial content.
To construct integrands at fixed post-Minkowskian order we propose a generalized Berends-Giele construction that minimizes the number of expensive tensor contraction operations.
We also discuss simplification and canonicalization strategies through ``serialization'' of worldline propagators and the application of ``selfie''-subdiagram partial-fraction identities, yielding a minimal canonical topology basis.
We further show that the graph--numerator representation naturally generalizes the notion of an integral family and its sectors, relevant for integration-by-parts (IBP) reduction: propagators are not constrained to a single master labeling, and each sector carries its own minimal set of independent scalar products, completing into an \emph{overcomplete} global family.
This setup minimizes the total number of required integral sectors through an optimal use of sector symmetries and enables independent optimization of propagator labeling for IBP reductions.
The algorithms are implemented in the \texttt{Rust} crate \texttt{FeynRIR}, which is currently able to efficiently compute integrands for the variation of the effective WEFT action, the equations of motion for the worldlines and the (radiation-reacted) impulse.
}

\begin{document}

\maketitle
\flushbottom

\section{Introduction}

Accurate predictions for the dynamics and radiation of compact binaries are essential for interpreting gravitational-wave observations~\cite{LIGOScientific:2016aoc,LIGOScientific:2026sit,LIGOScientific:2026wfs}.
The sensitivity of the next generation of ground-based~\cite{Reitze:2019iox,ET:2025xjr} and space-based detectors~\cite{LISA} will place increasingly stringent demands on analytic calculations of the relativistic two-body problem.
Two complementary perturbative expansions organize these calculations.
The post-Newtonian (PN) expansion in the orbital velocity and gravitational potential describes the slow-motion, weak-field regime, providing increasingly precise predictions for conservative dynamics, radiation reaction, and waveforms during inspiral~\cite{Damour:2014jta,Galley:2015kus,Maia:2017yok,Maia:2017gxn,Marchand:2017pir,Foffa:2019rdf,Foffa:2019yfl,Cho:2022syn,Blanchet:2023sbv,Foffa:2019eeb,Foffa:2019hrb,Blumlein:2019zku,Almeida:2021xwn,Blumlein:2020pyo,Bini:2020wpo,Bini:2020nsb,Bini:2020hmy,Blumlein:2021txj,Porto:2024cwd,Almeida:2024lbv,Cho:2021mqw,Amalberti:2024jaa,Blanchet:2026suq,Pratten:2026tup,Trestini:2026fvg,Blanchet:2026wwa,Bini:2026suo,Porto:2026fsd,Brunello:2025gpf,Brunello:2026anu}.
The post-Minkowskian (PM) expansion is an expansion in Newton's constant $G$ at fixed velocity and is naturally suited to scattering, retaining the full velocity dependence at each order.
Modern scattering-amplitude and worldline methods have driven rapid progress in this framework, expressing conservative and radiative observables in terms of loop integrals~\cite{Khalil:2022ylj,Bini:2024rsy,Damour:2016gwp,Damour:2017zjx,Bjerrum-Bohr:2018xdl,Cheung:2018wkq,Kosower:2018adc,Bern:2019nnu,Kalin:2019rwq,Kalin:2019inp,Cho:2021arx,Damour:2019lcq,Kalin:2020mvi,Kalin:2020fhe,Kalin:2020lmz,Liu:2021zxr,Jakobsen:2021zvh,Mougiakakos:2021ckm,DiVecchia:2021bdo,Kalin:2022hph,Jakobsen:2022psy,Dlapa:2021npj,Dlapa:2021vgp,Dlapa:2022lmu,Bern:2021dqo,Bern:2021yeh,Bini:2022enm,Damgaard:2023ttc,Dlapa:2023hsl,Jakobsen:2023ndj,Jakobsen:2023hig,Klemm:2024wtd,Frellesvig:2023bbf,Bern:2024vqs,Bern:2024adl,Buonanno:2024vkx,Dlapa:2024cje,Dlapa:2025biy,Cheung:2024byb,Bini:2024tft,Driesse:2024xad,Driesse:2024feo,Heissenberg:2025ocy,Heissenberg:2025fcr,Bini:2025vuk,Bern:2025wyd,Driesse:2026qiz,Jakobsen:2026ayw,Damgaard:2026whk,Castet:2026gao,Bini:2026vaq,Geralico:2026efi,Bini:2026jwc,Saavedra:2026ttk,Dlapa:2026oyq}.
Scattering data also provide access to bound motion through the \emph{boundary-to-bound} (B2B) correspondence~\cite{Kalin:2019rwq,Kalin:2019inp,Cho:2021arx}.
For the local conservative dynamics, analytic continuation of the radial action reconstructed from scattering data yields gauge-invariant bound-orbit observables such as the periastron advance and orbital frequencies.
The correspondence extends to radiated energy and angular momentum in the adiabatic approximation~\cite{Saketh:2021sri,Cho:2021arx}, with further developments for generic Kerr geodesics~\cite{Gonzo:2023goe} and tree-level bound waveforms~\cite{Adamo:2024oxy}.
At higher PM orders, nonlocal-in-time tail effects~\cite{Damour:2014jta,Galley:2015kus} require separate treatment when reconstructing dynamics valid for generic bound orbits~\cite{Dlapa:2021vgp,Dlapa:2023hsl,Dlapa:2024cje,Dlapa:2025biy,Dlapa:2026oyq}.
Related effective-one-body (EOB) constructions incorporate PM scattering information into bound-orbit Hamiltonians and, together with radiative information and numerical-relativity input, inspiral-merger-ringdown waveform models~\cite{Khalil:2022ylj,Buonanno:2024byg,Long:2025nmj}.

The worldline effective field theory (WEFT) approach, introduced in~\cite{Kalin:2020mvi}, maps the dynamics of binary compact objects onto the machinery of quantum field theory.
The worldline action couples massive point-particle sources to the gravitational field and admits a systematic PM expansion order by order.
This formalism has been extended to fully radiation-reacted (scattering) observables in~\cite{Kalin:2022hph}.
Adapting many of the advanced computational tools from quantum field theory to the WEFT and the closely related WQFT formalism~\cite{Jakobsen:2021smu,Jakobsen:2022psy} has led to increasingly higher-PM results for gravitational observables, in particular the impulse, the radiated momentum, and the gravitational waveform~\cite{Dlapa:2021npj,Dlapa:2021vgp,Dlapa:2022lmu,Dlapa:2024cje,Dlapa:2025biy,Dlapa:2026oyq,Jakobsen:2023ndj,Jakobsen:2023hig,Jakobsen:2023pvx,Klemm:2024wtd,Driesse:2024xad,Driesse:2026qiz}.

Within the WEFT framework, the construction of integrands at a given PM order is in principle a well-defined field-theoretic problem.
In practice, however, the combinatorial complexity grows factorially with the PM order.
Beyond the well-known explosion in the number of Feynman diagrams, each diagram carries a tensorial numerator whose size is governed by the worldline expansion: every vertex introduces a fresh algebraic structure, and higher-order \emph{iterations}---gravitons emitted and reabsorbed along the classical worldline---inject further nested tensor structures into the numerator.
Moreover, the \emph{representation} of the integrand directly influences the feasibility of the subsequent analytic integration.
In particular, a poor choice of the loop momentum labeling, \emph{integral families}, and integral ordering directly influences the algebraic size of the integrand and the feasibility of the subsequent analytic integration.
The standard tool chain of integration-by-parts (IBP) reduction~\cite{Chetyrkin:1981qh,Tkachov:1981wb} followed by the method of differential equations~\cite{Henn:2013pwa} is thus facing a dual bottleneck: the sheer size and number of basic scalar integrals, and the fact that their representation relies on arbitrary choices during the construction rather than on the demands of the reduction procedure.

To address both of these issues we developed \texttt{FeynRIR} (Feynman Rule Integrands in \texttt{Rust}),\footnote{\texttt{FeynRIR} is a powerful (successor of) \texttt{WoLF} (WorldLine Framework)~\cite{Kalin:2022hph,Neef:thesis} and adopts many of its algorithmic ideas.} a \texttt{Rust} library that implements a \emph{graph-centered} construction, simplification, and canonicalization algorithm for WEFT integrands, using the \texttt{Symbolica} crate for symbolic manipulations.
The central design principle is to treat Feynman diagrams not as monolithic algebraic expressions but as \emph{graph-numerator pairs}, and to build every pair recursively by gluing together smaller partial objects.
This has three important consequences for the construction:

\begin{enumerate}
  \item \textbf{Recursive multi-set construction.}
    The algorithm represents the set of partial graphs at each stage as a \emph{canonical multi-set}, grouping identical graphs under a single representative with a multiplicity count.
    Generalizing the Berends--Giele recursion from quantum chromodynamics to the WEFT context, the algorithm generates all partitions of the current multi-set and glues the partial graphs in each partition through a new graviton vertex.
    This multi-set bookkeeping automatically tracks symmetry factors and ensures that no partial graph is ever constructed twice.
    The recursive structure produces a directed acyclic graph (DAG) of dependencies which is used to schedule the computation and cache intermediate results across PM orders and across different observables.

  \item \textbf{Graph-numerator representation.}
    Every contribution in the integrand is represented as a pair consisting of a Feynman graph and an algebraic numerator.
    Graphical elements encode propagators, WL source vertices, the WL sink vertex, and worldline iterations, while the numerator carries only the \emph{minimal} algebraic information.
    The graph component carries enough structure to reconstruct the full Feynman integral, including momentum routing, $i0$ prescriptions, and proper-time dependence, without ever writing down the corresponding exponential or propagator factors explicitly.
    Throughout the entire construction no explicit choice of independent loop momenta or their labeling is required.
    The resulting representation is therefore generic and systematic, independent of any ad-hoc choices.

  \item \textbf{Canonicalization through partial-fraction identities.}
    After the tree-level integrand is assembled and worldline iterations have been recursively inserted, the resulting set of graph-numerator pairs contains large numbers of redundant topologies.
    \textit{Serialization} identities for WL propagators and \textit{selfie subdiagram} identities for graviton loops acting on the same worldline are used to \emph{planarize} WL propagators and fuse propagators, producing a minimal canonical basis of topologies.
    These identities reorganize the graph component; the numerator only requires the corresponding scalar coefficients and momentum relabeling.
\end{enumerate}

Beyond these construction-level choices, the graph-centered representation enables an important conceptual shift in how one thinks about \emph{integral families} and \emph{sectors}.
In the traditional IBP literature an integral family is defined by a fixed set of propagators and a single choice of loop-momentum routing, with sectors specified by power-counting strings over that fixed set.
This definition, in particular, forces one to introduce an arbitrary, fixed set of \emph{irreducible scalar products (ISPs)} to complete the set of physical propagators.
These ISPs are purely formal, do not correspond to any physical pole, and complicate the embedding of all graphs into the same family and the identification of symmetries, and introduce an arbitrariness in the form of IBP relations.
The graph-numerator representation naturally leads to a generalization of this notion: the propagator labeling is carried by the graph itself, and edges are not constrained to originate from a single master labeling.
Different graphs in the same SF sector may therefore carry different subsets of a global edge labeling, and each individual sector is completed by its own minimal set of independent scalar products (which may be optimized according to different desired criteria).
The collection of all required sectors thus defines an \emph{overcomplete} set of propagators, which is, crucially, overcomplete only globally but linearly independent within each individual sector.
This overcompleteness trivializes symmetries between sectors and allows propagator labelings to be optimized independently for syzygy generation and for IBP rule discovery.
We will show that this generalized notion of an integral family allows every SF sector at fixed PM order to be embedded into a single coherent family, with sector-to-subsector relations and symmetry mappings becoming algorithmically accessible.

In its present form \texttt{FeynRIR} implements the full recursive construction pipeline described in this publication.
The implementation features a fully generic Feynman-rule engine that supports arbitrary gauge choices and total-derivative terms for the bulk graviton vertices of General Relativity.
The variation of the effective action, worldline iterations at arbitrary PM order, and the impulse observable are available.
Canonicalization identities for serialization and selfie subdiagrams are implemented, as is an experimental implementation of the matching of the canonicalized graph-numerator pairs against a topology library.
The algorithm has been applied to generate the impulse integrands up to 5PM, taking less than two hours of computational time using 32 parallel worker threads.

The benchmarks in Sec.~\ref{sec:benchmarks} demonstrate the efficiency of this approach: individual construction phases take less than a second through 3PM, while the 4PM variation of the effective action is constructed in just 17.4\,s on a desktop machine using 32 worker threads.
Obtaining the 4PM impulse from the prepared action and trajectory corrections takes a further 10.2\,s.
At 5PM, the variation of the effective action is constructed in 62.82\,min on a cluster node, likewise using only 32 worker threads, followed by a further 36.58\,min to obtain the impulse from the prepared inputs.
To facilitate further analytic calculations and independent checks, we make the precomputed 1--4PM integrands for the variation of the effective action and the impulse publicly available at \url{https://gitlab.com/kaelingre/weft_integrands/}.

The paper is organized as follows.
Section~\ref{sec:construction} presents the recursive construction of graph-numerator pairs for the variation of the effective action and the insertion of worldline iterations in the PM expansion.
Section~\ref{sec:simplification} develops the graphical partial-fraction identities used to simplify and canonicalize these integrands.
Section~\ref{sec:ibp} organizes the resulting integrands into generalized integral families and sectors for subsequent IBP reduction, including the treatment of symmetries and cut worldline propagators.
Section~\ref{sec:feynrir} describes the \texttt{FeynRIR} library, its installation and output format, and benchmarks through 5PM.
We conclude in Sec.~\ref{sec:conclusions}.

\section{WEFT Integrand construction}\label{sec:construction}

Throughout, we work in dimensional regularization with spacetime dimension $D=4-2\epsilon$.

Using the full WEFT action, following the ideas in~\cite{Kalin:2022hph} (see also the derivations in App.~\ref{app:feynman_rules}), it is a simple task to derive Feynman rules for (minus) the in-in variation of the effective action
\begin{equation}\label{eq:varSeff}
  \cF_a^\mu[x](\sigma) \equiv \left.- \eta^{\mu\nu}\frac{\delta S_\textrm{eff}[x_+,x_-]}{\delta x_{a-}^\nu(\sigma)}\right|_{\substack{x_-=0\\x_+=x}}\,.
\end{equation}
Note that it is convenient to compute minus the variation of the effective action (without a factor of $i$), since this is what appears on the RHS of the equations of motion of the worldlines and what we integrate over to get the impulse $\Delta p$.

The necessary (Feynman) graphs are assembled from the following ingredients:
\begin{itemize}
\item Any number of \emph{WL source} vertices, which couple the bodies to the gravitational field,
  \begin{equation}
    \begin{tikzpicture}%
      [baseline={([yshift=-0.5ex]current bounding box.center)},font=\scriptsize]%
      \draw[directed graviton] (0,0) node[above=0.1] {$a$} -- node[left] {$k\downarrow$} (0,-1) node[below=0.1] {$\mu,\nu$};
      \fill (0,0) circle (0.04);
      \end{tikzpicture}
    = -\frac{i m_a \kappa}{2}\int \dd\tau e^{i k\cdot x_a(\tau)} \dot{x}_a^\mu(\tau) \dot{x}_a^\nu(\tau)
  \end{equation}
\item A single \emph{WL sink} vertex, which inserts the force obtained by varying the worldline,
  \begin{equation}
    \begin{tikzpicture}%
      [baseline={([yshift=-0.5ex]current bounding box.center)},font=\scriptsize]%
      \node[xSource] (A) at (0,0) {};
      \draw[directed graviton] (0,-1) node[below=0.1] {$\mu,\nu$} -- node[left] {$k\uparrow$}  (A) node[above=0.1] {$a,\sigma,\alpha$};
    \end{tikzpicture}
    = -\frac{im_a \kappa}{2} e^{-i k\cdot x_a(\sigma)} \begin{multlined}[t]
      \left[
      k^\alpha \dot{x}_a^\mu(\sigma)\dot{x}_a^\nu(\sigma)
      -k\cdot \dot{x}_a(\sigma) \eta^{\mu\alpha}\dot{x}_a^{\nu}(\sigma)
      -i\,\eta^{\mu\alpha}\ddot{x}_a^\nu(\sigma)\right.\\
      -\left.k\cdot\dot{x}_a(\sigma)\eta^{\nu\alpha}\dot{x}_a^\mu(\sigma)
      -i\,\eta^{\nu\alpha}\ddot{x}_a^\mu(\sigma) 
    \right]
    \end{multlined}
  \end{equation}
  This rule contains an additional factor of $i$ compared to the expression presented in~\cite{Dlapa:2023hsl}.
  This cancels the imaginary unit in the exponential of the effective action $e^{i\,\cS_\textrm{eff}}$ and introduces the convenient minus sign, such that these rules are truly computing minus the variation of the effective action.
\item Graviton vertices describe gravitational self-interactions, each with a single outgoing graviton propagator and the rest incoming
  \begin{align}
    &\begin{tikzpicture}%
      [baseline={([yshift=-0.5ex]current bounding box.center)},font=\scriptsize]%
      \draw[directed graviton] (0,0) -- (90:1);
      \draw[directed graviton] (210:1) -- (0,0);
      \draw[directed graviton] (330:1) -- (0,0);
    \end{tikzpicture}
    \,,
    &
    &\begin{tikzpicture}%
      [baseline={([yshift=-0.5ex]current bounding box.center)},font=\scriptsize]%
      \draw[directed graviton] (0,0) -- (90:1);
      \draw[directed graviton] (180:1) -- (0,0);
      \draw[directed graviton] (270:1) -- (0,0);
      \draw[directed graviton] (0:1) -- (0,0);
    \end{tikzpicture}
    \,,
    &
    &\begin{tikzpicture}%
      [baseline={([yshift=-0.5ex]current bounding box.center)},font=\scriptsize]%
      \draw[directed graviton] (0,0) -- (90:1);
      \draw[directed graviton] (162:1) -- (0,0);
      \draw[directed graviton] (234:1) -- (0,0);
      \draw[directed graviton] (306:1) -- (0,0);
      \draw[directed graviton] (18:1) -- (0,0);
    \end{tikzpicture}
    \,,
    &
    &\begin{tikzpicture}%
      [baseline={([yshift=-0.5ex]current bounding box.center)},font=\scriptsize]%
      \draw[directed graviton] (0,0) -- (90:1);
      \draw[directed graviton] (150:1) -- (0,0);
      \draw[directed graviton] (210:1) -- (0,0);
      \draw[directed graviton] (270:1) -- (0,0);
      \draw[directed graviton] (330:1) -- (0,0);
      \draw[directed graviton] (30:1) -- (0,0);
    \end{tikzpicture}
    \,, \dots
  \end{align}
  The \texttt{FeynRIR} library contains hard coded (algebraically minimal) expressions for these vertices, see e.g.~\cite{Kalin:2020mvi, Dlapa:2023hsl}.
  Furthermore, the crate also contains a module that allows the user to generate graviton Feynman rules for a generic gauge and with generic total derivative terms, see also App.~\ref{app:feynman_rules}.
  The parameters in these expressions may then be freely chosen by the user or even used symbolically.
\end{itemize}

\subsection{A recursive algorithm for the variation of the effective action}

We propose an efficient recursive algorithm for the computation of the variation of the effective action.
It can be seen as a generalization of the Berends-Giele construction~\cite{Berends:1987me} where we have two different types of ``external'' legs, i.e. the WL source vertices of the two black holes.
During the whole construction, we pair up partially constructed graphs with an algebraic expression for their numerators according to the Feynman rules.
Each partial graph has one distinguished vertex, which is the latest one that has been added to it.
At each step of the recursion we have a collection of such graph-numerator pairs, which are glued together step by step.

At a fixed PM order, the recursion has several starting configurations, one for each assignment of WL source vertices to the two black holes.
To be precise: At $n$PM order we need $n$ WL source vertices and one WL sink vertex.
In this recursive procedure, we ignore the WL sink vertex until the very end.
Hence, we distribute the $n$ WL source vertices in all possible ways to the two BHs.
E.g. at 4PM we have five ways to do that.
\begin{align}
  (1111)\,, & &(1112)\,, & &(1122)\,, & &(1222)\,, & &(2222)\,.
\end{align}
Given such a distribution, the starting point for the recursion is the collection of graph-numerator pairs where each WL source vertex is a partial graph paired with the vertex rule for this source.
Given a collection of graph-numerator pairs, at each step of the recursion we do the following:
\begin{itemize}
\item Generate all possible unique (unordered) partitions of the collection.
  Throw away the trivial partition, i.e. where each element of the partition is a singleton.
  By singleton we mean a single graph-numerator pair which in this step of the recursion is grouped only with itself and no gluing can happen.
  This is necessary in order not to end up with an infinite recursion.
  We also throw away partitions that contain an element which only groups together two or more objects that were treated as singletons in the last step of the recursion.
  The latter filter is tricky, but very important in order not to overcount (starting from 4PM on).
\item For each group of size $m\geq 2$ in the partition, glue together the $m$ partial graphs by connecting the distinguished vertex of each partial graph with a new $(m+1)$-valent graviton vertex through a graviton propagator.
  The $(m+1)$th leg of this vertex is kept with open indices and the new vertex is marked as the new distinguished vertex of the graph-numerator.
\item Perform the analog of the last step for the algebraic numerator expressions by multiplying them together with the propagator rules and the vertex rule for the new graviton vertex.
  Then contract the open indices.
  \texttt{FeynRIR} uses an optimized strategy by contracting edge-by-edge in order to never produce large intermediate expressions.
\end{itemize}
Finally, when the collection has only one element, the recursion terminates.
As a final step, we connect the latest distinguished vertex of this single component to a WL sink vertex and do once more the same for the numerator by multiplying it with a graviton propagator and the vertex rule for the sink.

Some technical comments are in order:
\begin{itemize}
\item We assign to each vertex and edge a unique numeric identifier.
  The unique identifier for graviton propagators is used to label their momentum.
  Note that we do not yet resolve momentum conservation rules.
  The unique identifier for the WL vertices is used to label the proper time in the vertex rule over which we will eventually integrate.
\item When gluing together graphs, the identifiers of one graph must be shifted up in order to make them unique again.
  The same shift must be applied to the loop momenta and the proper times to keep the graph and the numerator expression synchronized at any time.
\item For in-in integrands, one must use directed (arrowed) graviton propagators.
  We never write down algebraic expressions for the propagators or the exponential factors in the WL source or sink vertex rules as the graphs contain all information to reconstruct them.
  E.g. it is the arrowed edges of the graph that encode the $i0$ prescription.
\item Internally, we use \emph{multisets} to represent the collection of partial graphs and numerators at each stage of the algorithm.
  A multiset doesn't allow repeated identical elements, but assigns a counter to each element instead, which denotes how many times the corresponding element appears.
  This makes it much easier to keep track of identical configurations/partitions.
  It also trivializes the computation of the symmetry factor of a diagram, which is also computed recursively within the above algorithm.
\end{itemize}

\paragraph{Example.}
Let us work through an explicit example of this recursion.
Consider the 4PM variation of the effective action with the $(1122)$ starting configuration, i.e. two WL sources for each of the black holes (and one WL sink vertex for BH1).
We denote the distinguished vertex of each partial graph with a cross.
The starting point of the recursion is hence this collection of four partial graphs (and their numerator, which we will not spell out explicitly):
\newcommand{\pgOne}{{%
  \begin{tikzpicture}%
     [baseline={([yshift=-0.5ex]current bounding box.center)},font=\scriptsize]%
     \node[xSource] at (0,0) {} node[above=0.1] {1};%
  \end{tikzpicture}}}
\newcommand{\pgTwo}{%
  \begin{tikzpicture}%
     [baseline={([yshift=-0.5ex]current bounding box.center)},font=\scriptsize]%
     \node[xSource] at (0,0) {} node[above=0.1] {2};%
  \end{tikzpicture}}
\begin{equation}
  \left\{\pgOne,\pgOne,\pgTwo,\pgTwo\right\}
\end{equation}
\begingroup
\allowdisplaybreaks
\begin{itemize}
\item\textbf{Recursion 1: }$(A)$ to $(H)$ label the (required) partitions in this iteration.
  \begin{align*}
    & \textrm{label} & & \textrm{Gluing process} & & \textrm{symmetry factor} & & \textrm{(status)}\\
    &(A): & \left(\pgOne\pgOne\pgTwo\pgTwo\right) &\rightarrow
    \left\{
    \begin{tikzpicture}
      [baseline={([yshift=-0.5ex]current bounding box.center)},font=\scriptsize]
      \node (A) at (-0.5,0.5) {};
      \fill (A) circle (0.05);
      \node (B) at (0.5,0.5) {};
      \fill (B) circle (0.05);
      \node (C) at (-0.5,-0.5) {};
      \fill (C) circle (0.05);
      \node (D) at (0.5,-0.5) {};
      \fill (D) circle (0.05);
      \node[xSource] (E) at (0,0) {};
      \draw[directed graviton] (A) node[above=0.1] {1} -- (E);
      \draw[directed graviton] (B) node[above=0.1] {1} -- (E);
      \draw[directed graviton] (C) node[below=0.1] {2} -- (E);
      \draw[directed graviton] (D) node[below=0.1] {2} -- (E);
    \end{tikzpicture}
    \right\} & & \{2 \times 2\} & & \textrm{(done)}
    \\
    &(B): & \left(\pgOne\pgOne\pgTwo\right)\left(\pgTwo\right) &\rightarrow
    \left\{
    \begin{tikzpicture}
      [baseline={([yshift=-0.5ex]current bounding box.center)},font=\scriptsize]
      \node (A) at (-0.5,0.5) {};
      \fill (A) circle (0.05);
      \node (B) at (0.5,0.5) {};
      \fill (B) circle (0.05);
      \node (C) at (0,-0.5) {};
      \fill (C) circle (0.05);
      \node[xSource] (D) at (0,0) {};
      \draw[directed graviton] (A) node[above=0.1] {1} -- (D);
      \draw[directed graviton] (B) node[above=0.1] {1} -- (D);
      \draw[directed graviton] (C) node[below=0.1] {2} -- (D);
    \end{tikzpicture}
    ,
    \pgTwo
    \right\} & & \{2\}
    \\
    &(C): & \left(\pgOne\right)\left(\pgOne\pgTwo\pgTwo\right) &\rightarrow
    \left\{
    \pgOne,
    \begin{tikzpicture}
      [baseline={([yshift=-0.5ex]current bounding box.center)},font=\scriptsize]
      \node (A) at (0,0.5) {};
      \fill (A) circle (0.05);
      \node (B) at (-0.5,-0.5) {};
      \fill (B) circle (0.05);
      \node (C) at (0.5,-0.5) {};
      \fill (C) circle (0.05);
      \node[xSource] (D) at (0,0) {};
      \draw[directed graviton] (A) node[above=0.1] {1} -- (D);
      \draw[directed graviton] (B) node[below=0.1] {2} -- (D);
      \draw[directed graviton] (C) node[below=0.1] {2} -- (D);
    \end{tikzpicture}
    \right\} & & \{2\}
    \\
    &(D): & \left(\pgOne\pgOne\right)\left(\pgTwo\pgTwo\right) &\rightarrow
    \left\{
    \begin{tikzpicture}
      [baseline={([yshift=-0.5ex]current bounding box.center)},font=\scriptsize]
      \node (A) at (-0.5,0.5) {};
      \fill (A) circle (0.05);
      \node (B) at (0.5,0.5) {};
      \fill (B) circle (0.05);
      \node[xSource] (C) at (0,0) {};
      \draw[directed graviton] (A) node[above=0.1] {1} -- (C);
      \draw[directed graviton] (B) node[above=0.1] {1} -- (C);
    \end{tikzpicture}
    ,
    \begin{tikzpicture}
      [baseline={([yshift=-0.5ex]current bounding box.center)},font=\scriptsize]
      \node (A) at (-0.5,-0.5) {};
      \fill (A) circle (0.05);
      \node (B) at (0.5,-0.5) {};
      \fill (B) circle (0.05);
      \node[xSource] (C) at (0,0) {};
      \draw[directed graviton] (A) node[below=0.1] {2} -- (C);
      \draw[directed graviton] (B) node[below=0.1] {2} -- (C);
    \end{tikzpicture}
    \right\} & & \{2,2\}
    \\
    &(E): & \left(\pgOne\pgTwo\right)\left(\pgOne\pgTwo\right) &\rightarrow
    \left\{
    \begin{tikzpicture}
      [baseline={([yshift=-0.5ex]current bounding box.center)},font=\scriptsize]
      \node (A) at (0,0.5) {};
      \fill (A) circle (0.05);
      \node (B) at (0,-0.5) {};
      \fill (B) circle (0.05);
      \node[xSource] (C) at (0,0) {};
      \draw[directed graviton] (A) node[above=0.1] {1} -- (C);
      \draw[directed graviton] (B) node[below=0.1] {2} -- (C);
    \end{tikzpicture},
    \begin{tikzpicture}
      [baseline={([yshift=-0.5ex]current bounding box.center)},font=\scriptsize]
      \node (A) at (0,0.5) {};
      \fill (A) circle (0.05);
      \node (B) at (0,-0.5) {};
      \fill (B) circle (0.05);
      \node[xSource] (C) at (0,0) {};
      \draw[directed graviton] (A) node[above=0.1] {1} -- (C);
      \draw[directed graviton] (B) node[below=0.1] {2} -- (C);
    \end{tikzpicture}
    \right\} & & \{1,1\}
    \\
    &(F): & \left(\pgOne\pgOne\right)\left(\pgTwo\right)\left(\pgTwo\right) &\rightarrow
    \left\{
    \begin{tikzpicture}
      [baseline={([yshift=-0.5ex]current bounding box.center)},font=\scriptsize]
      \node (A) at (-0.5,0.5) {};
      \fill (A) circle (0.05);
      \node (B) at (0.5,0.5) {};
      \fill (B) circle (0.05);
      \node[xSource] (C) at (0,0) {};
      \draw[directed graviton] (A) node[above=0.1] {1} -- (C);
      \draw[directed graviton] (B) node[above=0.1] {1} -- (C);
    \end{tikzpicture}
    ,\pgTwo,\pgTwo
    \right\} & & \{2,1,1\}
    \\
    &(G): & \left(\pgTwo\pgTwo\right)\left(\pgOne\right)\left(\pgOne\right) &\rightarrow
    \left\{
    \begin{tikzpicture}
      [baseline={([yshift=-0.5ex]current bounding box.center)},font=\scriptsize]
      \node (A) at (-0.5,-0.5) {};
      \fill (A) circle (0.05);
      \node (B) at (0.5,-0.5) {};
      \fill (B) circle (0.05);
      \node[xSource] (C) at (0,0) {};
      \draw[directed graviton] (A) node[below=0.1] {2} -- (C);
      \draw[directed graviton] (B) node[below=0.1] {2} -- (C);
    \end{tikzpicture}
    ,\pgOne,\pgOne
    \right\} & & \{2,1,1\}
    \\
    &(H): & \left(\pgOne\pgTwo\right)\left(\pgOne\right)\left(\pgTwo\right) &\rightarrow
    \left\{
    \begin{tikzpicture}
      [baseline={([yshift=-0.5ex]current bounding box.center)},font=\scriptsize]
      \node (A) at (0,0.5) {};
      \fill (A) circle (0.05);
      \node (B) at (0,-0.5) {};
      \fill (B) circle (0.05);
      \node[xSource] (C) at (0,0) {};
      \draw[directed graviton] (A) node[above=0.1] {1} -- (C);
      \draw[directed graviton] (B) node[below=0.1] {2} -- (C);
    \end{tikzpicture}
    ,\pgOne,\pgTwo
    \right\} & & \{1,1,1\}
  \end{align*}
  Note that we ignore the all-singleton partition $\left(\pgOne\right)\left(\pgOne\right)\left(\pgTwo\right)\left(\pgTwo\right)$.
\item\textbf{Recursion 2: }
  \begin{align*}
    & & & \textrm{Process} & & \textrm{symmetry factor} & & \textrm{(status)}\\
    &(B): & \left(
    \begin{tikzpicture}
      [baseline={([yshift=-0.5ex]current bounding box.center)},font=\scriptsize]
      \node (A) at (-0.5,0.5) {};
      \fill (A) circle (0.05);
      \node (B) at (0.5,0.5) {};
      \fill (B) circle (0.05);
      \node (C) at (0,-0.5) {};
      \fill (C) circle (0.05);
      \node[xSource] (D) at (0,0) {};
      \draw[directed graviton] (A) node[above=0.1] {1} -- (D);
      \draw[directed graviton] (B) node[above=0.1] {1} -- (D);
      \draw[directed graviton] (C) node[below=0.1] {2} -- (D);
    \end{tikzpicture}
    \pgTwo
    \right)
    &\rightarrow
    \left\{
    \begin{tikzpicture}
      [baseline={([yshift=-0.5ex]current bounding box.center)},font=\scriptsize]
      \node (A) at (-0.5,0.75) {};
      \fill (A) circle (0.05);
      \node (B) at (0.5,0.75) {};
      \fill (B) circle (0.05);
      \node (C) at (-0.5,-0.75) {};
      \fill (C) circle (0.05);
      \node (D) at (1,-0.75) {};
      \fill (D) circle (0.05);
      \node[xSource] (E) at (0.5,-0.375) {};
      \draw[directed graviton] (A) node[above=0.1] {1} -- (0,0);
      \draw[directed graviton] (B) node[above=0.1] {1} -- (0,0);
      \draw[directed graviton] (C) node[below=0.1] {2} -- (0,0);
      \draw[directed graviton] (D) node[below=0.1] {2} -- (E);
      \draw[directed graviton] (0,0) -- (E);
    \end{tikzpicture}
    \right\} & & \{2\times 1\} & & \textrm{(done)}\\
    &(C): & \left(
    \pgOne
    \begin{tikzpicture}
      [baseline={([yshift=-0.5ex]current bounding box.center)},font=\scriptsize]
      \node (A) at (0,0.5) {};
      \fill (A) circle (0.05);
      \node (B) at (-0.5,-0.5) {};
      \fill (B) circle (0.05);
      \node (C) at (0.5,-0.5) {};
      \fill (C) circle (0.05);
      \node[xSource] (D) at (0,0) {};
      \draw[directed graviton] (A) node[above=0.1] {1} -- (D);
      \draw[directed graviton] (B) node[below=0.1] {2} -- (D);
      \draw[directed graviton] (C) node[below=0.1] {2} -- (D);
    \end{tikzpicture}
    \right)
    &\rightarrow
    \left\{
    \begin{tikzpicture}
      [baseline={([yshift=-0.5ex]current bounding box.center)},font=\scriptsize]
      \node (A) at (-0.5,0.75) {};
      \fill (A) circle (0.05);
      \node (B) at (-0.5,-0.75) {};
      \fill (B) circle (0.05);
      \node (C) at (0.5,-0.75) {};
      \fill (C) circle (0.05);
      \node (D) at (1,0.75) {};
      \fill (D) circle (0.05);
      \node[xSource] (E) at (0.5,0.375) {};
      \draw[directed graviton] (A) node[above=0.1] {1} -- (0,0);
      \draw[directed graviton] (B) node[below=0.1] {2} -- (0,0);
      \draw[directed graviton] (C) node[below=0.1] {2} -- (0,0);
      \draw[directed graviton] (D) node[above=0.1] {1} -- (E);
      \draw[directed graviton] (0,0) -- (E);
    \end{tikzpicture}
    \right\} & & \{2\times1\}& & \textrm{(done)}
    \\
    &(D): & \left(
    \begin{tikzpicture}
      [baseline={([yshift=-0.5ex]current bounding box.center)},font=\scriptsize]
      \node (A) at (-0.5,0.5) {};
      \fill (A) circle (0.05);
      \node (B) at (0.5,0.5) {};
      \fill (B) circle (0.05);
      \node[xSource] (C) at (0,0) {};
      \draw[directed graviton] (A) node[above=0.1] {1} -- (C);
      \draw[directed graviton] (B) node[above=0.1] {1} -- (C);
    \end{tikzpicture}
    \begin{tikzpicture}
      [baseline={([yshift=-0.5ex]current bounding box.center)},font=\scriptsize]
      \node (A) at (-0.5,-0.5) {};
      \fill (A) circle (0.05);
      \node (B) at (0.5,-0.5) {};
      \fill (B) circle (0.05);
      \node[xSource] (C) at (0,0) {};
      \draw[directed graviton] (A) node[below=0.1] {2} -- (C);
      \draw[directed graviton] (B) node[below=0.1] {2} -- (C);
    \end{tikzpicture}
    \right)
    &\rightarrow
    \left\{
    \begin{tikzpicture}
      [baseline={([yshift=-0.5ex]current bounding box.center)},font=\scriptsize]
      \node (A) at (-0.5,1) {};
      \fill (A) circle (0.05);
      \node (B) at (0.5,1) {};
      \fill (B) circle (0.05);
      \draw[directed graviton] (A) node[above=0.1] {1} -- (0,0.5);
      \draw[directed graviton] (B) node[above=0.1] {1} -- (0,0.5);
      \node (C) at (-0.5,-1) {};
      \fill (C) circle (0.05);
      \node (D) at (0.5,-1) {};
      \fill (D) circle (0.05);
      \draw[directed graviton] (C) node[below=0.1] {2} -- (0,-0.5);
      \draw[directed graviton] (D) node[below=0.1] {2} -- (0,-0.5);
      \node[xSource] (E) at (0,0) {};
      \draw[directed graviton] (0,0.5)  -- (E);
      \draw[directed graviton] (0,-0.5)  -- (E);
    \end{tikzpicture}
    \right\} & & \{(2\times 2) \times 1\} & & \textrm{(done)}\\
    &(E): & \left(
    \begin{tikzpicture}
      [baseline={([yshift=-0.5ex]current bounding box.center)},font=\scriptsize]
      \node (A) at (0,0.5) {};
      \fill (A) circle (0.05);
      \node (B) at (0,-0.5) {};
      \fill (B) circle (0.05);
      \node[xSource] (C) at (0,0) {};
      \draw[directed graviton] (A) node[above=0.1] {1} -- (C);
      \draw[directed graviton] (B) node[below=0.1] {2} -- (C);
    \end{tikzpicture}
    \quad
    \begin{tikzpicture}
      [baseline={([yshift=-0.5ex]current bounding box.center)},font=\scriptsize]
      \node (A) at (0,0.5) {};
      \fill (A) circle (0.05);
      \node (B) at (0,-0.5) {};
      \fill (B) circle (0.05);
      \node[xSource] (C) at (0,0) {};
      \draw[directed graviton] (A) node[above=0.1] {1} -- (C);
      \draw[directed graviton] (B) node[below=0.1] {2} -- (C);
    \end{tikzpicture}
    \right)
    &\rightarrow
    \left\{
    \begin{tikzpicture}
      [baseline={([yshift=-0.5ex]current bounding box.center)},font=\scriptsize]
      \node (A) at (-0.75,0.5) {};
      \fill (A) circle (0.05);
      \node (B) at (-0.75,-0.5) {};
      \fill (B) circle (0.05);
      \draw[directed graviton] (A) node[above=0.1] {1} -- (-0.75,0);
      \draw[directed graviton] (B) node[below=0.1] {2} -- (-0.75,0);
      \node (C) at (0.75,0.5) {};
      \fill (C) circle (0.05);
      \node (D) at (0.75,-0.5) {};
      \fill (D) circle (0.05);
      \draw[directed graviton] (C) node[above=0.1] {1} -- (0.75,0);
      \draw[directed graviton] (D) node[below=0.1] {2} -- (0.75,0);
      \node[xSource] (E) at (0,0) {};
      \draw[directed graviton] (-0.75,0) -- (E);
      \draw[directed graviton] (0.75,0) -- (E);
    \end{tikzpicture}
    \right\} & & \{(1\times 1) \times 2\} & & \textrm{(done)}\\
    &(F1): & 
    \left(
    \begin{tikzpicture}
      [baseline={([yshift=-0.5ex]current bounding box.center)},font=\scriptsize]
      \node (A) at (-0.5,0.5) {};
      \fill (A) circle (0.05);
      \node (B) at (0.5,0.5) {};
      \fill (B) circle (0.05);
      \node[xSource] (C) at (0,0) {};
      \draw[directed graviton] (A) node[above=0.1] {1} -- (C);
      \draw[directed graviton] (B) node[above=0.1] {1} -- (C);
    \end{tikzpicture}
    \pgTwo\quad \pgTwo
    \right)
    &\rightarrow
    \left\{
    \begin{tikzpicture}
      [baseline={([yshift=-0.5ex]current bounding box.center)},font=\scriptsize]
      \node (A) at (-0.5,0.75) {};
      \fill (A) circle (0.05);
      \node (B) at (0.5,0.75) {};
      \fill (B) circle (0.05);
      \draw[directed graviton] (A) node[above=0.1] {1} -- (0,0.25);
      \draw[directed graviton] (B) node[above=0.1] {1} -- (0,0.25);
      \node (C) at (-0.5,-0.75) {};
      \fill (C) circle (0.05);
      \node (D) at (0.5,-0.75) {};
      \fill (D) circle (0.05);
      \node[xSource] (E) at (0,-0.25) {};
      \draw[directed graviton] (C) node[below=0.1] {2} -- (E);
      \draw[directed graviton] (D) node[below=0.1] {2} -- (E);
      \draw[directed graviton] (0,0.25) -- (E);
    \end{tikzpicture}
    \right\} & & \{(2\times 1 \times 1) \times 2\} & & \textrm{(done)}\\
    &(F2): & 
    \left(
    \begin{tikzpicture}
      [baseline={([yshift=-0.5ex]current bounding box.center)},font=\scriptsize]
      \node (A) at (-0.5,0.5) {};
      \fill (A) circle (0.05);
      \node (B) at (0.5,0.5) {};
      \fill (B) circle (0.05);
      \node[xSource] (C) at (0,0) {};
      \draw[directed graviton] (A) node[above=0.1] {1} -- (C);
      \draw[directed graviton] (B) node[above=0.1] {1} -- (C);
    \end{tikzpicture}
    \pgTwo\right)\left(\pgTwo\right)
    &\rightarrow
    \left\{
    \begin{tikzpicture}
      [baseline={([yshift=-0.5ex]current bounding box.center)},font=\scriptsize]
      \node (A) at (-0.5,0.75) {};
      \fill (A) circle (0.05);
      \node (B) at (0.5,0.75) {};
      \fill (B) circle (0.05);
      \draw[directed graviton] (A) node[above=0.1] {1} -- (0,0.25);
      \draw[directed graviton] (B) node[above=0.1] {1} -- (0,0.25);
      \node (C) at (0,-0.75) {};
      \fill (C) circle (0.05);
      \node[xSource] (E) at (0,-0.25) {};
      \draw[directed graviton] (C) node[below=0.1] {2} -- (E);
      \draw[directed graviton] (0,0.25) -- (E);
    \end{tikzpicture},
    \pgTwo
    \right\} & & \{(2\times1)\times 1, 1\times 1\} \\
    &(G1): & 
    \left(
    \begin{tikzpicture}
      [baseline={([yshift=-0.5ex]current bounding box.center)},font=\scriptsize]
      \node (A) at (-0.5,-0.5) {};
      \fill (A) circle (0.05);
      \node (B) at (0.5,-0.5) {};
      \fill (B) circle (0.05);
      \node[xSource] (C) at (0,0) {};
      \draw[directed graviton] (A) node[below=0.1] {2} -- (C);
      \draw[directed graviton] (B) node[below=0.1] {2} -- (C);
    \end{tikzpicture}
    \pgOne\quad \pgOne
    \right)
    &\rightarrow
    \left\{
    \begin{tikzpicture}
      [baseline={([yshift=-0.5ex]current bounding box.center)},font=\scriptsize]
      \node (A) at (-0.5,-0.75) {};
      \fill (A) circle (0.05);
      \node (B) at (0.5,-0.75) {};
      \fill (B) circle (0.05);
      \draw[directed graviton] (A) node[below=0.1] {2} -- (0,-0.25);
      \draw[directed graviton] (B) node[below=0.1] {2} -- (0,-0.25);
      \node (C) at (-0.5,0.75) {};
      \fill (C) circle (0.05);
      \node (D) at (0.5,0.75) {};
      \fill (D) circle (0.05);
      \node[xSource] (E) at (0,0.25) {};
      \draw[directed graviton] (C) node[above=0.1] {1} -- (E);
      \draw[directed graviton] (D) node[above=0.1] {1} -- (E);
      \draw[directed graviton] (0,-0.25) -- (E);
    \end{tikzpicture}
    \right\} & & \{(2\times 1 \times 1)\times 2\} & & \textrm{(done)}\\
    &(G2): & 
    \left(
    \begin{tikzpicture}
      [baseline={([yshift=-0.5ex]current bounding box.center)},font=\scriptsize]
      \node (A) at (-0.5,-0.5) {};
      \fill (A) circle (0.05);
      \node (B) at (0.5,-0.5) {};
      \fill (B) circle (0.05);
      \node[xSource] (C) at (0,0) {};
      \draw[directed graviton] (A) node[below=0.1] {2} -- (C);
      \draw[directed graviton] (B) node[below=0.1] {2} -- (C);
    \end{tikzpicture}
    \pgOne\right)\left(\pgOne\right)
    &\rightarrow
    \left\{
    \begin{tikzpicture}
      [baseline={([yshift=-0.5ex]current bounding box.center)},font=\scriptsize]
      \node (A) at (-0.5,-0.75) {};
      \fill (A) circle (0.05);
      \node (B) at (0.5,-0.75) {};
      \fill (B) circle (0.05);
      \draw[directed graviton] (A) node[below=0.1] {2} -- (0,-0.25);
      \draw[directed graviton] (B) node[below=0.1] {2} -- (0,-0.25);
      \node (C) at (0,0.75) {};
      \fill (C) circle (0.05);
      \node[xSource] (E) at (0,0.25) {};
      \draw[directed graviton] (C) node[above=0.1] {1} -- (E);
      \draw[directed graviton] (0,-0.25) -- (E);
    \end{tikzpicture},
    \pgOne
    \right\} & & \{(2\times 1)\times 1, 1 \times 1\}\\
    &(H1): &
    \left(
    \begin{tikzpicture}
      [baseline={([yshift=-0.5ex]current bounding box.center)},font=\scriptsize]
      \node (A) at (0,0.5) {};
      \fill (A) circle (0.05);
      \node (B) at (0,-0.5) {};
      \fill (B) circle (0.05);
      \node[xSource] (C) at (0,0) {};
      \draw[directed graviton] (A) node[above=0.1] {1} -- (C);
      \draw[directed graviton] (B) node[below=0.1] {2} -- (C);
    \end{tikzpicture}
    \quad\pgOne\quad\pgTwo
    \right)
    &\rightarrow
    \left\{
    \begin{tikzpicture}
      [baseline={([yshift=-0.5ex]current bounding box.center)},font=\scriptsize]
      \node (A) at (-0.5,0.5) {};
      \fill (A) circle (0.05);
      \node (B) at (-0.5,-0.5) {};
      \fill (B) circle (0.05);
      \draw[directed graviton] (A) node[above=0.1] {1} -- (-0.5,0);
      \draw[directed graviton] (B) node[below=0.1] {2} -- (-0.5,0);
      \node (C) at (0.5,0.5) {};
      \fill (C) circle (0.05);
      \node (D) at (0.5,-0.5) {};
      \fill (D) circle (0.05);
      \node[xSource] (E) at (0.5,0) {};
      \draw[directed graviton] (C) node[above=0.1] {1} -- (E);
      \draw[directed graviton] (D) node[below=0.1] {2} -- (E);
      \draw[directed graviton] (-0.5,0) -- (E);
    \end{tikzpicture}
    \right\} & & \{(1\times 1 \times 1)\times 1\}& & \textrm{(done)}\\
    &(H2): &
    \left(
    \begin{tikzpicture}
      [baseline={([yshift=-0.5ex]current bounding box.center)},font=\scriptsize]
      \node (A) at (0,0.5) {};
      \fill (A) circle (0.05);
      \node (B) at (0,-0.5) {};
      \fill (B) circle (0.05);
      \node[xSource] (C) at (0,0) {};
      \draw[directed graviton] (A) node[above=0.1] {1} -- (C);
      \draw[directed graviton] (B) node[below=0.1] {2} -- (C);
    \end{tikzpicture}
    \quad\pgOne\right)\left(\pgTwo\right)
    &\rightarrow
    \left\{
    \begin{tikzpicture}
      [baseline={([yshift=-0.5ex]current bounding box.center)},font=\scriptsize]
      \node (A) at (-0.5,0.5) {};
      \fill (A) circle (0.05);
      \node (B) at (-0.5,-0.5) {};
      \fill (B) circle (0.05);
      \draw[directed graviton] (A) node[above=0.1] {1} -- (-0.5,0);
      \draw[directed graviton] (B) node[below=0.1] {2} -- (-0.5,0);
      \node (C) at (0.5,0.5) {};
      \fill (C) circle (0.05);
      \node[xSource] (E) at (0,0.25) {};
      \draw[directed graviton] (C) node[above=0.1] {1} -- (E);
      \draw[directed graviton] (-0.5,0) -- (E);
    \end{tikzpicture}
    ,\pgTwo
    \right\} & & \{(1\times 1)\times 1, 1\times 1\}\\
    &(H3): &
    \left(
    \begin{tikzpicture}
      [baseline={([yshift=-0.5ex]current bounding box.center)},font=\scriptsize]
      \node (A) at (0,0.5) {};
      \fill (A) circle (0.05);
      \node (B) at (0,-0.5) {};
      \fill (B) circle (0.05);
      \node[xSource] (C) at (0,0) {};
      \draw[directed graviton] (A) node[above=0.1] {1} -- (C);
      \draw[directed graviton] (B) node[below=0.1] {2} -- (C);
    \end{tikzpicture}
    \quad\pgTwo\right)\left(\pgOne\right)
    &\rightarrow
    \left\{
    \begin{tikzpicture}
      [baseline={([yshift=-0.5ex]current bounding box.center)},font=\scriptsize]
      \node (A) at (-0.5,0.5) {};
      \fill (A) circle (0.05);
      \node (B) at (-0.5,-0.5) {};
      \fill (B) circle (0.05);
      \draw[directed graviton] (A) node[above=0.1] {1} -- (-0.5,0);
      \draw[directed graviton] (B) node[below=0.1] {2} -- (-0.5,0);
      \node (C) at (0.5,-0.5) {};
      \fill (C) circle (0.05);
      \node[xSource] (E) at (0,-0.25) {};
      \draw[directed graviton] (C) node[below=0.1] {2} -- (E);
      \draw[directed graviton] (-0.5,0) -- (E);
    \end{tikzpicture}
    ,\pgOne
    \right\} & & \{(1\times 1)\times 1, 1\times 1\}
  \end{align*}
  Note that there exists one more partition for the cases $(F)$, $(G)$, and $(H)$, which we ignore since one group contains only singletons.
\item\textbf{Recursion 3: }
  \begin{align*}
    & & & \textrm{Process} & & \textrm{symmetry factor} & & \textrm{(status)}\\
    &(F2): &
    \left(
    \begin{tikzpicture}
      [baseline={([yshift=-0.5ex]current bounding box.center)},font=\scriptsize]
      \node (A) at (-0.5,0.75) {};
      \fill (A) circle (0.05);
      \node (B) at (0.5,0.75) {};
      \fill (B) circle (0.05);
      \draw[directed graviton] (A) node[above=0.1] {1} -- (0,0.25);
      \draw[directed graviton] (B) node[above=0.1] {1} -- (0,0.25);
      \node (C) at (0,-0.75) {};
      \fill (C) circle (0.05);
      \node[xSource] (E) at (0,-0.25) {};
      \draw[directed graviton] (C) node[below=0.1] {2} -- (E);
      \draw[directed graviton] (0,0.25) -- (E);
    \end{tikzpicture}
    \pgTwo
    \right)
    &\rightarrow
    \left\{
    \begin{tikzpicture}
      [baseline={([yshift=-0.5ex]current bounding box.center)},font=\scriptsize]
      \node (A) at (-0.5,0.75) {};
      \fill (A) circle (0.05);
      \node (B) at (0.5,0.75) {};
      \fill (B) circle (0.05);
      \draw[directed graviton] (A) node[above=0.1] {1} -- (0,0.25);
      \draw[directed graviton] (B) node[above=0.1] {1} -- (0,0.25);
      \node (C) at (-0.5,-0.75) {};
      \fill (C) circle (0.05);
      \node (D) at (1,-0.75) {};
      \fill (D) circle (0.05);
      \draw[directed graviton] (C) node[below=0.1] {2} -- (0,-0.25);
      \draw[directed graviton] (0,0.25) -- (0,-0.25);
      \node[xSource] (E) at (0.5,-0.5) {};
      \draw[directed graviton] (D) node[below=0.1] {2} -- (E);
      \draw[directed graviton] (0,-0.25) -- (E);
    \end{tikzpicture}
    \right\} & & \{(2\times 1)\times 1\}& & \textrm{(done)}\\
    &(G2): &
    \left(
    \begin{tikzpicture}
      [baseline={([yshift=-0.5ex]current bounding box.center)},font=\scriptsize]
      \node (A) at (-0.5,-0.75) {};
      \fill (A) circle (0.05);
      \node (B) at (0.5,-0.75) {};
      \fill (B) circle (0.05);
      \draw[directed graviton] (A) node[below=0.1] {2} -- (0,-0.25);
      \draw[directed graviton] (B) node[below=0.1] {2} -- (0,-0.25);
      \node (C) at (0,0.75) {};
      \fill (C) circle (0.05);
      \node[xSource] (E) at (0,0.25) {};
      \draw[directed graviton] (C) node[above=0.1] {1} -- (E);
      \draw[directed graviton] (0,-0.25) -- (E);
    \end{tikzpicture}
    \pgOne
    \right)
    &\rightarrow
    \left\{
    \begin{tikzpicture}
      [baseline={([yshift=-0.5ex]current bounding box.center)},font=\scriptsize]
      \node (A) at (-0.5,-0.75) {};
      \fill (A) circle (0.05);
      \node (B) at (0.5,-0.75) {};
      \fill (B) circle (0.05);
      \draw[directed graviton] (A) node[below=0.1] {2} -- (0,-0.25);
      \draw[directed graviton] (B) node[below=0.1] {2} -- (0,-0.25);
      \node (C) at (-0.5,0.75) {};
      \fill (C) circle (0.05);
      \node (D) at (1,0.75) {};
      \fill (D) circle (0.05);
      \draw[directed graviton] (C) node[above=0.1] {1} -- (0,0.25);
      \draw[directed graviton] (0,-0.25) -- (0,0.25);
      \node[xSource] (E) at (0.5,0.5) {};
      \draw[directed graviton] (D) node[above=0.1] {1} -- (E);
      \draw[directed graviton] (0,0.25) -- (E);
    \end{tikzpicture}
    \right\} & & \{(2\times 1) \times 1\}& & \textrm{(done)}\\
    &(H2): &
    \left(
    \begin{tikzpicture}
      [baseline={([yshift=-0.5ex]current bounding box.center)},font=\scriptsize]
      \node (A) at (-0.5,0.5) {};
      \fill (A) circle (0.05);
      \node (B) at (-0.5,-0.5) {};
      \fill (B) circle (0.05);
      \draw[directed graviton] (A) node[above=0.1] {1} -- (-0.5,0);
      \draw[directed graviton] (B) node[below=0.1] {2} -- (-0.5,0);
      \node (C) at (0.5,0.5) {};
      \fill (C) circle (0.05);
      \node[xSource] (E) at (0,0.25) {};
      \draw[directed graviton] (C) node[above=0.1] {1} -- (E);
      \draw[directed graviton] (-0.5,0) -- (E);
    \end{tikzpicture}
    \pgTwo
    \right)
    &\rightarrow
    \left\{
    \begin{tikzpicture}
      [baseline={([yshift=-0.5ex]current bounding box.center)},font=\scriptsize]
      \node (A) at (-0.5,0.5) {};
      \fill (A) circle (0.05);
      \node (B) at (-0.5,-0.5) {};
      \fill (B) circle (0.05);
      \draw[directed graviton] (A) node[above=0.1] {1} -- (-0.5,0);
      \draw[directed graviton] (B) node[below=0.1] {2} -- (-0.5,0);
      \node (C) at (0.5,0.5) {};
      \fill (C) circle (0.05);
      \draw[directed graviton] (C) node[above=0.1] {1} -- (0,0.25);
      \draw[directed graviton] (-0.5,0) -- (0,0.25);
      \node (D) at (0.5,-0.5) {};
      \fill (D) circle (0.05);
      \node[xSource] (E) at (0.25,-0.175) {};
      \draw[directed graviton] (D) node[below=0.1] {2} -- (E);
      \draw[directed graviton] (0,0.25) -- (E);
    \end{tikzpicture}
    \right\} & & \{(1\times 1)\times 1\}& & \textrm{(done)}\\
    &(H3): &
    \left(
    \begin{tikzpicture}
      [baseline={([yshift=-0.5ex]current bounding box.center)},font=\scriptsize]
      \node (A) at (-0.5,-0.5) {};
      \fill (A) circle (0.05);
      \node (B) at (-0.5,0.5) {};
      \fill (B) circle (0.05);
      \draw[directed graviton] (A) node[below=0.1] {2} -- (-0.5,0);
      \draw[directed graviton] (B) node[above=0.1] {1} -- (-0.5,0);
      \node (C) at (0.5,-0.5) {};
      \fill (C) circle (0.05);
      \node[xSource] (E) at (0,-0.25) {};
      \draw[directed graviton] (C) node[below=0.1] {2} -- (E);
      \draw[directed graviton] (-0.5,0) -- (E);
    \end{tikzpicture}
    \pgOne
    \right)
    &\rightarrow
    \left\{
    \begin{tikzpicture}
      [baseline={([yshift=-0.5ex]current bounding box.center)},font=\scriptsize]
      \node (A) at (-0.5,-0.5) {};
      \fill (A) circle (0.05);
      \node (B) at (-0.5,0.5) {};
      \fill (B) circle (0.05);
      \draw[directed graviton] (A) node[below=0.1] {2} -- (-0.5,0);
      \draw[directed graviton] (B) node[above=0.1] {1} -- (-0.5,0);
      \node (C) at (0.5,-0.5) {};
      \fill (C) circle (0.05);
      \draw[directed graviton] (C) node[below=0.1] {2} -- (0,-0.25);
      \draw[directed graviton] (-0.5,0) -- (0,-0.25);
      \node (D) at (0.5,0.5) {};
      \fill (D) circle (0.05);
      \node[xSource] (E) at (0.25,0.175) {};
      \draw[directed graviton] (D) node[above=0.1] {1} -- (E);
      \draw[directed graviton] (0,-0.25) -- (E);
    \end{tikzpicture}
    \right\} & & \{(1\times 1) \times 1\} & & \textrm{(done)}
  \end{align*}
\end{itemize}
Finally, we need to connect all the diagrams to the unique WL sink vertex for WL 1 (say we want to compute the variation w.r.t. WL 1).
This leads to the known 12 diagrams at this order (divided by their symmetry factor):
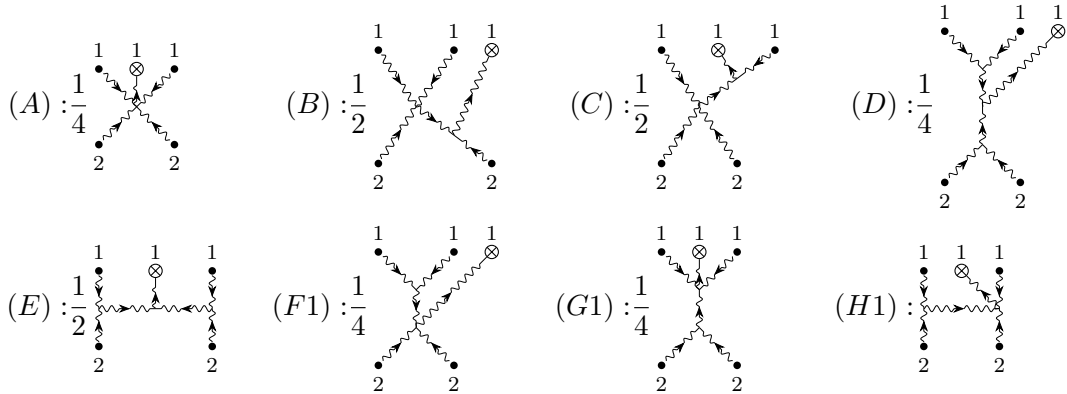
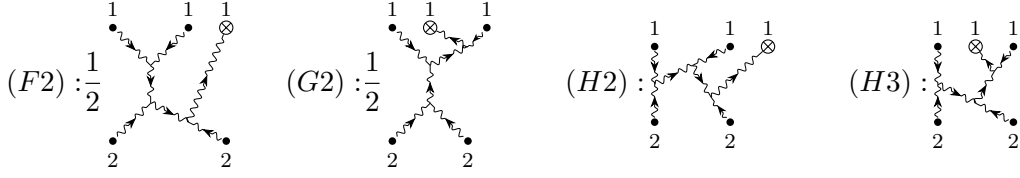
\begin{align*}
  (A): & \frac{1}{4}\begin{tikzpicture}
      [baseline={([yshift=-0.5ex]current bounding box.center)},font=\scriptsize]
      \node (A) at (-0.5,0.5) {};
      \fill (A) circle (0.05);
      \node (B) at (0.5,0.5) {};
      \fill (B) circle (0.05);
      \node (C) at (-0.5,-0.5) {};
      \fill (C) circle (0.05);
      \node (D) at (0.5,-0.5) {};
      \fill (D) circle (0.05);
      \draw[directed graviton] (A) node[above=0.1] {1} -- (0,0);
      \draw[directed graviton] (B) node[above=0.1] {1} -- (0,0);
      \draw[directed graviton] (C) node[below=0.1] {2} -- (0,0);
      \draw[directed graviton] (D) node[below=0.1] {2} -- (0,0);
      \node[xSource] (E) at (0,0.5) {};
      \draw[directed graviton] (0,0) -- (E) node[above=0.1] {1};
    \end{tikzpicture}
  &
  (B): & \frac{1}{2}\begin{tikzpicture}
      [baseline={([yshift=-0.5ex]current bounding box.center)},font=\scriptsize]
      \node (A) at (-0.5,0.75) {};
      \fill (A) circle (0.05);
      \node (B) at (0.5,0.75) {};
      \fill (B) circle (0.05);
      \node (C) at (-0.5,-0.75) {};
      \fill (C) circle (0.05);
      \node (D) at (1,-0.75) {};
      \fill (D) circle (0.05);
      \draw[directed graviton] (A) node[above=0.1] {1} -- (0,0);
      \draw[directed graviton] (B) node[above=0.1] {1} -- (0,0);
      \draw[directed graviton] (C) node[below=0.1] {2} -- (0,0);
      \draw[directed graviton] (D) node[below=0.1] {2} -- (0.5,-0.375);
      \draw[directed graviton] (0,0) -- (0.5,-0.375);
      \node[xSource] (E) at (1,0.75) {};
      \draw[directed graviton] (0.5,-0.375) -- (E) node[above=0.1] {1};
    \end{tikzpicture}
  &
  (C): & \frac{1}{2}\begin{tikzpicture}
      [baseline={([yshift=-0.5ex]current bounding box.center)},font=\scriptsize]
      \node (A) at (-0.5,0.75) {};
      \fill (A) circle (0.05);
      \node (B) at (-0.5,-0.75) {};
      \fill (B) circle (0.05);
      \node (C) at (0.5,-0.75) {};
      \fill (C) circle (0.05);
      \node (D) at (1,0.75) {};
      \fill (D) circle (0.05);
      \draw[directed graviton] (A) node[above=0.1] {1} -- (0,0);
      \draw[directed graviton] (B) node[below=0.1] {2} -- (0,0);
      \draw[directed graviton] (C) node[below=0.1] {2} -- (0,0);
      \draw[directed graviton] (D) node[above=0.1] {1} -- (0.5,0.375);
      \draw[directed graviton] (0,0) -- (0.5,0.375);
      \node[xSource] (E) at (0.25,0.75) {};
      \draw[directed graviton] (0.5,0.375) -- (E) node[above=0.1] {1};
    \end{tikzpicture}
  &
  (D): & \frac{1}{4}\begin{tikzpicture}
      [baseline={([yshift=-0.5ex]current bounding box.center)},font=\scriptsize]
      \node (A) at (-0.5,1) {};
      \fill (A) circle (0.05);
      \node (B) at (0.5,1) {};
      \fill (B) circle (0.05);
      \draw[directed graviton] (A) node[above=0.1] {1} -- (0,0.5);
      \draw[directed graviton] (B) node[above=0.1] {1} -- (0,0.5);
      \node (C) at (-0.5,-1) {};
      \fill (C) circle (0.05);
      \node (D) at (0.5,-1) {};
      \fill (D) circle (0.05);
      \draw[directed graviton] (C) node[below=0.1] {2} -- (0,-0.5);
      \draw[directed graviton] (D) node[below=0.1] {2} -- (0,-0.5);
      \draw[directed graviton] (0,0.5)  -- (0,0);
      \draw[directed graviton] (0,-0.5)  -- (0,0);
      \node[xSource] (E) at (1,1) {};
      \draw[directed graviton] (0,0)  -- (E) node[above=0.1] {1};
    \end{tikzpicture}\\
  (E): & \frac{1}{2}\begin{tikzpicture}
      [baseline={([yshift=-0.5ex]current bounding box.center)},font=\scriptsize]
      \node (A) at (-0.75,0.5) {};
      \fill (A) circle (0.05);
      \node (B) at (-0.75,-0.5) {};
      \fill (B) circle (0.05);
      \draw[directed graviton] (A) node[above=0.1] {1} -- (-0.75,0);
      \draw[directed graviton] (B) node[below=0.1] {2} -- (-0.75,0);
      \node (C) at (0.75,0.5) {};
      \fill (C) circle (0.05);
      \node (D) at (0.75,-0.5) {};
      \fill (D) circle (0.05);
      \draw[directed graviton] (C) node[above=0.1] {1} -- (0.75,0);
      \draw[directed graviton] (D) node[below=0.1] {2} -- (0.75,0);
      \draw[directed graviton] (-0.75,0) -- (0,0);
      \draw[directed graviton] (0.75,0) -- (0,0);
      \node[xSource] (E) at (0,0.5) {};
      \draw[directed graviton] (0,0) -- (E) node[above=0.1] {1};
    \end{tikzpicture}
  &
  (F1): & \frac{1}{4}\begin{tikzpicture}
      [baseline={([yshift=-0.5ex]current bounding box.center)},font=\scriptsize]
      \node (A) at (-0.5,0.75) {};
      \fill (A) circle (0.05);
      \node (B) at (0.5,0.75) {};
      \fill (B) circle (0.05);
      \draw[directed graviton] (A) node[above=0.1] {1} -- (0,0.25);
      \draw[directed graviton] (B) node[above=0.1] {1} -- (0,0.25);
      \node (C) at (-0.5,-0.75) {};
      \fill (C) circle (0.05);
      \node (D) at (0.5,-0.75) {};
      \fill (D) circle (0.05);
      \draw[directed graviton] (C) node[below=0.1] {2} -- (0,-0.25);
      \draw[directed graviton] (D) node[below=0.1] {2} -- (0,-0.25);
      \draw[directed graviton] (0,0.25) -- (0,-0.25);
      \node[xSource] (E) at (1,0.75) {};
      \draw[directed graviton] (0,-0.25) -- (E) node[above=0.1] {1};
    \end{tikzpicture}
  &
  (G1): & \frac{1}{4}\begin{tikzpicture}
      [baseline={([yshift=-0.5ex]current bounding box.center)},font=\scriptsize]
      \node (A) at (-0.5,-0.75) {};
      \fill (A) circle (0.05);
      \node (B) at (0.5,-0.75) {};
      \fill (B) circle (0.05);
      \draw[directed graviton] (A) node[below=0.1] {2} -- (0,-0.25);
      \draw[directed graviton] (B) node[below=0.1] {2} -- (0,-0.25);
      \node (C) at (-0.5,0.75) {};
      \fill (C) circle (0.05);
      \node (D) at (0.5,0.75) {};
      \fill (D) circle (0.05);
      \draw[directed graviton] (C) node[above=0.1] {1} -- (0,0.25);
      \draw[directed graviton] (D) node[above=0.1] {1} -- (0,0.25);
      \draw[directed graviton] (0,-0.25) -- (0,0.25);
      \node[xSource] (E) at (0,0.75) {};
      \draw[directed graviton] (0,0.25) -- (E) node[above=0.1] {1};
    \end{tikzpicture}
  &
  (H1): & \begin{tikzpicture}
      [baseline={([yshift=-0.5ex]current bounding box.center)},font=\scriptsize]
      \node (A) at (-0.5,0.5) {};
      \fill (A) circle (0.05);
      \node (B) at (-0.5,-0.5) {};
      \fill (B) circle (0.05);
      \draw[directed graviton] (A) node[above=0.1] {1} -- (-0.5,0);
      \draw[directed graviton] (B) node[below=0.1] {2} -- (-0.5,0);
      \node (C) at (0.5,0.5) {};
      \fill (C) circle (0.05);
      \node (D) at (0.5,-0.5) {};
      \fill (D) circle (0.05);
      \draw[directed graviton] (C) node[above=0.1] {1} -- (0.5,0);
      \draw[directed graviton] (D) node[below=0.1] {2} -- (0.5,0);
      \draw[directed graviton] (-0.5,0) -- (0.5,0);
      \node[xSource] (E) at (0,0.5) {};
      \draw[directed graviton] (0.5,0) -- (E) node[above=0.1] {1};
    \end{tikzpicture}\\
  (F2): & \frac{1}{2}\begin{tikzpicture}
      [baseline={([yshift=-0.5ex]current bounding box.center)},font=\scriptsize]
      \node (A) at (-0.5,0.75) {};
      \fill (A) circle (0.05);
      \node (B) at (0.5,0.75) {};
      \fill (B) circle (0.05);
      \draw[directed graviton] (A) node[above=0.1] {1} -- (0,0.25);
      \draw[directed graviton] (B) node[above=0.1] {1} -- (0,0.25);
      \node (C) at (-0.5,-0.75) {};
      \fill (C) circle (0.05);
      \node (D) at (1,-0.75) {};
      \fill (D) circle (0.05);
      \draw[directed graviton] (C) node[below=0.1] {2} -- (0,-0.25);
      \draw[directed graviton] (0,0.25) -- (0,-0.25);
      \draw[directed graviton] (D) node[below=0.1] {2} -- (0.5,-0.5);
      \draw[directed graviton] (0,-0.25) -- (0.5,-0.5);
      \node[xSource] (E) at (1,0.75) {};
      \draw[directed graviton] (0.5,-0.5) -- (E) node[above=0.1] {1};
    \end{tikzpicture}
  &
  (G2): & \frac{1}{2}\begin{tikzpicture}
      [baseline={([yshift=-0.5ex]current bounding box.center)},font=\scriptsize]
      \node (A) at (-0.5,-0.75) {};
      \fill (A) circle (0.05);
      \node (B) at (0.5,-0.75) {};
      \fill (B) circle (0.05);
      \draw[directed graviton] (A) node[below=0.1] {2} -- (0,-0.25);
      \draw[directed graviton] (B) node[below=0.1] {2} -- (0,-0.25);
      \node (C) at (-0.5,0.75) {};
      \fill (C) circle (0.05);
      \node (D) at (0.75,0.75) {};
      \fill (D) circle (0.05);
      \draw[directed graviton] (C) node[above=0.1] {1} -- (0,0.25);
      \draw[directed graviton] (0,-0.25) -- (0,0.25);
      \draw[directed graviton] (D) node[above=0.1] {1} -- (0.5,0.5);
      \draw[directed graviton] (0,0.25) -- (0.5,0.5);
      \node[xSource] (E) at (0,0.75) {};
      \draw[directed graviton] (0.5,0.5) -- (E) node[above=0.1] {1};
    \end{tikzpicture}
  &
  (H2): & \begin{tikzpicture}
      [baseline={([yshift=-0.5ex]current bounding box.center)},font=\scriptsize]
      \node (A) at (-0.5,0.5) {};
      \fill (A) circle (0.05);
      \node (B) at (-0.5,-0.5) {};
      \fill (B) circle (0.05);
      \draw[directed graviton] (A) node[above=0.1] {1} -- (-0.5,0);
      \draw[directed graviton] (B) node[below=0.1] {2} -- (-0.5,0);
      \node (C) at (0.5,0.5) {};
      \fill (C) circle (0.05);
      \draw[directed graviton] (C) node[above=0.1] {1} -- (0,0.25);
      \draw[directed graviton] (-0.5,0) -- (0,0.25);
      \node (D) at (0.5,-0.5) {};
      \fill (D) circle (0.05);
      \draw[directed graviton] (D) node[below=0.1] {2} -- (0.25,-0.175);
      \draw[directed graviton] (0,0.25) -- (0.25,-0.175);
      \node[xSource] (E) at (1,0.5) {};
      \draw[directed graviton] (0.25,-0.175) -- (E) node[above=0.1] {1};
    \end{tikzpicture}
  &
  (H3): & \begin{tikzpicture}
      [baseline={([yshift=-0.5ex]current bounding box.center)},font=\scriptsize]
      \node (A) at (-0.5,-0.5) {};
      \fill (A) circle (0.05);
      \node (B) at (-0.5,0.5) {};
      \fill (B) circle (0.05);
      \draw[directed graviton] (A) node[below=0.1] {2} -- (-0.5,0);
      \draw[directed graviton] (B) node[above=0.1] {1} -- (-0.5,0);
      \node (C) at (0.5,-0.5) {};
      \fill (C) circle (0.05);
      \draw[directed graviton] (C) node[below=0.1] {2} -- (0,-0.25);
      \draw[directed graviton] (-0.5,0) -- (0,-0.25);
      \node (D) at (0.5,0.5) {};
      \fill (D) circle (0.05);
      \draw[directed graviton] (D) node[above=0.1] {1} -- (0.25,0.175);
      \draw[directed graviton] (0,-0.25) -- (0.25,0.175);
      \node[xSource] (E) at (0,0.5) {};
      \draw[directed graviton] (0.25,0.175) -- (E) node[above=0.1] {1};
    \end{tikzpicture}
\end{align*}
\endgroup

\subsubsection{Details of the implementation in \texttt{FeynRIR}}

\texttt{FeynRIR} optimizes the workflow by minimizing the number of contraction/gluing steps.
This is achieved by pre-computing a recursive multiset representation of the diagrams at a fixed order.
During this pre-computation \texttt{FeynRIR} stores every unique partial diagram in a lookup table which it then uses to construct a dependency DAG (directed acyclic graph) of each partial diagram.
It is then easy to compute every unique contraction only a single time by building up every unique partial diagram following the dependency DAG.
The results for each partial diagram (a node in the DAG) are stored and are re-used at every higher PM order.
The DAG also renders the synchronization of parallel threads working on contractions trivial.

At each stage of the actual integrand construction, having at hand the dependency DAG, we represent a single term/contribution by a pair of a graph and a numerator.
The graph does not only encode the propagators but also other factors which we do not explicitly represent algebraically in the numerator.
The interpretation of the pieces of a graph at this stage is:
\begin{itemize}
\item WL source vertex ($a=1,2$):
  \begin{equation}\label{eq:source_final}
    \begin{tikzpicture}
      [baseline={([yshift=-0.5ex]current bounding box.center)},font=\scriptsize]
      \draw[directed graviton] (0,0) node[above=0.1] {$a$} -- node[left] {$p_j\downarrow$} (0,-0.5);
      \fill (0,0) circle (0.05);
    \end{tikzpicture}
    = i\, m_a \kappa \int \dd\tau_n e^{i\, p_j \cdot x_a(\tau_n)}\,,
  \end{equation}
  where $n$ is the identifier of the WL source vertex and $j$ the identifier of the outgoing graviton propagator.
\item WL sink vertex ($a=1,2$):
  \begin{equation}
    \begin{tikzpicture}
      [baseline={([yshift=-0.5ex]current bounding box.center)},font=\scriptsize]
      \node[xSource] (A) at (0,0) {};
      \draw[directed graviton] (0,-0.7) -- node[left] {$p_j\uparrow$} (A) node[above=0.1] {$a$};
    \end{tikzpicture}
    = i\, m_a \kappa e^{-i\, p_j \cdot x_a(\tau_n)}\,,
  \end{equation}
  where $n$ is the identifier of the WL sink vertex and $j$ the identifier of the incoming graviton propagator.
  Note that $\tau_n = \sigma$ in eq.~\eqref{eq:varSeff}.
\item Graviton propagator:
  \begin{equation}
    \begin{tikzpicture}
      [baseline={([yshift=-0.5ex]current bounding box.center)},font=\scriptsize]
      \draw[directed graviton] (0,0) -- node[above] {$\overset{p_j}{\rightarrow}$}(1,0);
    \end{tikzpicture}
    = \int\frac{\dd^D p_j}{(2\pi)^D} \frac{i}{(p_j^0 + i0)^2 - \bp_j^2}\,,
  \end{equation}
  where $j$ is the identifier of the graviton propagator.
\item Graviton vertex:
  \begin{equation}
    \begin{tikzpicture}
      [baseline={([yshift=-0.5ex]current bounding box.center)},font=\scriptsize]
      \node (A) at (0,0) {};
      \fill (A) circle (0.05);
      \draw[directed graviton] (A) -- (0:0.7) node[right] {$p_{j_1}$};
      \draw[directed graviton] (45:0.7) node[above right] {$p_{j_2}$} -- (A);
      \draw[directed graviton] (90:0.7) node[above] {$p_{j_3}$} -- (A);
      \draw[directed graviton] (135:0.7) node[above left] {$p_{j_4}$} -- (A);
      \fill (180:0.5) circle (0.02);
      \fill (220:0.5) circle (0.02);
      \fill (260:0.5) circle (0.02);
      \fill (300:0.5) circle (0.02);
    \end{tikzpicture}
    = i \kappa^{n-2}(2\pi)^D \delta^{(D)}(p_{j_1} - p_{j_2} - p_{j_3} - p_{j_4} - \dots)\,,
  \end{equation}
  where $j_1$ is the identifier of the outgoing graviton propagator, and $j_2$, $j_3$, $j_4$, \dots are the identifiers of the incoming graviton propagators, and $n$ is the valence of the vertex.
\end{itemize}
The rest of the Feynman rules that are associated with a graph-numerator pair are stored as an algebraic expression that will henceforth be called the \emph{numerator}.
For the algebraic expression of each numerator the following conventions are used:
\begin{itemize}
\item Proper times in WL arguments are labeled by their associated vertex identifier $n$ for WL sources.
\item Loop momenta are labeled with the corresponding edge identifier.
  We do not resolve any momentum conserving delta functions at this point in any way.
\item The single open Lorentz index due to the WL sink vertex is labeled by a dedicated \texttt{Symbolica} variable \texttt{mu}, corresponding to $\mu$ in eq.~\eqref{eq:varSeff}.
\item The terms corresponding to a total time derivative in the rule for the WL sink vertex are flagged with the \texttt{Symbolica} symbol \texttt{flagtd}.
  When integrating the corresponding proper time $\sigma$ in eq.~\eqref{eq:varSeff} over full $\mathbb{R}$, as is for example the case in the computation for the impulse, these terms can be safely set to zero.
  When the variation of the effective action is used for the derivation of the equations of motion (required for the iterations) this variable will be set to one since these terms contribute.
\end{itemize}

\paragraph{Simplification.}
Consider the identity
\begin{equation}
  \begin{aligned}
    \int\dd\tau e^{i\, k\cdot x_a(\tau)} k\cdot \dot{x}_a(\tau) \dot{x}_a^\nu(\tau) &= -i \int\dd\tau \partial_\tau\left(e^{i\, k\cdot x_a(\tau)}\right) \dot{x}_a^\nu(\tau)\\
    &=i \int\dd\tau e^{i\, k\cdot x_a(\tau)} \ddot{x}_a^\nu(\tau)\,,
  \end{aligned}
\end{equation}
where we used integration by parts to arrive at the second line.
Such combinations will often appear in the integrand stemming from a WL source vertex contracted to a graviton vertex through a graviton propagator.
Using this identity whenever possible is beneficial for the computational efficiency as it trades two velocities $\dot{x}_a$, which are of order $\cO(G^0)$, for a single acceleration of order $\cO(G^1)$.

This identity may be applied in \texttt{FeynRIR} through a simple replacement rule on the numerator factor of the effective action before expansion of the WLs.

\subsection{Iterations and expansion of WLs}\label{sec:iterations}

Given an expression for the variation of the effective action, the next step is to expand the WLs in powers of $G$ and replace each order by the corresponding iteration.
We continue to represent each contribution as a graph-numerator pair.
This requires us to keep track of which vertex a given WL that we expand belongs to such that we can attach the graph of the iteration at the correct spot in the action graph.

\subsubsection{Representing iterations graphically}
Let us first describe how iterations are represented in this language.
Iterations are also (sums of) graph-numerator pairs, where all WLs have already been replaced by iterations.
They may have a WL propagator attached to a main sink.
To be precise, a term in an iteration contributing to $\delta^{(n)} x_a^\mu(\sigma)$ has zero, one, or two proper time derivatives acting on it, which may be graphically represented as
\begin{align}\label{eq:graphical_iterations}
  \delta^{(n)}\ddot{x}_a^\mu(\sigma) &\simeq \begin{tikzpicture}
    [baseline={([yshift=-0.5ex]current bounding box.center)},font=\scriptsize]
    \node[iter] (A) at (0,0) {};
    \draw[directed graviton] (0,-0.7) -- node[right] {$\uparrow p_m$} (A) node[above=0.1] {$a$};
    \draw[pattern = north east lines] (0,-0.9) circle (0.2);
    \draw[quark] (60:0.7) -- (A);
    \draw[pattern = north east lines] (60:0.9) circle (0.2);
    \draw[quark] (30:0.7) -- (A);
    \draw[pattern = north east lines] (30:0.9) circle (0.2);
    \fill (20:0.6) circle (0.02);
    \fill (10:0.6) circle (0.02);
    \fill (0:0.6) circle (0.02);
  \end{tikzpicture}\,,
  &
  \delta^{(n)}\dot{x}_a^\mu(\sigma) &\simeq \begin{tikzpicture}
    [baseline={([yshift=-0.5ex]current bounding box.center)},font=\scriptsize]
    \node[xSource] (A) at (0,0) {};
    \node[iter] (B) at (-0.7,0) {};
    \draw[quark] (A) node[above = 0.1] {$a$} -- (B);
    \draw[directed graviton] (0,-0.7) -- node[right] {$\uparrow p_m$} (A);
    \draw[pattern = north east lines] (0,-0.9) circle (0.2);
    \draw[quark] (60:0.7) -- (A);
    \draw[pattern = north east lines] (60:0.9) circle (0.2);
    \draw[quark] (30:0.7) -- (A);
    \draw[pattern = north east lines] (30:0.9) circle (0.2);
    \fill (20:0.6) circle (0.02);
    \fill (10:0.6) circle (0.02);
    \fill (0:0.6) circle (0.02);
  \end{tikzpicture}\,,
  &
  \delta^{(n)}x_a^\mu(\sigma) &\simeq \begin{tikzpicture}
    [baseline={([yshift=-0.5ex]current bounding box.center)},font=\scriptsize]
    \node[xSource] (A) at (0,0) {};
    \node[iter] (B) at (-0.7,0) {};
    \draw[double,quark] (A) node[above = 0.1] {$a$} -- (B);
    \draw[directed graviton] (0,-0.7) -- node[right] {$\uparrow p_m$} (A);
    \draw[pattern = north east lines] (0,-0.9) circle (0.2);
    \draw[quark] (60:0.7) -- (A);
    \draw[pattern = north east lines] (60:0.9) circle (0.2);
    \draw[quark] (30:0.7) -- (A);
    \draw[pattern = north east lines] (30:0.9) circle (0.2);
    \fill (20:0.6) circle (0.02);
    \fill (10:0.6) circle (0.02);
    \fill (0:0.6) circle (0.02);
  \end{tikzpicture}
\end{align}
The empty circle is going to be fused with the WL vertex into which the iteration is plugged in and has no further (mathematical) interpretation attached to it.
All the blobs represent expressions which already had iterations plugged in.
For graph-numerator pairs where iterations have been replaced, we have new graphical ingredients, the WL propagators, which represent the following mathematical expressions:
\begin{align}\label{eq:wl_prop}
  \begin{tikzpicture}
    [baseline={([yshift=-0.5ex]current bounding box.center)},font=\scriptsize]
    \node[xSource] (A) at (0,0) {};
    \draw[quark] (A) node[above=0.1] {$a$} -- (0.7,0);
  \end{tikzpicture}
  &= \frac{1}{m_a}\frac{i}{k\cdot u_a + i0}\,,
  &
  \begin{tikzpicture}
    [baseline={([yshift=-0.5ex]current bounding box.center)},font=\scriptsize]
    \node[xSource] (A) at (0,0) {};
    \draw[double,quark] (A) node[above=0.1] {$a$} -- (0.7,0);
  \end{tikzpicture}
  &= \frac{1}{m_a}\frac{i}{(k\cdot u_a + i0)^2}\,,
\end{align}
where $k$ is the sum of all inflowing momenta (energies) into the WL sink vertex, through the (unique) graviton propagator and through WL propagators via iterations.
Diagrams with iterations plugged in also require a different interpretation of the WL source and sink vertices since the WL field in the exponential has been expanded:
\begin{itemize}
\item WL source vertex ($a=1,2$):
  \begin{equation}\label{eq:source_vertex_final}
    \begin{tikzpicture}
      [baseline={([yshift=-0.5ex]current bounding box.center)},font=\scriptsize]
      \draw[directed graviton] (0,0) node[above=0.1] {$a$} -- node[left] {$p_j\downarrow$} (0,-0.5);
      \fill (0,0) circle (0.05);
    \end{tikzpicture}
    = i\, m_a \kappa \int \dd\tau_n e^{i\, p_j \cdot (b_a + u_a \tau_n)}\,,
  \end{equation}
  where $n$ is the identifier of the WL source vertex and $j$ the identifier of the outgoing graviton propagator.
\item WL sink vertex ($a=1,2$):
  \begin{equation}\label{eq:sink_vertex_final}
    \begin{tikzpicture}
      [baseline={([yshift=-0.5ex]current bounding box.center)},font=\scriptsize]
      \node[xSource] (A) at (0,0) {};
      \draw[directed graviton] (0,-0.7) -- node[left] {$p_j\uparrow$} (A) node[above=0.1] {$a$};
    \end{tikzpicture}
    = i\, m_a \kappa e^{-i\, p_j \cdot (b_a + u_a \tau_n)}\,,
  \end{equation}
  where $n$ is the identifier of the WL sink vertex and $j$ the identifier of the incoming graviton propagator.
  Note that the proper time argument of the final WL sink vertex is equal to $\sigma$, the proper time argument of an iteration $\delta^{(n)}x_a^\mu(\sigma)$.
\end{itemize}

\subsubsection{Computing and inserting iterations}\label{sec:iteration_insertion}
Let us get to the actual strategy of plugging in iterations into variations of the effective action to produce higher order iterations or integrands for observables.
From a practical point of view, we distribute the available orders of $G$ in all possible ways to the WL vertices (sources and sinks) and then compute the appropriate series coefficient of each vertex individually.
Say, we need the series coefficient of a vertex at some order $G^m$.
Each vertex itself has WLs appearing at two places, in the exponent and the numerator.
Expanding a WL formally,
\begin{equation}
  x_a^\mu(\tau) = b_a^\mu + u_a^\mu \tau + \sum_{n=1}^m \delta^{(n)}x_a^\mu(\tau) G^n\,,
\end{equation}
and plugging this into a WL vertex leads, at order $G^m$, to terms of the form
\begin{equation}
  \delta^{(n_1)}x_a^{(d_1),\mu_1}(\tau) \delta^{(n_2)}x_a^{(d_2),\mu_2}(\tau) \cdots \delta^{(n_s)}x_a^{(d_s),\mu_s}(\tau)\,,
\end{equation}
where $(d_i)=0,1,2$ denotes derivatives w.r.t. the proper time.
An important observation is that whenever two such contributions agree up to permutations of the pairs $(n_i,d_i)$ they lead to the same diagram upon their insertion and we optimally collect them into the same contribution from the outset.
E.g. $\delta^{(2)}\dot{x}_1^\mu(\tau) \delta^{(3)}\dot{x}_1^\nu(\tau) \delta^{(1)}\ddot{x}_1^\rho(\tau)$ and $\delta^{(2)}\dot{x}_1^\mu(\tau) \delta^{(1)}\ddot{x}_1^\nu(\tau) \delta^{(3)}\dot{x}_1^\rho(\tau)$ contribute to the same graph.

In order to expand the exponential factor, the following formula collects the contributions at a given power $G^n$ in an optimal way:
\begin{equation}
  e^{\sum_{n=0}^\infty c_n G^n} = e^{c_0}\sum_{n=0}^\infty G^n \sum_{\sigma \in \mathcal{P}(n)} \prod_m \frac{1}{\sigma^m!} c_{\sigma_m}^{\sigma^m}\,,
\end{equation}
where $\mathcal{P}(n)$ is the set of all partitions of the integer $n$, written as $n=\sum_m \sigma_m \sigma^m$ with mutually different $\sigma_m$.
In other words, the number $\sigma_m$ appears $\sigma^m$ times in the partition of $n$.

Having performed the expansion in $G$, the actual insertion of the iterations is straightforward:
\begin{itemize}
\item Each term (graph-numerator pair) in an iteration is glued to the expanded graph-numerator pair at the correct insertion point and produces a new term in the answer.
  This requires an update of the vertex and edge identifiers and correspondingly to all the momentum and proper time labels.
\item In this process, the ``external'' proper time argument of the iteration must be replaced by the proper time argument of the insertion point.
  This does not actually require any work, but is simply a change of the interpretation of a WL sink vertex that has an outgoing WL propagator: It no longer is evaluated at the proper time associated to its vertex identifier but at the proper time associated to the last WL vertex following the outgoing WL propagators iteratively.
\end{itemize}
\texttt{FeynRIR} takes care of all the non-trivial combinatorics involved in this process automatically.

For example, plugging in the tree iteration into one WL vertex of the tree variation of the eff. action has a graphical contribution of the form
\begin{equation}
  \begin{tikzpicture}
    [baseline={([yshift=-0.5ex]current bounding box.center)},font=\scriptsize]
    \node[xSource] (A) at (0,-0.7) {};
    \draw[directed graviton] (0,0) -- (A);
    \fill (0,0) circle (0.05);
  \end{tikzpicture}
  \times
  \begin{tikzpicture}
    [baseline={([yshift=-0.5ex]current bounding box.center)},font=\scriptsize]
    \node[xSource] (A) at (0,0) {};
    \draw[directed graviton] (0,-0.7) -- (A);
    \fill (0,-0.7) circle (0.05);
    \node[iter] (B) at (-0.7,0) {};
    \draw[quark] (A) -- (B);
  \end{tikzpicture}
  =
  \begin{tikzpicture}
    [baseline={([yshift=-0.5ex]current bounding box.center)},font=\scriptsize]
    \node[xSource] (A) at (0,-0.7) {};
    \draw[directed graviton] (0,0) -- node[right=0.1] {$p_1$} (A) node[below=0.1] {$2,\sigma$};
    \fill (0,0) circle (0.05);
    \node[xSource] (C) at (0.7,0) {};
    \draw[directed graviton] (0.7,-0.7) node[below=0.1] {$2,\tau_2$} -- node[right=0.1] {$p_2$} (C) node[above=0.1] {1};
    \fill (0.7,-0.7) circle (0.05);
    \node[iter] (B) at (0,0) {};
    \draw[quark] (C) -- (B) node[above=0.1] {$1,\tau_1$};
  \end{tikzpicture}
  \propto \int\dd\tau_1\dd\tau_2 \frac{e^{i\, (p_2-p_1)\cdot(b_2-b_1) + i\, p_2 \cdot u_2 \tau_2 + i\, (p_1-p_2)\cdot u_1 \tau_1 - i\, p_1\cdot u_2 \sigma}}{(p_2\cdot u_1 + i0)((p_1^0+i0)^2-\bp_1^2)((p_2^0+i0)^2-\bp_2^2)}\,,
\end{equation}
where $\sigma$ is the ``external'' proper time from eq.~\eqref{eq:varSeff} and we have dropped some overall factor of masses and imaginary $i$.

\paragraph{Factors of the imaginary unit.}
It is possible to make the numerator free of any explicit factors of $i$, by adhering to the following conventions.
Instead of working with $\ddot{x}_a^\mu(\tau)$, $\dot{x}_a^\mu(\tau)$, and $x_a^\mu(\tau)$ one may work with $i\ddot{x}_a^\mu(\tau)$, $\dot{x}_a^\mu(\tau)$, and $i x_a^\mu(\tau)$.
With this rewriting, every Feynman rule carries one overall factor of $i$, but is otherwise free of imaginary units.
When ``solving'' the equations of motion for the WLs (i.e. computing iterations) one finds effective rules of the form
\begin{equation}
  \begin{aligned}
    i\,\ddot{x}_a^\mu(\tau) &\rightarrow i \times 1\,,\\
    \dot{x}_a^\mu(\tau) &\rightarrow \frac{1\times i}{k \cdot u_a \pm i0}\,,\\
    i\,x_a^\mu(\tau) &\rightarrow \frac{i\times i^2}{(k \cdot u_a \pm i0)^2}\,,
  \end{aligned}
\end{equation}
which consistently produce a single overall factor of $i$.
Our definition of the WL propagator in Eq.~\eqref{eq:wl_prop} already conveniently absorbs this factor in the graph-numerator language.
Any sign must be absorbed into the numerator factor.

For $\ddot{x}$ no WL propagator is introduced in the graph, which might be troublesome concerning the corresponding factor of $i$.
But instead a (sink) WL vertex is fused into another one.
Hence, we implicitly drop a second factor of $i$ from the merged vertex and it is sufficient to adapt the sign of the numerator and never explicitly introduce these factors in the first place.
Note that the solution of the equations of motion also introduces a factor of $1/m_a$, which we also absorbed into the definition of the WL propagator in Eq.~\eqref{eq:wl_prop}.
In the case of $\ddot{x}$ this factor cancels against a factor of $m_a$ in the fused vertex and the whole diagram correctly encodes all factors of $i$ and $m_a$.

The same bookkeeping is required when a selfie identity contracts a WL propagator and fuses its two endpoint vertices, see sec.~\ref{sec:selfies}.
For vertices with $h_1$ and $h_2$ attached graviton legs, the fused vertex carries $i m_a\kappa^{h_1+h_2}$, as in eqs.~\eqref{eq:export_wl_nodes} and~\eqref{eq:export_wl_sink}.
Stripping off the phases, integration measures and energy denominators, the ratio of the scalar factors before and after fusion is
\begin{equation}\label{eq:wl_fusion_prefactor}
  \frac{(i m_a\kappa^{h_1})(i/m_a)(i m_a\kappa^{h_2})}
       {i m_a\kappa^{h_1+h_2}}=-1\,.
\end{equation}
Increasing the power of a surviving WL propagator does not introduce another factor of $i$ or $1/m_a$.
Thus each such fusion requires a minus sign multiplying the numerator; the powers of $\kappa$ are retained by the fused vertex.

\section{Simplifying and canonicalizing the integrand}\label{sec:simplification}

We propose two types of partial fraction identities that simplify the integrand representation.
A first version serializes several WL propagators flowing into the same WL vertex.
A second identity can be applied for WL propagators connected to a ``selfie'' subdiagram.
Both relations have an intuitive graphical representation and only require scalar coefficients and momentum relabeling in the numerator.

\subsection{Serialize WL propagators}

A first partial fraction identity, observed in~\cite{Driesse:2024xad} to simplify the 5PM impulse integrand, can be used to decrease the number of topologies related to the WL propagator structure.
We use them in a way such that WL propagators are ``serialized'' or ``planarized''.
A simple example provides a first hint at the generic pattern.
Consider the identity
\begin{equation}
  \frac{1}{(\ell_1\cdot u_2 + i0)(\ell_2\cdot u_2 + i0)} = \frac{1}{(\ell_1+\ell_2)\cdot u_2 + i0}\left[
    \frac{1}{\ell_1\cdot u_2 + i0} + \frac{1}{\ell_2\cdot u_2 + i0}
  \right]\,.
\end{equation}
This identity, and generalizations thereof, may be applied systematically to \emph{serialize} several WL propagators flowing into the same vertex.
Let us start with the simplest case of two iterations that are plugged into the same (source or sink) vertex.
The above partial fraction identity corresponds diagrammatically to an identity of the form
\begin{equation}
  \begin{tikzpicture}
    [baseline={([yshift=-0.5ex]current bounding box.center)},scale=0.8]
    \node[xSource] (A) at (0,-1) {};
    \draw[directed graviton] (0,0) -- node[left] {$\ell_1\!\downarrow$} (A);
    \node[xSource] (B) at (1,-1) {};
    \draw[directed graviton] (1,0) -- node[left] {$\ell_2\!\downarrow$} (B);
    \draw[quark] (B) -- (2,-1);
    \draw[quark] (A) .. controls (0.25,-1.5) and (1.75,-1.5) .. (2,-1);
    \fill (2,-1) circle (0.05);
  \end{tikzpicture}
  =
  \begin{tikzpicture}
    [baseline={([yshift=-0.5ex]current bounding box.center)},scale=0.8]
    \node[xSource] (A) at (0,-1) {};
    \draw[directed graviton] (0,0) -- node[left] {$\ell_1\!\downarrow$} (A);
    \node[xSource] (B) at (1,-1) {};
    \draw[directed graviton] (1,0) -- node[left] {$\ell_2\!\downarrow$} (B);
    \draw[quark] (B) -- (2,-1);
    \draw[quark] (A) -- (B);
    \draw[white,quark] (A) .. controls (0.25,-1.5) and (1.75,-1.5) .. (2,-1);
    \fill (2,-1) circle (0.05);
  \end{tikzpicture}
  +
  \begin{tikzpicture}
    [baseline={([yshift=-0.5ex]current bounding box.center)},scale=0.8]
    \node[xSource] (A) at (0,-1) {};
    \draw[directed graviton] (0,0) -- node[left] {$\ell_2\!\downarrow$} (A);
    \node[xSource] (B) at (1,-1) {};
    \draw[directed graviton] (1,0) -- node[left] {$\ell_1\!\downarrow$} (B);
    \draw[quark] (B) -- (2,-1);
    \draw[quark] (A) -- (B);
    \draw[white,quark] (A) .. controls (0.25,-1.5) and (1.75,-1.5) .. (2,-1);
    \fill (2,-1) circle (0.05);
  \end{tikzpicture}\,.
\end{equation}
This identity can be applied anywhere such subdiagrams appear.

A systematic recursive application of this identity planarizes/serializes any number of incoming WL iterations into the same vertex.
A more involved example is
\begin{equation}
  \begin{aligned}
    \begin{tikzpicture}
      [baseline={([yshift=-0.5ex]current bounding box.center)},scale=0.8]
      \node[xSource] (A) at (0,-1) {};
      \draw[directed graviton] (0,0) -- node[left] {$\ell_2\!\downarrow$} (A);
      \node[xSource] (B) at (1,-1) {};
      \draw[directed graviton] (1,0) -- node[left] {$\ell_3\!\downarrow$} (B);
      \node[xSource] (C) at (-1,-1) {};
      \draw[directed graviton] (-1,0) -- node[left] {$\ell_1\!\downarrow$} (C);
      \draw[quark] (B) -- (2,-1);
      \draw[quark] (A) .. controls (0.25,-1.5) and (1.75,-1.5) .. (2,-1);
      \draw[quark] (C) .. controls (-0.75,-1.75) and (1.75,-1.75) .. (2,-1);
      \fill (2,-1) circle (0.05);
    \end{tikzpicture}
    &=
    \begin{tikzpicture}
      [baseline={([yshift=-0.5ex]current bounding box.center)},scale=0.8]
      \node[xSource] (A) at (0,-1) {};
      \draw[directed graviton] (0,0) -- node[left] {$\ell_2\!\downarrow$} (A);
      \node[xSource] (B) at (1,-1) {};
      \draw[directed graviton] (1,0) -- node[left] {$\ell_3\!\downarrow$} (B);
      \node[xSource] (C) at (-1,-1) {};
      \draw[directed graviton] (-1,0) -- node[left] {$\ell_1\!\downarrow$} (C);
      \draw[quark] (B) -- (2,-1);
      \draw[quark] (A) -- (B);
      \draw[quark] (C) -- (A);
      \fill (2,-1) circle (0.05);
      \draw[white,quark] (A) .. controls (0.25,-1.5) and (1.75,-1.5) .. (2,-1);
      \draw[white,quark] (C) .. controls (-0.75,-1.75) and (1.75,-1.75) .. (2,-1);
    \end{tikzpicture}
    +
    \begin{tikzpicture}
      [baseline={([yshift=-0.5ex]current bounding box.center)},scale=0.8]
      \node[xSource] (A) at (0,-1) {};
      \draw[directed graviton] (0,0) -- node[left] {$\ell_3\!\downarrow$} (A);
      \node[xSource] (B) at (1,-1) {};
      \draw[directed graviton] (1,0) -- node[left] {$\ell_2\!\downarrow$} (B);
      \node[xSource] (C) at (-1,-1) {};
      \draw[directed graviton] (-1,0) -- node[left] {$\ell_1\!\downarrow$} (C);
      \draw[quark] (B) -- (2,-1);
      \draw[quark] (A) -- (B);
      \draw[quark] (C) -- (A);
      \fill (2,-1) circle (0.05);
      \draw[white,quark] (A) .. controls (0.25,-1.5) and (1.75,-1.5) .. (2,-1);
      \draw[white,quark] (C) .. controls (-0.75,-1.75) and (1.75,-1.75) .. (2,-1);
    \end{tikzpicture}
    +
    \begin{tikzpicture}
      [baseline={([yshift=-0.5ex]current bounding box.center)},scale=0.8]
      \node[xSource] (A) at (0,-1) {};
      \draw[directed graviton] (0,0) -- node[left] {$\ell_1\!\downarrow$} (A);
      \node[xSource] (B) at (1,-1) {};
      \draw[directed graviton] (1,0) -- node[left] {$\ell_3\!\downarrow$} (B);
      \node[xSource] (C) at (-1,-1) {};
      \draw[directed graviton] (-1,0) -- node[left] {$\ell_2\!\downarrow$} (C);
      \draw[quark] (B) -- (2,-1);
      \draw[quark] (A) -- (B);
      \draw[quark] (C) -- (A);
      \fill (2,-1) circle (0.05);
      \draw[white,quark] (A) .. controls (0.25,-1.5) and (1.75,-1.5) .. (2,-1);
      \draw[white,quark] (C) .. controls (-0.75,-1.75) and (1.75,-1.75) .. (2,-1);
    \end{tikzpicture}\\
    &\quad+
    \begin{tikzpicture}
      [baseline={([yshift=-0.5ex]current bounding box.center)},scale=0.8]
      \node[xSource] (A) at (0,-1) {};
      \draw[directed graviton] (0,0) -- node[left] {$\ell_3\!\downarrow$} (A);
      \node[xSource] (B) at (1,-1) {};
      \draw[directed graviton] (1,0) -- node[left] {$\ell_1\!\downarrow$} (B);
      \node[xSource] (C) at (-1,-1) {};
      \draw[directed graviton] (-1,0) -- node[left] {$\ell_2\!\downarrow$} (C);
      \draw[quark] (B) -- (2,-1);
      \draw[quark] (A) -- (B);
      \draw[quark] (C) -- (A);
      \fill (2,-1) circle (0.05);
      \draw[white,quark] (A) .. controls (0.25,-1.5) and (1.75,-1.5) .. (2,-1);
      \draw[white,quark] (C) .. controls (-0.75,-1.75) and (1.75,-1.75) .. (2,-1);
    \end{tikzpicture}
    +
    \begin{tikzpicture}
      [baseline={([yshift=-0.5ex]current bounding box.center)},scale=0.8]
      \node[xSource] (A) at (0,-1) {};
      \draw[directed graviton] (0,0) -- node[left] {$\ell_1\!\downarrow$} (A);
      \node[xSource] (B) at (1,-1) {};
      \draw[directed graviton] (1,0) -- node[left] {$\ell_2\!\downarrow$} (B);
      \node[xSource] (C) at (-1,-1) {};
      \draw[directed graviton] (-1,0) -- node[left] {$\ell_3\!\downarrow$} (C);
      \draw[quark] (B) -- (2,-1);
      \draw[quark] (A) -- (B);
      \draw[quark] (C) -- (A);
      \fill (2,-1) circle (0.05);
      \draw[white,quark] (A) .. controls (0.25,-1.5) and (1.75,-1.5) .. (2,-1);
      \draw[white,quark] (C) .. controls (-0.75,-1.75) and (1.75,-1.75) .. (2,-1);
    \end{tikzpicture}
    +
    \begin{tikzpicture}
      [baseline={([yshift=-0.5ex]current bounding box.center)},scale=0.8]
      \node[xSource] (A) at (0,-1) {};
      \draw[directed graviton] (0,0) -- node[left] {$\ell_2\!\downarrow$} (A);
      \node[xSource] (B) at (1,-1) {};
      \draw[directed graviton] (1,0) -- node[left] {$\ell_1\!\downarrow$} (B);
      \node[xSource] (C) at (-1,-1) {};
      \draw[directed graviton] (-1,0) -- node[left] {$\ell_3\!\downarrow$} (C);
      \draw[quark] (B) -- (2,-1);
      \draw[quark] (A) -- (B);
      \draw[quark] (C) -- (A);
      \fill (2,-1) circle (0.05);
      \draw[white,quark] (A) .. controls (0.25,-1.5) and (1.75,-1.5) .. (2,-1);
      \draw[white,quark] (C) .. controls (-0.75,-1.75) and (1.75,-1.75) .. (2,-1);
    \end{tikzpicture}
  \end{aligned}
\end{equation}
Some useful generalizations for higher powers of WL propagators are given by
\tikzset{
  triple/.style={
    preaction={draw, line width=2.5pt, color=black},   
    preaction={draw, line width=1.75pt, color=white},   
    draw, line width=0.5pt, color=black                
  }
}%
\begin{equation}
  \begin{aligned}
    \begin{tikzpicture}
      [baseline={([yshift=-0.5ex]current bounding box.center)},scale=0.8]
      \node[xSource] (A) at (0,-1) {};
      \draw[directed graviton] (0,0) -- node[left] {$\ell_1\!\downarrow$} (A);
      \node[xSource] (B) at (1,-1) {};
      \draw[directed graviton] (1,0) -- node[left] {$\ell_2\!\downarrow$} (B);
      \draw[quark,double] (B) -- (2,-1);
      \draw[quark] (A) .. controls (0.25,-1.5) and (1.75,-1.5) .. (2,-1);
      \fill (2,-1) circle (0.05);
    \end{tikzpicture}
    &=
    \begin{tikzpicture}
      [baseline={([yshift=-0.5ex]current bounding box.center)},scale=0.8]
      \node[xSource] (A) at (0,-1) {};
      \draw[directed graviton] (0,0) -- node[left] {$\ell_1\!\downarrow$} (A);
      \node[xSource] (B) at (1,-1) {};
      \draw[directed graviton] (1,0) -- node[left] {$\ell_2\!\downarrow$} (B);
      \draw[quark,double] (B) -- (2,-1);
      \draw[quark] (A) -- (B);
      \draw[white,quark] (A) .. controls (0.25,-1.5) and (1.75,-1.5) .. (2,-1);
      \fill (2,-1) circle (0.05);
    \end{tikzpicture}
    +
    \begin{tikzpicture}
      [baseline={([yshift=-0.5ex]current bounding box.center)},scale=0.8]
      \node[xSource] (A) at (0,-1) {};
      \draw[directed graviton] (0,0) -- node[left] {$\ell_2\!\downarrow$} (A);
      \node[xSource] (B) at (1,-1) {};
      \draw[directed graviton] (1,0) -- node[left] {$\ell_1\!\downarrow$} (B);
      \draw[quark,double] (B) -- (2,-1);
      \draw[quark] (A) -- (B);
      \draw[white,quark] (A) .. controls (0.25,-1.5) and (1.75,-1.5) .. (2,-1);
      \fill (2,-1) circle (0.05);
    \end{tikzpicture}
    +
    \begin{tikzpicture}
      [baseline={([yshift=-0.5ex]current bounding box.center)},scale=0.8]
      \node[xSource] (A) at (0,-1) {};
      \draw[directed graviton] (0,0) -- node[left] {$\ell_2\!\downarrow$} (A);
      \node[xSource] (B) at (1,-1) {};
      \draw[directed graviton] (1,0) -- node[left] {$\ell_1\!\downarrow$} (B);
      \draw[quark] (B) -- (2,-1);
      \draw[quark,double] (A) -- (B);
      \draw[white,quark] (A) .. controls (0.25,-1.5) and (1.75,-1.5) .. (2,-1);
      \fill (2,-1) circle (0.05);
    \end{tikzpicture}\,,\\
    \begin{tikzpicture}
      [baseline={([yshift=-0.5ex]current bounding box.center)},scale=0.8]
      \node[xSource] (A) at (0,-1) {};
      \draw[directed graviton] (0,0) -- node[left] {$\ell_1\!\downarrow$} (A);
      \node[xSource] (B) at (1,-1) {};
      \draw[directed graviton] (1,0) -- node[left] {$\ell_2\!\downarrow$} (B);
      \draw[quark,double] (B) -- (2,-1);
      \draw[quark,double] (A) .. controls (0.25,-1.5) and (1.75,-1.5) .. (2,-1);
      \fill (2,-1) circle (0.05);
    \end{tikzpicture}
    &=
    2\begin{tikzpicture}
      [baseline={([yshift=-0.5ex]current bounding box.center)},scale=0.8]
      \node[xSource] (A) at (0,-1) {};
      \draw[directed graviton] (0,0) -- node[left] {$\ell_1\!\downarrow$} (A);
      \node[xSource] (B) at (1,-1) {};
      \draw[directed graviton] (1,0) -- node[left] {$\ell_2\!\downarrow$} (B);
      \draw[quark,triple] (B) -- (2,-1);
      \draw[quark] (A) -- (B);
      \draw[white,quark] (A) .. controls (0.25,-1.5) and (1.75,-1.5) .. (2,-1);
      \fill (2,-1) circle (0.05);
    \end{tikzpicture}
    +
    2\begin{tikzpicture}
      [baseline={([yshift=-0.5ex]current bounding box.center)},scale=0.8]
      \node[xSource] (A) at (0,-1) {};
      \draw[directed graviton] (0,0) -- node[left] {$\ell_2\!\downarrow$} (A);
      \node[xSource] (B) at (1,-1) {};
      \draw[directed graviton] (1,0) -- node[left] {$\ell_1\!\downarrow$} (B);
      \draw[quark,triple] (B) -- (2,-1);
      \draw[quark] (A) -- (B);
      \draw[white,quark] (A) .. controls (0.25,-1.5) and (1.75,-1.5) .. (2,-1);
      \fill (2,-1) circle (0.05);
    \end{tikzpicture}
    +
    \begin{tikzpicture}
      [baseline={([yshift=-0.5ex]current bounding box.center)},scale=0.8]
      \node[xSource] (A) at (0,-1) {};
      \draw[directed graviton] (0,0) -- node[left] {$\ell_1\!\downarrow$} (A);
      \node[xSource] (B) at (1,-1) {};
      \draw[directed graviton] (1,0) -- node[left] {$\ell_2\!\downarrow$} (B);
      \draw[quark,double] (B) -- (2,-1);
      \draw[quark,double] (A) -- (B);
      \draw[white,quark] (A) .. controls (0.25,-1.5) and (1.75,-1.5) .. (2,-1);
      \fill (2,-1) circle (0.05);
    \end{tikzpicture}
    +
    \begin{tikzpicture}
      [baseline={([yshift=-0.5ex]current bounding box.center)},scale=0.8]
      \node[xSource] (A) at (0,-1) {};
      \draw[directed graviton] (0,0) -- node[left] {$\ell_2\!\downarrow$} (A);
      \node[xSource] (B) at (1,-1) {};
      \draw[directed graviton] (1,0) -- node[left] {$\ell_1\!\downarrow$} (B);
      \draw[quark,double] (B) -- (2,-1);
      \draw[quark,double] (A) -- (B);
      \draw[white,quark] (A) .. controls (0.25,-1.5) and (1.75,-1.5) .. (2,-1);
      \fill (2,-1) circle (0.05);
    \end{tikzpicture}\,,
  \end{aligned}
\end{equation}
where double and triple lines stand for squared and cubed WL propagators.

In \texttt{FeynRIR} the following generic formula is used, serializing any number of WL propagators:
\begin{equation}
  \prod_{i=1}^n \frac{1}{a_i^{m_i}} = \prod_{i=1}^n \frac{(-1)^{m_i-1}}{(m_i-1)!} \partial_{a_i}^{m_i-1} \sum_{\pi\in\cS_n}\prod_{j=1}^n \frac{1}{A_j(\pi)}\,,
\end{equation}
with $\cS_n$ the set of permutations of $\{1,\dots,n\}$ and $A_j(\pi)=\sum_{k=1}^j a_{\pi(k)}$.
In practice we only need the version with $m_i=1$ or $m_i=2$, i.e. we only take derivatives w.r.t. denominators that appear squared on the LHS.
Note that the final sum in the form presented here may contain the same combination of denominators several times.

Optimally, one may want to collect those into a single new diagram with a combinatorial coefficient.
Hence, \texttt{FeynRIR} implements a closed form version of this formula, where each denominator combination only appears once, assuming $m_i=1$ or $m_i=2$,
\begin{equation}
\prod_{i=1}^n \frac{1}{a_i^{m_i}}=\sum_{\pi \in S_n}\ \sum_{\substack{h \ge 0 \\ \sum_k h_k = t}}
\left[\prod_{k=1}^n \big(R_k(\pi) - t + H_k(h)\big)^{\underline{h_k}}\right]
\prod_{k=1}^n \frac{1}{A_k(\pi)^{1+h_k}},
\end{equation}
where $T=\{i:m_i=2\}$, $t=|T|$,
\begin{align}
A_k(\pi) & =\sum_{j=1}^k a_{\pi(j)}\,, & 
R_k(\pi) &= \#\{i\in T:\mathrm{pos}_\pi(i)\le k\} \,, &
H_k(h) &= \sum_{l=k}^n h_l,
\end{align}
and the falling factorial $x^{\underline r}=x(x-1)\cdots(x-r+1)$.
Note that in these conventions and since all the WL propagators are incoming, they all carry the same $+i0$ contour deformation.
Hence, also every propagator on the RHS will inherit the same sign.
Furthermore, the exponential factor that is associated to the combination of the (sink and source) WL vertices (see eqs.\eqref{eq:source_vertex_final} and \eqref{eq:sink_vertex_final}) is unchanged and we are indeed allowed to apply this graphical identity to every subgraph in a graph-numerator pair.

\subsection{Selfie subdiagrams}\label{sec:selfies}

The second identity handles a problem related to ``selfie'' subdiagrams.
A selfie subdiagram is defined as a subgraph consisting only of graviton propagators and vertices, and WL (source or sink) vertices of a single black hole.

\subsubsection{Examples}

Consider the following (sub)graph of an observable or an iteration:
\begin{equation}
  \begin{tikzpicture}
    [baseline={([yshift=-0.5ex]current bounding box.center)},scale=0.9,font=\scriptsize]
    \node[xSource] (A) at (0,0) {};
    \draw[directed graviton] (0,-1) -- node[left] {$\ell_1$} (A);
    \draw[quark] (A) -- node[above=0.1] {$\ell_1\cdot u_1$} (1,0);
    \fill (1,0) circle (0.05);
    \node[xSource] (B) at (2,0) {};
    \draw[directed graviton] (1,0) .. controls (1,-0.6) and (2,-0.6) .. node[below=0.1] {$\ell_2$} (B);
    \draw[quark] (B) -- node[above=0.1] {$\ell_2\cdot u_1$} (3,0);
    \fill (3,0) circle (0.05);
    \draw[directed graviton] (3,0) -- (3,-1);
  \end{tikzpicture}\,.
\end{equation}
The WL sink and source vertices at the start and end of the left WL propagator encode an expression
\begin{equation}
  \propto \int \dd\tau e^{i (\ell_2-\ell_1)\cdot (b_1 + u_1 \tau)}\,.
\end{equation}
Performing this proper time integral introduces the delta function $\delta((\ell_1-\ell_2)\cdot u_1)$.
Enforcing manifest energy conservation in the diagram, this delta function can be drawn in the diagram as
\begin{equation}\label{eq:selfie_subdiagram_ex}
  \begin{aligned}
  \begin{tikzpicture}
    [baseline={([yshift=-0.5ex]current bounding box.center)},scale=0.9,font=\scriptsize]
    \node[xSource] (A) at (0,0) {};
    \draw[directed graviton] (0,-1) -- node[left] {$\ell_1$} (A);
    \draw[quark] (A) -- node[above] {$\ell_1\cdot u_1$} (1,0);
    \fill (1,0) circle (0.05);
    \node[xSource] (B) at (2,0) {};
    \draw[directed graviton] (1,0) .. controls (1,-0.6) and (2,-0.6) .. node[below=0.1] {$\ell_2$} (B);
    \draw[quark] (B) -- node[above] {$\ell_2\cdot u_1$} (3,0);
    \fill (3,0) circle (0.05);
    \draw[directed graviton] (3,0) -- (3,-1);
    \draw[dashed] (1,0) .. controls (1,0.8) and (3,0.8) .. node[above] {$\delta((\ell_1-\ell_2)\cdot u_1)$}(3,0);
  \end{tikzpicture}
  =
  \begin{tikzpicture}
    [baseline={([yshift=-0.5ex]current bounding box.center)},scale=0.9,font=\scriptsize]
    \node[xSource] (A) at (0,0) {};
    \draw[directed graviton] (0,-1) -- node[left] {$\ell_1$} (A);
    \draw[quark] (A) -- node[above] {$\ell_1\cdot u_1$} (1,0);
    \fill (1,0) circle (0.05);
    \node[xSource] (B) at (2,0) {};
    \draw[directed graviton] (1,0) .. controls (1,-0.6) and (2,-0.6) .. node[below=0.1] {$\ell_2$} (B);
    \draw[quark] (B) -- node[above] {$\ell_1\cdot u_1$} (3,0);
    \fill (3,0) circle (0.05);
    \draw[directed graviton] (3,0) -- (3,-1);
    \draw[dashed] (1,0) -- (B);
  \end{tikzpicture}
  &=-
  \begin{tikzpicture}
    [baseline={([yshift=-0.5ex]current bounding box.center)},scale=0.9,font=\scriptsize]
    \node[xSource] (A) at (0,0) {};
    \draw[directed graviton] (0,-1) -- node[left] {$\ell_1$} (A);
    \draw[quark,double] (A) -- node[above] {$\ell_1\cdot u_1$} (1,0);
    \fill (1,0) circle (0.05);
    \node[xSource] (B) at (2,0) {};
    \draw[directed graviton] (1,0) .. controls (1,-0.6) and (1.93,-0.6) .. node[below=0.1] {$\ell_2$} ([xshift=-2,yshift=2]B.south);
    \draw[directed graviton] ([xshift=2,yshift=2]B.south) -- (2.07,-1);
    \draw[dashed] (1,0) -- (B);
  \end{tikzpicture}
  \\
  &=-
  \begin{tikzpicture}
    [baseline={([yshift=-0.5ex]current bounding box.center)},scale=0.9,font=\scriptsize]
    \node[xSource] (A) at (0,0) {};
    \draw[directed graviton] (-0.07,-1) -- node[left] {$\ell_1$} ([xshift=-2,yshift=2]A.south);
    \node[xSource] (B) at (1,0) {};
    \draw[directed graviton] ([xshift=2,yshift=2]A) .. controls (0.07,-0.6) and (1,-0.6) .. node[below=0.1] {$\ell_2$} (B);
    \draw[quark,double] (B) -- node[above] {$\ell_1\cdot u_1$} (2,0);
    \fill (2,0) circle (0.05);
    \draw[directed graviton] (2,0) -- (2,-1);
    \draw[dashed] (A) -- (B);
  \end{tikzpicture}\,,
  \end{aligned}
\end{equation}
where we first used the delta function to equate the energies of the two WL propagators, then fused them since they also have the same $i0$ prescription.
Either choice of fused graph carries a minus sign.\footnote{The fusion removes one WL vertex and one WL propagator, whose factors $(i m_a)(i/m_a)=-1$ must be retained in the numerator.
The powers of $\kappa$ are carried by the fused vertex, while increasing the power of the surviving propagator introduces no additional factor of $i$.
See the discussion of imaginary factors in Sec.~\ref{sec:iteration_insertion}, in particular eq.~\eqref{eq:wl_fusion_prefactor}.}

The same mechanism is observed for any selfie graviton subdiagram that has several WL propagators attached to it.
In order to embed the propagators of such a diagram into an integral family as is commonly done in an IBP integral reduction, these WL propagators must be fused.
The reason is that they carry linearly dependent energies once the delta functions (from the WL source vertices in the selfie subdiagram) are taken into account and are, thus, as propagators linearly dependent.

Note that these diagrams do not represent topsector topologies.
E.g.
\begin{equation}
  \begin{tikzpicture}
    [baseline={([yshift=-0.5ex]current bounding box.center)},scale=0.9,font=\scriptsize]
    \node[xSource] (A) at (0,0) {};
    \draw[directed graviton] (0,-1) -- node[left] {$\ell_1$} (A);
    \draw[quark,double] (A) -- node[above] {$\ell_1\cdot u_1$} (1,0);
    \fill (1,0) circle (0.05);
    \node[xSource] (B) at (2,0) {};
    \draw[directed graviton] (1,0) .. controls (1,-0.6) and (1.93,-0.6) .. node[below=0.1] {$\ell_2$} ([xshift=-2,yshift=2]B.south);
    \draw[directed graviton] ([xshift=2,yshift=2]B.south) -- (2.07,-1);
    \draw[dashed] (1,0) -- (B);
  \end{tikzpicture}
  <
  \begin{tikzpicture}
    [baseline={([yshift=-0.5ex]current bounding box.center)},scale=0.9,font=\scriptsize]
    \node[xSource] (A) at (0,0) {};
    \draw[directed graviton] (0,-1) -- node[left] {$\ell_1$} (A);
    \draw[quark,double] (A) -- node[above] {$\ell_1\cdot u_1$} (1,0);
    \fill (1,0) circle (0.05);
    \draw[directed graviton] (1,0) .. controls (1,-0.6) and (1.8,-0.5) .. node[below=0.1] {$\ell_2$} (2,-0.5);
    \draw[directed graviton] (2,0) -- (2,-0.5);
    \draw[directed graviton] (2,-0.5) -- (2,-1);
    \draw[dashed] (1,0) -- (2,0);
    \fill (2,0) circle (0.05);
    \fill (2,-0.5) circle (0.05);
  \end{tikzpicture}\,.
\end{equation}
The challenge is to systematically use these identities to bring diagrams/topologies into a canonical form.
E.g. one has to decide which of the two graphs on the RHS in eq.~\eqref{eq:selfie_subdiagram_ex} one wants as the unique representative for this topology.
For many examples it is possible to make a local unique choice, often depending on the direction of the arrows.
But there are symmetric examples where one would have to take into account the whole rest of the diagrams to decide, for example for the symmetric subdiagram
\begin{equation}
  \begin{tikzpicture}
    [baseline={([yshift=-0.5ex]current bounding box.center)},scale=0.9,font=\scriptsize]
    \node[xSource] (A) at (0,0) {};
    \draw[directed graviton] (0,-1) -- (A);
    \draw[quark] (A) -- (1,0);
    \fill (1,0) circle (0.05);
    
    \node[xSource] (B) at (4,0) {};
    \draw[directed graviton] (4,-1) -- (B);
    \draw[quark] (B) --(3,0);
    \fill (3,0) circle (0.05);

    \node[xSource] (C) at (2,0) {};
    \draw[directed graviton] (2,-0.7) -- (C);

    \draw[directed graviton] (1,0) .. controls (1,-0.5) and (1.5,-0.7) .. (2,-0.7);

    \draw[directed graviton] (3,0) .. controls (3,-0.5) and (2.5,-0.7) .. (2,-0.7);

    \draw[dashed] (1,0) -- (C);
    \draw[dashed] (3,0) -- (C);

    \fill (2,-0.7) circle (0.05);
  \end{tikzpicture}\,.
\end{equation}

We also conjecture here that the sign of the $i0$ of these linearly dependent WL propagators is irrelevant.
In this simple case of a one-loop selfie subdiagram, this is proven by the fact that the selfie loop integration produces a result proportional to the linear propagator (with a fixed $i0$ sign according to the direction of the graviton propagator w.r.t. to the energy flow) to some fractional ($D$-dependent) power.
In the relevant radiation regions of the integral, this term ``overrules'' the sign of the $i0$ from the integer-powered WL propagators, see e.g.~\cite{Dlapa:2023hsl}.
This means that these two propagators can even be fused in case they have an opposite sign of the $i0$.
We do not have a generic proof of this behavior for high-loop selfies with a more complicated energy flow.

The whole situation is more complex for high-loop selfie subdiagrams with more than two in- or out-flowing WL propagators as, e.g., in this graph here:
\begin{equation}
  \begin{tikzpicture}
    [baseline={([yshift=-0.5ex]current bounding box.center)},scale=0.9,font=\scriptsize]
    \node[xSource] (A) at (0,0) {};
    \draw[directed graviton] (0,-1) -- (A);
    \draw[quark] (A) -- (1,0);
    \fill (1,0) circle (0.05);
    
    \node[xSource] (B) at (4,0) {};
    \draw[directed graviton] (4,-1) -- (B);
    \draw[quark] (B) --(3,0);
    \fill (3,0) circle (0.05);

    \node[xSource] (C) at (2,0) {};
    \draw[directed graviton] (2,-0.7) -- (C);

    \draw[directed graviton] (1,0) .. controls (1,-0.5) and (1.5,-0.7) .. (2,-0.7);

    \draw[directed graviton] (3,0) .. controls (3,-0.5) and (2.5,-0.7) .. (2,-0.7);

    \draw[dashed] (1,0) -- (C);
    \draw[dashed] (3,0) -- (C);

    \fill (2,-0.7) circle (0.05);

    \draw[quark] (C) .. controls (2,0.5) and (5,0.5) .. (5,0);
    \draw[directed graviton] (5,0) -- (5,-1);

    \fill (5,0) circle (0.05);
  \end{tikzpicture}\,.
\end{equation}
The WL propagator energies are linearly dependent, i.e. sum up to zero due to energy conservation.
In order to embed these propagators into an integral family one must use partial fraction identities to erase one of the WL propagators.
This leads to identities that can be represented as
\begin{equation}
  \begin{aligned}
    \begin{tikzpicture}
      [baseline={([yshift=-0.5ex]current bounding box.center)},scale=0.9,font=\scriptsize]
        \node[xSource] (A) at (0,0) {};
        \draw[directed graviton] (0,-1) -- (A);
        \draw[quark] (A) -- node[below] {$\ell_1\cdot u_1$} (1,0);
        \fill (1,0) circle (0.05);
        
        \node[xSource] (B) at (4,0) {};
        \draw[directed graviton] (4,-1) -- (B);
        \draw[quark] (B) -- node[below] {$\ell_2\cdot u_1$} (3,0);
        \fill (3,0) circle (0.05);

        \node[xSource] (C) at (2,0) {};
        \draw[directed graviton] (2,-0.7) -- (C);

        \draw[directed graviton] (1,0) .. controls (1,-0.5) and (1.5,-0.7) .. (2,-0.7);

        \draw[directed graviton] (3,0) .. controls (3,-0.5) and (2.5,-0.7) .. (2,-0.7);

        \draw[dashed] (1,0) -- (C);
        \draw[dashed] (3,0) -- (C);

        \fill (2,-0.7) circle (0.05);

        \draw[quark] (C) .. controls (2,0.5) and (5,0.5) .. node[above] {$(\ell_1+\ell_2)\cdot u_1$} (5,0);
        \draw[directed graviton] (5,0) -- (5,-1);

        \fill (5,0) circle (0.05);
    \end{tikzpicture}
    &=-
    \begin{tikzpicture}
      [baseline={([yshift=-0.5ex]current bounding box.center)},scale=0.9,font=\scriptsize]
        \node[xSource] (A) at (0,0) {};
        \draw[directed graviton] (0,-1) -- (A);
        \draw[quark] (A) -- (1,0);
        \fill (1,0) circle (0.05);
        
        \node[xSource] (B) at (3,0) {};
        \draw[directed graviton] (4,-1) -- (B);

        \node[xSource] (C) at (2,0) {};
        \draw[directed graviton] (2,-0.7) -- (C);

        \draw[directed graviton] (1,0) .. controls (1,-0.5) and (1.5,-0.7) .. (2,-0.7);

        \draw[directed graviton] (B) .. controls (3,-0.5) and (2.5,-0.7) .. (2,-0.7);

        \draw[dashed] (1,0) -- (C);
        \draw[dashed] (B) -- (C);

        \fill (2,-0.7) circle (0.05);

        \draw[quark,double] (C) .. controls (2,0.5) and (5,0.5) .. (5,0);
        \draw[directed graviton] (5,0) -- (5,-1);

        \fill (5,0) circle (0.05);
    \end{tikzpicture}
    & &-
    \begin{tikzpicture}
      [baseline={([yshift=-0.5ex]current bounding box.center)},scale=0.9,font=\scriptsize]
        \node[xSource] (A) at (1,0) {};
        \draw[directed graviton] (0,-1) -- (A);
        
        \node[xSource] (B) at (4,0) {};
        \draw[directed graviton] (4,-1) -- (B);
        \draw[quark] (B) --(3,0);
        \fill (3,0) circle (0.05);

        \node[xSource] (C) at (2,0) {};
        \draw[directed graviton] (2,-0.7) -- (C);

        \draw[directed graviton] (A) .. controls (1,-0.5) and (1.5,-0.7) .. (2,-0.7);

        \draw[directed graviton] (3,0) .. controls (3,-0.5) and (2.5,-0.7) .. (2,-0.7);

        \draw[dashed] (A) -- (C);
        \draw[dashed] (3,0) -- (C);

        \fill (2,-0.7) circle (0.05);

        \draw[quark, double] (C) .. controls (2,0.5) and (5,0.5) .. (5,0);
        \draw[directed graviton] (5,0) -- (5,-1);

        \fill (5,0) circle (0.05);
    \end{tikzpicture}\,,\\
    \frac{1}{(\ell_1\cdot u_1)(\ell_2\cdot u_1)((\ell_1+\ell_2)\cdot u_1)} &= \frac{1}{(\ell_1\cdot u_1)((\ell_1+\ell_2)\cdot u_1)^2} & & + \frac{1}{(\ell_2\cdot u_1)((\ell_1+\ell_2)\cdot u_1)^2}
  \end{aligned}
\end{equation}
We also conjecture here that the $i0$ signs of the linear propagators are insignificant and one can fuse/split in any desired way.

\subsubsection{Generic treatment}

We propose a generally applicable strategy for treating the linearly dependent WL propagators flowing in and out of selfie subdiagrams.
The generic graphical structure of a selfie subdiagram is
\begin{equation}
  \begin{tikzpicture}
    [baseline={([yshift=-0.5ex]current bounding box.center)},scale=0.8]
    \fill[blob] (0,0) circle (0.7);
    \node[fill=white] at (0,0) {selfie};
    \draw[quark] (180:1.5) -- (180:0.7);
    \draw[quark] (135:1.5) -- (135:0.7);
    \draw[quark] (225:1.5) -- (225:0.7);
    \draw[dotted] (100:1) arc[start angle=100, end angle=80, radius=1];
    \draw[dotted] (260:1) arc[start angle=260, end angle=280, radius=1];
    \draw[quark] (0:0.7) -- (0:1.5);
  \end{tikzpicture}\,,
\end{equation}
where the middle blob is a tree diagram consisting only of WL vertices of one WL, say WL 1, and graviton propagators.
Also note that several of these WL propagators may be attached to the same WL vertex.
Due to energy conservation the arguments of the linear WL propagators fulfill
\begin{equation}
  \sum_{i\in\textrm{in}} \ell_i\cdot u_1 = \sum_{j\in\textrm{out}}\ell_j\cdot u_1\,.
\end{equation}
In a proper WEFT diagram we have at most a single outgoing WL propagator.

Let us first consider the case where there is exactly one outgoing WL propagator and call it $\ell_0$.
The scalar partial-fraction identity and its graphical counterpart, including the vertex and propagator prefactors, read
\begin{equation}
  \begin{aligned}
    \frac{1}{\ell_0\cdot u_1}\prod_{i\in\textrm{in}}\frac{1}{\ell_i\cdot u_1} &=
    \frac{1}{(\ell_0\cdot u_1)^2} \sum_{j\in\textrm{in}} \prod_{\substack{i\in\textrm{in}\\i\neq j}}\frac{1}{\ell_i\cdot u_1}\\
    \begin{tikzpicture}
      [baseline={([yshift=-0.5ex]current bounding box.center)},scale=0.8]
      \fill[blob] (0,0) circle (0.7);
      \node[fill=white] at (0,0) {selfie};
      \draw[quark] (180:1.5) -- (180:0.7);
      \draw[quark] (135:1.5) -- (135:0.7);
      \draw[quark] (225:1.5) -- (225:0.7);
      \draw[dotted] (100:1) arc[start angle=100, end angle=80, radius=1];
      \draw[dotted] (260:1) arc[start angle=260, end angle=280, radius=1];
      \draw[quark] (0:0.7) -- (0:1.5);
    \end{tikzpicture}
    &=-
    \begin{tikzpicture}
      [baseline={([yshift=-0.5ex]current bounding box.center)},scale=0.8]
      \fill[blob] (0,0) circle (0.7);
      \node[fill=white] at (0,0) {selfie};
      \draw[quark] (180:1.5) -- (180:0.7);
      \draw[quark,black!20!white] (135:1.5) -- (135:0.7);
      \draw[quark] (225:1.5) -- (225:0.7);
      \draw[dotted] (100:1) arc[start angle=100, end angle=80, radius=1];
      \draw[dotted] (260:1) arc[start angle=260, end angle=280, radius=1];
      \draw[quark,double] (0:0.7) -- (0:1.5);
    \end{tikzpicture}
    -
    \begin{tikzpicture}
      [baseline={([yshift=-0.5ex]current bounding box.center)},scale=0.8]
      \fill[blob] (0,0) circle (0.7);
      \node[fill=white] at (0,0) {selfie};
      \draw[quark,black!20!white] (180:1.5) -- (180:0.7);
      \draw[quark] (135:1.5) -- (135:0.7);
      \draw[quark] (225:1.5) -- (225:0.7);
      \draw[dotted] (100:1) arc[start angle=100, end angle=80, radius=1];
      \draw[dotted] (260:1) arc[start angle=260, end angle=280, radius=1];
      \draw[quark,double] (0:0.7) -- (0:1.5);
    \end{tikzpicture}
    -
    \begin{tikzpicture}
      [baseline={([yshift=-0.5ex]current bounding box.center)},scale=0.8]
      \fill[blob] (0,0) circle (0.7);
      \node[fill=white] at (0,0) {selfie};
      \draw[quark] (180:1.5) -- (180:0.7);
      \draw[quark] (135:1.5) -- (135:0.7);
      \draw[quark,black!20!white] (225:1.5) -- (225:0.7);
      \draw[dotted] (100:1) arc[start angle=100, end angle=80, radius=1];
      \draw[dotted] (260:1) arc[start angle=260, end angle=280, radius=1];
      \draw[quark,double] (0:0.7) -- (0:1.5);
    \end{tikzpicture}
    -\dots\,,
  \end{aligned}
\end{equation}
where a faded WL represents a \emph{contracted} WL.
Each term in the graphical identity carries the fusion minus sign explained in eq.~\eqref{eq:wl_fusion_prefactor}; the scalar denominator identity does not include vertex or propagator prefactors.
The formal generalized version to higher powers of this equation is
\begin{equation}
  \prod_{i=0}^{n}\frac{1}{a_i^{m_i}} =
  \sum_{j=1}^{n}\sum_{\substack{0 \le c_i \le m_i-1 \\ 0<i \neq j}}\binom{m_j-1+\sum_{i\ne j}c_i}{c_1,\dots,\widehat{c_j},\dots,c_n,m_j-1}\frac{1}{a_0^{m_0+m_j+\sum_{i\ne j}c_i}}\prod_{i\ne j}\frac{1}{a_i^{m_i-c_i}}\,,
\end{equation}
where $a_0=\sum_{i=1}^n a_i$ and a hat denotes an omitted element.
Note that both sums and the product over $i\ne j$ do not include $i=0$ but range from $1$ to $n$ excluding $j$.
This identity can be derived by recursively multiplying every term appearing in the current expression by $1=\left(\sum_{i=1}^n a_i\right)/a_0$, expanding and canceling the $a_i$.
Starting from the expression on the left hand side of the equation and applying this procedure to every term that still contains at least one power of each $a_i$ in the denominator will eventually land on the expression on the right hand side.
The multinomial factor simply counts how many times a certain combination corresponds to the final expression of different branches in this algorithm.
\texttt{FeynRIR} applies this identity to graph-numerator pairs with the relevant structures and multiplies each resulting numerator by the scalar coefficient, including the fusion minus sign.

In all other cases, where no outgoing WL propagator is present, \texttt{FeynRIR} decides to democratically symmetrize over all choices of ``special'' legs.
In principle, that might even be the better version in general, since most (probably even all) of these topologies are anyway present and need to be considered for integration.
Note that one needs to be careful about the $i0$ prescription since the energy conservation might introduce minus signs.

Note that after this step, we may have duplicate diagrams in our collection and one may, after an appropriate canonicalization of the graph representation, re-organize an integrand into graph-numerator pairs of mutually distinct graphs.
\texttt{FeynRIR} contains helper functions to achieve this and fuse the corresponding terms into a single graph-numerator pair.
The exponentials remain unchanged by these operations; the corresponding scalar coefficients and fusion signs are included in the numerator.

Finally, we note that in the case of an insertion of an acceleration $\delta^{(n)}\ddot{x}_a^\mu$ as depicted in~\eqref{eq:graphical_iterations}, several graviton propagators might meet at the same WL (source or sink) vertex.
This construct needs to be interpreted as a single big selfie subdiagram.
For example, we may have
\begin{equation}
  \begin{tikzpicture}
    [baseline={([yshift=-0.5ex]current bounding box.center)},scale=0.9,font=\scriptsize]
    \node[xSource] (A) at (0,0) {};
    \draw[directed graviton] (0,-1) -- (A);
    \draw[quark] (A) -- (1,0);
    \fill (1,0) circle (0.05);
    \node (B) at (2,0) {};
    \fill (B) circle (0.05);
    \draw[directed graviton] (1,0) .. controls (1,-0.6) and (2,-0.6) ..  (B);
    \node[xSource] (C) at (3,0) {};
    \draw[quark] (C) -- (4,0);
    \draw[directed graviton] (B) .. controls (2,-0.6) and (3,-0.6) .. (C);
    \fill (4,0) circle (0.05);
    \draw[directed graviton] (4,0) -- (4,-1);
  \end{tikzpicture}\,,
\end{equation}
where the two graviton propagators in the middle need to be considered a single selfie subdiagram.\footnote{
The attentive reader may have noticed that the first graviton propagator points to a WL source vertex.
This is not a mistake.
The reason is that at this vertex a WL sink and a WL source vertex have been merged due to the absence of a WL propagator.
The choice of a surviving WL source vertex is correct since it carries a proper time integration according to equation~\eqref{eq:source_final}, whereas the integration would be absent for a WL sink vertex.}

\subsection{Proper time integrals}\label{sec:proper-time-integrals}

At this stage of the integrand construction, each WL source vertex still encodes a proper time integration.
When computing the impulse from the variation of the effective action through
\begin{equation}
  \Delta p_a^\mu = - \eta^{\mu\nu} \int_{-\infty}^\infty \dd\sigma \left.\frac{\delta\cS_\textrm{eff}[x_+,x_-]}{\delta x_{a-}^\nu(\sigma)}\right|_{\substack{x_-=0\\x_+=x}}\,,
\end{equation}
there is also a final proper time integral associated to the final WL sink vertex with proper time $\sigma$.

Since the proper time only appears in the exponent, each such integral amounts to a Fourier transform of the form
\begin{equation}
  \int_{-\infty}^\infty \dd\tau e^{i\, P.u_a \tau} = \hat\delta(P.u_a)\,,
\end{equation}
where $\hat\delta(x) = 2\pi \delta(x)$.
The momentum $P$ is the total loop momentum flow at the corresponding WL vertex: The momentum of the outgoing graviton momentum minus the sum of all incoming momenta (energies) through WL iterations and gravitons.
The latter is possible through fused WL vertices due to insertion of iterations of $\ddot{x}$.
These delta functions may be used to simplify the algebraic expression of the numerator factor.

Due to symmetries and the matching to integral families as discussed in the next section, it is beneficial to perform any proper time integrals after having matched to a family.
As we will see, this makes it much easier to manifest symmetries, have a consistent loop momentum labeling across all diagrams, and trivialize the mapping to a fixed integral family.
Most of this relies on the fact that we can continue working with tree diagrams, as only these delta functions (interpreted as cut WL propagators) will close the loops.
We will come back to these ideas in Sec.~\ref{sec:ibp_proper_time}.

\section{Toward IBP reduction}\label{sec:ibp}

Representing an integrand as a graph-numerator pair also leads to a novel approach to write it in terms of an integral family.
The novelty consists of a generalization of the notion of an integral family compared to the standard IBP literature.
We will show that this allows the embedding of each self-force (SF) sector into a single (generalized) integral family, while at the same time almost trivializing symmetries and modifying the notion of a \emph{sector}.
Note that this setup is especially beneficial if one targets a subsequent \emph{symbolic} integral reduction (see e.g.~\cite{Lee:2012cn,Lee:2013mka,Dlapa:2026oyq,Feng:2026imq,delaCruz:2026mas}), which acts independently on each (generalized) integral sector.

\paragraph{SF sectors:} The integrand for the impulse $\Delta p_a^\mu$ naturally splits into independent SF sectors.\footnote{This was for example observed in~\cite{Damour:2019lcq} and was dubbed \emph{mass polynomiality}}
In our Post-Minkowskian WEFT setup the mass structure is trivial: Each WL vertex introduces a mass $m_a$, whereas each WL propagator introduces an inverse mass $m_a^{-1}$.
Since the number of WL vertices (of one type) is always larger than the number of WL propagators the final expression is trivially a polynomial.
Furthermore, the homogeneous degree of the particle masses at $n$PM is given by $n+1$.
Each coefficient of a mass monomial is independent and is called an SF sector.
The $m$SF sector corresponds to the mass monomial $m_1^{m+1}m_2^{n-m}$.
We also introduce the notation of $m\overline{\textrm{SF}}$ which corresponds to a mass scaling of $m_1^{n-m}m_2^{m+1}$.

\subsection{Invitation: 3PM (2-loop), 0SF example}\label{sec:topo_example}

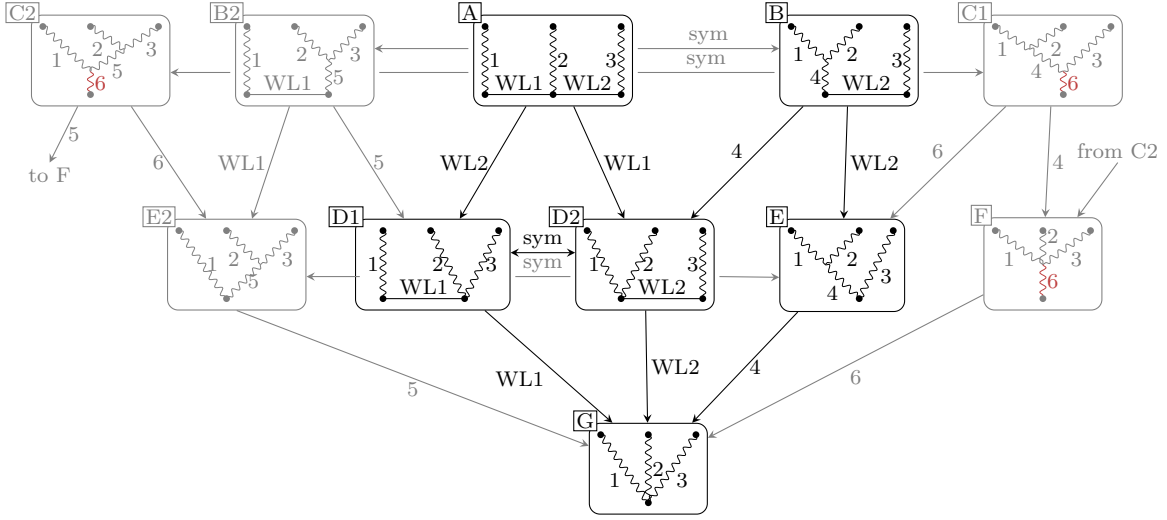
\begin{figure}[h]
  \begin{tikzpicture}%
  [baseline={([yshift=-0.5ex]current bounding box.center)},font=\scriptsize,scale=0.9]%
  \begin{scope}[shift={(-0.5,0)}, local bounding box=boxA]
    \draw[graviton] (0,0) -- node[right] {1} (0,-1);
    \draw[graviton] (1,0) -- node[right] {2} (1,-1);
    \draw[graviton] (2,0) -- node[left] {3} (2,-1);
    \draw (0,-1) -- node[above] {WL1} (1,-1);
    \draw (1,-1) -- node[above] {WL2} (2,-1);
    \fill (0,0) circle (0.05);
    \fill (1,0) circle (0.05);
    \fill (2,0) circle (0.05);
    \fill (0,-1) circle (0.05);
    \fill (1,-1) circle (0.05);
    \fill (2,-1) circle (0.05);
  \end{scope}

  \node[
    draw,
    rounded corners,
    fit=(boxA),
    inner sep=3pt
  ] (A) {};

  \node[
    anchor=south east,
    fill=white,
    inner sep=1pt,
    draw
  ] at (boxA.north west) {A};

  \begin{scope}[shift={(4,0)}, local bounding box=boxB]
    \draw[graviton] (0,0) -- node[below left] {1} (0.5,-0.5);
    \draw[graviton] (1,0) -- node[below right] {2} (0.5,-0.5);
    \draw[graviton] (0.5,-0.5) -- node[left] {4} (0.5,-1);
    \draw[graviton] (1.7,0) -- node[left] {3} (1.7,-1);
    \draw (0.5,-1) -- node[above] {WL2} (1.7,-1);
    \fill (0,0) circle (0.05);
    \fill (1,0) circle (0.05);
    \fill (1.7,0) circle (0.05);
    \fill (0.5,-1) circle (0.05);
    \fill (1.7,-1) circle (0.05);
  \end{scope}

  \node[
    draw,
    rounded corners,
    fit=(boxB),
    inner sep=3pt
  ] (B) {};

  \node[
    anchor=south east,
    fill=white,
    inner sep=1pt,
    draw
  ] at (boxB.north west) {B};

  \begin{scope}[shift={(7,0)}, local bounding box=boxC1]
    \draw[graviton,gray] (0,0) -- node[below left] {1} (0.5,-0.33);
    \draw[graviton,gray] (1,0) -- node[below right] {2} (0.5,-0.33);
    \draw[graviton,gray] (1.7,0) -- node[below right] {3} (1,-0.66);
    \draw[graviton,gray] (0.5,-0.33) -- node[below left] {4} (1,-0.66);
    \draw[graviton,gray!50!red] (1,-0.66) -- node[right] {6} (1,-1);
    \fill[gray] (0,0) circle (0.05);
    \fill[gray] (1,0) circle (0.05);
    \fill[gray] (1.7,0) circle (0.05);
    \fill[gray] (1,-1) circle (0.05);
  \end{scope}

  \node[
    draw=gray,
    rounded corners,
    fit=(boxC1),
    inner sep=3pt
  ] (C1) {};

  \node[
    anchor=south east,
    fill=white,
    inner sep=1pt,
    draw=gray,
    text=gray
  ] at (boxC1.north west) {C1};

  \begin{scope}[shift={(-2,-3)}, local bounding box=boxD1]
    \draw[graviton] (0,0) -- node[left] {1} (0,-1);
    \draw[graviton] (0.7,0) -- node[left] {2} (1.2,-1);
    \draw[graviton] (1.7,0) -- node[right] {3} (1.2,-1);
    \draw (0,-1) -- node[above] {WL1} (1.2,-1);
    \fill (0,0) circle (0.05);
    \fill (0.7,0) circle (0.05);
    \fill (1.7,0) circle (0.05);
    \fill (0,-1) circle (0.05);
    \fill (1.2,-1) circle (0.05);
  \end{scope}

  \node[
    draw,
    rounded corners,
    fit=(boxD1),
    inner sep=3pt
  ] (D1) {};

  \node[
    anchor=south east,
    fill=white,
    inner sep=1pt,
    draw
  ] at (boxD1.north west) {D1};

  \begin{scope}[shift={(1,-3)}, local bounding box=boxD2]
    \draw[graviton] (0,0) -- node[left] {1} (0.5,-1);
    \draw[graviton] (1,0) -- node[right] {2} (0.5,-1);
    \draw[graviton] (1.7,0) -- node[left] {3} (1.7,-1);
    \draw (0.5,-1) -- node[above] {WL2} (1.7,-1);
    \fill (0,0) circle (0.05);
    \fill (1,0) circle (0.05);
    \fill (1.7,0) circle (0.05);
    \fill (0.5,-1) circle (0.05);
    \fill (1.7,-1) circle (0.05);
  \end{scope}

  \node[
    draw,
    rounded corners,
    fit=(boxD2),
    inner sep=3pt
  ] (D2) {};

  \node[
    anchor=south east,
    fill=white,
    inner sep=1pt,
    draw
  ] at (boxD2.north west) {D2};

  \begin{scope}[shift={(4,-3)}, local bounding box=boxE]
    \draw[graviton] (0,0) -- node[below left] {1} (0.5,-0.5);
    \draw[graviton] (1,0) -- node[below right] {2} (0.5,-0.5);
    \draw[graviton] (1.5,0) -- node[below right] {3} (1,-1);
    \draw[graviton] (0.5,-0.5) -- node[below left] {4} (1,-1);
    \fill (0,0) circle (0.05);
    \fill (1,0) circle (0.05);
    \fill (1.5,0) circle (0.05);
    \fill (1,-1) circle (0.05);
  \end{scope}

  \node[
    draw,
    rounded corners,
    fit=(boxE),
    inner sep=3pt
  ] (E) {};

  \node[
    anchor=south east,
    fill=white,
    inner sep=1pt,
    draw
  ] at (boxE.north west) {E};

  \begin{scope}[shift={(7,-3)}, local bounding box=boxF]
    \draw[graviton,gray] (0,0) -- node[below left] {1} (0.7,-0.5);
    \draw[graviton,gray] (0.7,0) -- node[above right] {2} (0.7,-0.5);
    \draw[graviton,gray] (1.4,0) -- node[below right] {3} (0.7,-0.5);
    \draw[graviton,gray!50!red] (0.7,-0.5) -- node[right] {6} (0.7,-1);
    \fill[gray] (0,0) circle (0.05);
    \fill[gray] (0.7,0) circle (0.05);
    \fill[gray] (1.4,0) circle (0.05);
    \fill[gray] (0.7,-1) circle (0.05);
  \end{scope}

  \node[
    draw=gray,
    rounded corners,
    fit=(boxF),
    inner sep=3pt
  ] (F) {};

  \node[
    anchor=south east,
    fill=white,
    inner sep=1pt,
    draw=gray,
    text=gray
  ] at (boxF.north west) {F};

  \begin{scope}[shift={(1.2,-6)}, local bounding box=boxG]
    \draw[graviton] (0,0) -- node[below left] {1} (0.7,-1);
    \draw[graviton] (0.7,0) -- node[right] {2} (0.7,-1);
    \draw[graviton] (1.4,0) -- node[below right] {3} (0.7,-1);
    \fill (0,0) circle (0.05);
    \fill (0.7,0) circle (0.05);
    \fill (1.4,0) circle (0.05);
    \fill (0.7,-1) circle (0.05);
  \end{scope}

  \node[
    draw,
    rounded corners,
    fit=(boxG),
    inner sep=3pt
  ] (G) {};

  \node[
    anchor=south east,
    fill=white,
    inner sep=1pt,
    draw
  ] at (boxG.north west) {G};

  \begin{scope}[shift={(-4,0)}, local bounding box=boxB2]
    \draw[graviton,gray] (0,0) -- node[right] {1} (0,-1);
    \draw[graviton,gray] (0.7,0) -- node[below left] {2} (1.2,-0.5);
    \draw[graviton,gray] (1.7,0) -- node[below right] {3} (1.2,-0.5);
    \draw[graviton,gray] (1.2,-0.5) -- node[right] {5} (1.2,-1);
    \draw[gray] (0,-1) -- node[above] {WL1} (1.2,-1);
    \fill[gray] (0,0) circle (0.05);
    \fill[gray] (0.7,0) circle (0.05);
    \fill[gray] (1.7,0) circle (0.05);
    \fill[gray] (0,-1) circle (0.05);
    \fill[gray] (1.2,-1) circle (0.05);
  \end{scope}

  \node[
    draw=gray,
    rounded corners,
    fit=(boxB2),
    inner sep=3pt
  ] (B2) {};

  \node[
    anchor=south east,
    fill=white,
    inner sep=1pt,
    draw=gray,
    text=gray
  ] at (boxB2.north west) {B2};

  \begin{scope}[shift={(-5,-3)}, local bounding box=boxE2]
    \draw[graviton,gray] (0,0) -- node[right] {1} (0.7,-1);
    \draw[graviton,gray] (0.7,0) -- node[below left] {2} (1.2,-0.5);
    \draw[graviton,gray] (1.7,0) -- node[below right] {3} (1.2,-0.5);
    \draw[graviton,gray] (1.2,-0.5) -- node[right] {5} (0.7,-1);
    \fill[gray] (0,0) circle (0.05);
    \fill[gray] (0.7,0) circle (0.05);
    \fill[gray] (1.7,0) circle (0.05);
    \fill[gray] (0.7,-1) circle (0.05);
  \end{scope}

  \node[
    draw=gray,
    rounded corners,
    fit=(boxE2),
    inner sep=3pt
  ] (E2) {};

  \node[
    anchor=south east,
    fill=white,
    inner sep=1pt,
    draw=gray,
    text=gray
  ] at (boxE2.north west) {E2};

  \begin{scope}[shift={(-7,0)}, local bounding box=boxC2]
    \draw[graviton,gray] (0,0) -- node[below left] {1} (0.7,-0.66);
    \draw[graviton,gray] (0.7,0) -- node[below left] {2} (1.2,-0.33);
    \draw[graviton,gray] (1.7,0) -- node[below right] {3} (1.2,-0.33);
    \draw[graviton,gray] (1.2,-0.33) -- node[below right] {5} (0.7,-0.66);
    \draw[graviton,gray!50!red] (0.7,-0.66) -- node[right] {6} (0.7,-1);
    \fill[gray] (0,0) circle (0.05);
    \fill[gray] (0.7,0) circle (0.05);
    \fill[gray] (1.7,0) circle (0.05);
    \fill[gray] (0.7,-1) circle (0.05);
  \end{scope}

  \node[
    draw=gray,
    rounded corners,
    fit=(boxC2),
    inner sep=3pt
  ] (C2) {};

  \node[
    anchor=south east,
    fill=white,
    inner sep=1pt,
    draw=gray,
    text=gray
  ] at (boxC2.north west) {C2};

  \draw[->] (A) -- node[left] {WL2} (D1);
  \draw[->] (A) -- node[right] {WL1} (D2);
  \draw[<->] ([yshift=5]D1.east) -- node[above] {sym} ([yshift=5]D2.west);
  \draw[->] (B) -- node[above left] {4} (D2);
  \draw[->] (B) -- node[right] {WL2} (E);
  \draw[->,gray] (C1) -- node[above left] {6} (E);
  \draw[->,gray] (C1) -- node[right] {4} (F);
  \draw[->] (D1) -- node[below left] {WL1} ([xshift=-15]G.north);
  \draw[->] (D2) -- node[right] {WL2} (G);
  \draw[->] (E) -- node[right] {4} (G);
  \draw[->,gray] (F) -- node[below right] {6} (G);

  \draw[->,gray] (B2) -- node[right] {5} (D1);
  \draw[->,gray] (B2) -- node[left] {WL1} (E2);
  \draw[->,gray] (E2.south) -- node[below] {5} (G);
  \draw[<-,gray] ([yshift=5]B2.east) -- ([xshift=-2,yshift=5]A.west);
  \draw[->,gray] ([xshift=2,yshift=5]A.east) -- node[above] {sym} ([yshift=5]B.west);
  \draw[<-,gray] ([yshift=-5]C2.east) -- ([xshift=-2,yshift=-5]B2.west);
  \draw[gray] ([yshift=-5,xshift=2]B2.east) -- ([yshift=-5,xshift=-2]A.west);
  \draw[gray] ([yshift=-5,xshift=2]A.east) -- node[above] {sym} ([yshift=-5,xshift=-2]B.west);
  \draw[->,gray] ([xshift=2,yshift=-5]B.east) -- ([yshift=-5]C1.west);
  \draw[->,gray] (C2) -- node[left] {6} (E2);
  \draw[->,gray] (C2) -- node[right] {5} (-6.9,-2) node[below] {to F};
  \draw[->,gray] (8.8,-2) node[above] {from C2} -- (F);
  \draw[<-,gray] ([yshift=-5]E2.east) -- ([xshift=2,yshift=-5]D1.west);
  \draw[gray] ([xshift=2,yshift=-5]D1.east) -- node[above] {sym} ([xshift=-2,yshift=-5]D2.west);
  \draw[->,gray] ([xshift=2,yshift=-5]D2.east) -- ([yshift=-5]E.west);
\end{tikzpicture}
  \caption{All required sectors for the 3PM 0SF/0$\overline{\textrm{SF}}$ are depicted in black.
  The grayed-out configurations are not needed for different reasons: either symmetries on the integrand before applying IBP relations make them obsolete or they shrink to a subtopology once one contracts the ``external'' momentum $q$ edge corresponding to label 6.
  The vertical arrows between the graphs indicate the contracted edge.
  Horizontal edges show (sector) symmetry relations.
  }
  \label{fig:topo_3PM_0SF}
\end{figure}

Let us consider the 3PM integrand in the 0SF or $0\overline{\textrm{SF}}$ sector.
For an IBP reduction (ignoring symmetry relations for the moment) the $i0$ prescription does not matter and we get the integral \emph{topologies} by simply dropping the arrows from all graphs.
Up to relabelings, this leads to the topology graphs presented in Fig.~\ref{fig:topo_3PM_0SF}.

In order to make contact with the concept of an integral family (and its sectors) we need to associate labels to each edge in a topology graph, corresponding to a sector.
Meaning that there may be several sectors corresponding to the same topology graph with a different labeling (related by a symmetry relation).
This is done consistently (w.r.t. edge contractions) in the example in Fig.~\ref{fig:topo_3PM_0SF}.

The first important observation is that we can pick one canonical labeling per topology and map all graphs in the integrand to its canonicalized version to label the momenta in the same way.
A second observation is that some graphs contain edges corresponding to the ``external'' momentum exchange of the two worldlines $q$ (marked in red in the figure).
The latter edge may be contracted as it is not representing a propagator relevant for IBPs.
Finally, there are other configurations corresponding to relabelings of these graphs -- that we do not even show here -- that correspond to sectors of the \emph{same} family.
We will see below that many non-zero sectors are missing from the list.

In our example, we choose topology graphs A, B, D1, E, G as the canonical representatives for the needed topologies.
When performing an IBP reduction we will also have to introduce the topology D2 since it is introduced by contracting WL1 in graph A and may appear in an intermediate step of the IBP reduction.
Since we do not want to use (sector) symmetry relations during the IBP reduction as this would interfere with the $i0$ assignment we never relate integrals in the D2 sector to those in D1 during the IBP reduction.
The sectors corresponding to the topologies B2, E2, and F will never be introduced and are, hence, never needed.

There is also a way to trivialize the D2 sector if one goes the symbolic IBP route.
Say, we have derived symbolic IBP reduction rules for sector D1.
Those rules can be trivially mapped to D2 using the symmetry relation.
The application of the rules -- which is the second step of the reduction procedure -- will still distinguish between the two sectors, but the symbolic rules need only be derived for one representative of a symmetry equivalence class.
We conclude that we need to derive symbolic reduction rules only for the 5 sectors A, B, D1, E, G.

We note that one may choose a different loop momentum labeling for each graph/sector as it is trivial to map from sector to subsector, i.e. one must apply a simple linear transformation to the numerator ISPs.

\paragraph{Comparison to traditional integral family.}
We can assign explicit loop momenta to the graphs and explicitly map to the more traditional concept of an integral family.
Let us assign the following momenta to the edges:
\begin{align}
  1&\rightarrow \ell_1 \,, & 2&\rightarrow \ell_1+\ell_2 - q \,, & 3&\rightarrow \ell_2 \,, & 4&\rightarrow \ell_2-q\,, & 5&\rightarrow\ell_1-q\,, & 6&\rightarrow q\,.
\end{align}
Upon correct direction assignment, and closing the loops by performing the proper time integrals,\footnote{In this example there is an obvious, unique choice for the delta functions on the upper worldline such that all the topologies are planar and the labeling is consistent.} we have the following set of propagators
\begin{align}
  \{\delta(\ell_1\cdot u_1),\delta(\ell_2\cdot u_1)\,|\,\ell_1\cdot u_2,\ell_2\cdot u_2\,|\,-\ell_1^2,-\ell_2^2,-(\ell_1-q)^2,-(\ell_2-q)^2,-(\ell_1+\ell_2-q)^2\}\,,
\end{align}
already familiar from e.g.~\cite{Dlapa:2023hsl}.
The graphs in Fig.~\ref{fig:topo_3PM_0SF} then correspond to the following sector strings:
\begin{equation}
  \begin{aligned}
    A &\rightarrow 11|11|11001 \,, & B &\rightarrow 11|01|11011 \,, & D1 &\rightarrow 11|10|11001 \,,\\
    \textcolor{gray}{D2} &\textcolor{gray}{\rightarrow 11|01|11001} \,, &  E = C1 &\rightarrow 11|00|11011 \,, &  G = F &\rightarrow 11|00|11001 \,,\\
    \textcolor{gray}{B2} &\textcolor{gray}{\rightarrow 11|10|11101} \,, &  \textcolor{gray}{E2 = C2} &\textcolor{gray}{\rightarrow 11|00|11101} \,.
  \end{aligned}
\end{equation}
We emphasize that with this approach, symbolic reduction rules for only 5 out of the 40 non-zero sectors of this family are required for a complete integral reduction.

\subsection{Identifying integral topologies and sectors}

The seemingly random choices and simplifications discussed in the previous section can be systematically generalized to any ``family'' of integrals, or rather to any SF sector at fixed PM order.
The following algorithm describes a systematic procedure to identify a minimal set of required sectors.
To be precise, the first set $I_1$ consists of sectors for which we require symbolic IBP reduction rules and a second set $I_2$ which contains all the sectors that appear in intermediate steps of the actual reduction.
The complement of the set $I_1$ within $I_2$ contains sectors for which we can trivially obtain symbolic IBP reduction rules through a symmetry relation.

\begin{enumerate}
\item Take all the graphs in the integrand and remove the arrows, rendering them \emph{topology graphs}.
\item Contract any edge that corresponds to the momentum transfer $q$ (easily identified as an edge that splits the graph into two parts such that each part only contains WL vertices of one WL).
\item From this set, choose all the graphs that are not subtopologies of any other graph (i.e. that cannot be obtained by contracting edges from another one).
\item Canonicalize these top-level topologies (and deduplicate the set).
\item Transform them to (top-level) sectors by labeling them consistently and introduce subsectors.
  Register all the (sub)sectors obtained in the following subprocedure in a directed acyclic graph (DAG) where edges correspond to contraction of propagators.
  Each edge carries a node and edge map recording the relation between sector and subsector.
  \begin{enumerate}
  \item Start with any top-level topology and label it in any desired way to get a starting top-level sector.
  \item Contract (recursively) the edges in all possible orders to get all labeled sub-topologies, i.e. subsectors.
    Ignore any zero-sector in the process, i.e. terminate the recursion at zero-sectors.
  \item Choose any not-yet labeled top-level topology that maps to an already obtained subsector by contracting a single edge and label it such that it is consistent with this contraction.
    There are two possibilities for the single edge that does not receive a labeling through this condition: \emph{i)} contracting another edge may lead to an already obtained subsector where the labeling agrees which fixes the labeling of the last leg, or \emph{ii)} one must introduce a new label otherwise.
  \item Repeat steps (b) and (c) for this new top-level sector until no top-level topologies are left over.
  \end{enumerate}
  \item All the (sub)sectors obtained up until here form the set $I_2$.
  \item Identify symmetry relations among these sectors and consistently choose one representative per symmetry equivalence class, remove the rest to obtain the set $I_1$.
    This can be done similarly as the recursive strategy in step 5 by starting with one choice of representative as far up in the DAG as possible and choosing representatives consistently downward in the DAG, making sure previous choices are respected.
\end{enumerate}

Some comments are in order.
\begin{itemize}
  \item\textbf{Zero sectors.}
    We conjecture the following graphical criterion to identify zero sectors in our WEFT integral families.
    Consider any maximal subgraph made of WL propagators only.
    If after removing all those WL propagators (but not the WL vertices) from the graph any of the disconnected remaining graphs does not have vertices of both WLs, the graph corresponds to a scaleless integral and can be safely ignored in dimensional regularization.
    The reason is that the affected disconnected subgraph factorizes and is independent of $q$, and hence scaleless.
  \item\textbf{Loop momentum assignment.}
    Each sector may have a different choice of loop momenta assignment (on top of the edge labeling).
    This makes the IBP reduction from sector to subsector slightly more involved as one needs to properly map the ISPs into each other.
    This is though nothing but a simple linear transformation.
  \item\textbf{Integral ordering.}
    Every unique sector in the set $I_1$ may have a different (symbolic) integral ordering.
    The additional sectors in $I_2$ related to another sector in $I_1$ must inherit the integral ordering through the symmetry such that the symmetry-mapped reduction rules stay valid.
  \item\textbf{ISPs. }
    This treatment of families/sectors does not make any statement about the choice of ISPs.
    As mentioned before, each sector may even choose a different set of ISPs to complete the set of independent dot products.
    At higher loop order this will become an essential feature, since the combination of all (required) sectors will define an overcomplete (linearly dependent) set of true propagators.
    Each individual sector will still have linearly independent propagators, which need to be (individually) completed by a linearly independent set of ISPs.
    Allowing for this overcompleteness is the reason why all the integrals of a fixed SF sector fit into the same family, which they would not in the traditional definition of an integral family.
  \item\textbf{Syzygies. }
    Closely related to the previous point, we observe that syzygy generators can be derived independently of the choice of ISPs.
    They only depend on the choice of propagator labeling.
    Hence, the propagator labeling may be optimized w.r.t. the syzygies whereas the freedom in choice of ISPs can be used to optimize the (finding of) symbolic reduction rules.
    Unfortunately, the integral ordering affects both, which makes it harder to choose it in an optimal way.
\end{itemize}

\texttt{FeynRIR} provides an experimental feature, a \emph{topology library}, which implements parts of this strategy and produces the DAG of non-zero topologies from a given WEFT impulse integrand.
\subsection{Proper time integrals and cut WL propagators}\label{sec:ibp_proper_time}

As already emphasized, up until this point it is not necessary to perform the proper time integrals and introduce explicit cut WL propagators (delta functions).
It is even beneficial to never introduce them in a graph-numerator or a graph representing a topology directly: Finding symmetries, graph isomorphisms and mapping to topologies remains a tree graph isomorphism problem which can very efficiently be solved by graph canonization.
Furthermore, one does not face the problem of inconsistent choices of delta functions.
I.e., there are often many different choices for the delta functions due to identities of the form $\delta(x)\delta(y) = \delta(x)\delta(x+y)$.

Hence, we choose to introduce the delta functions as a separate layer on top of the unique integral topologies only.
Having access to a dependency DAG for sectors as constructed in the previous section there is a simple algorithm to consistently introduce delta functions in all topologies of a fixed integral family.
Most conveniently, one starts with a planar, top-level \emph{ladder} topology graph, e.g. topology A in the example of Sec.~\ref{sec:topo_example}.
For such a topology there is a unique planar choice for the cut WL propagators, making it look like an actual ladder:
\begin{equation}
  \begin{aligned}
    &\begin{tikzpicture}
      [baseline={([yshift=-0.5ex]current bounding box.center)},scale=0.9,font=\scriptsize]
      \draw[graviton] (0,0) -- (0,-1);
      \draw[graviton] (1,0) -- (1,-1);
      \draw[graviton] (2,0) -- (2,-1);
      \draw[graviton] (3,0) -- (3,-1);
      \draw[graviton] (4,0) -- (4,-1);

      \draw[graviton] (6,0) -- (6,-1);

      \draw (0,0) -- (1,0);
      \draw (1,-1) -- (2,-1);
      \draw (2,-1) -- (3,-1);
      \draw (3,0) -- (4,0);
      \draw (4,-1) -- (4.4,-1);
      \draw[dotted] (4.4,-1) -- (5,-1);
      \draw (5.6,0) -- (6,0);
      \draw[dotted] (5,0) -- (5.6,0);
      \draw[dotted] (4.7,-0.5) -- (5.3,-0.5);

      \fill (0,-1) circle (0.04);
      \fill (2,0) circle (0.04);
      \fill (6,-1) circle (0.04);
    \end{tikzpicture}\\
    &\qquad\qquad\qquad\,\,\,\Downarrow\\
    &\begin{tikzpicture}
      [baseline={([yshift=-0.5ex]current bounding box.center)},scale=0.9,font=\scriptsize]
      \draw[graviton] (0,0) -- (0,-1);
      \draw[graviton] (1,0) -- (1,-1);
      \draw[graviton] (2,0) -- (2,-1);
      \draw[graviton] (3,0) -- (3,-1);
      \draw[graviton] (4,0) -- (4,-1);

      \draw[graviton] (6,0) -- (6,-1);

      \draw (0,0) -- (1,0);
      \draw (1,-1) -- (2,-1);
      \draw (2,-1) -- (3,-1);
      \draw (3,0) -- (4,0);
      \draw (4,-1) -- (4.4,-1);
      \draw[dotted] (4.4,-1) -- (5,-1);
      \draw (5.6,0) -- (6,0);
      \draw[dotted] (5,0) -- (5.6,0);
      \draw[dotted] (4.7,-0.5) -- (5.3,-0.5);
      \draw[dashed] (0,-1) -- (1,-1);
      \draw[dashed] (1,0) -- (2,0);
      \draw[dashed] (2,0) -- (3,0);
      \draw[dashed] (3,-1) -- (4,-1);
      \draw[dashed] (4,0) -- (4.5,0);
      \draw[dashed] (5.5,-1) -- (6,-1);

      \fill (0,-1) circle (0.04);
      \fill (2,0) circle (0.04);
      \fill (6,-1) circle (0.04);
    \end{tikzpicture}\,.
  \end{aligned}
\end{equation}
Importantly, we graphically introduce and interpret the cut WL propagators in such a way that the ``energy'' $\sum_i \ell_i\cdot u_a$ is conserved at each WL vertex.

Following the DAG's edges up and down one may then -- using the node map -- introduce all cut WL propagators such that they are globally consistent in the whole family (including their label).
Note that when following up an edge of the DAG that introduces a new WL propagator, we always attach the possible delta function that was connected to the WL vertex that is being split into two such that at each end of the chain of WL propagators there is maximally one delta function attached and no delta function is attached in the middle of such a chain.
E.g.
\begin{equation}
  \begin{tikzpicture}
    [baseline={([yshift=-0.5ex]current bounding box.center)},scale=0.9,font=\scriptsize]
    \draw[graviton] (0,0) node[above] {1} -- (0,-1);
    \draw[graviton] (0,0) -- (1,-1);
    \draw[graviton] (2,0) node[above] {2} -- (2,-1);
    \draw (0,0) -- (-0.4,0);
    \draw[dotted] (-0.4,0) -- (-1,0);
    \draw (0,-1) -- (-0.4,-1);
    \draw[dotted] (-0.4,-1) -- (-1,-1);
    \draw (1,-1) -- (2,-1);
    \draw[dotted] (2,0) -- (2.6, 0);
    \draw[dotted] (2,-1) -- (2.6, -1);

    \draw[dashed] (0,0) -- (2,0);
    \draw[dashed] (0,-1) -- (2,-1);
  \end{tikzpicture}
  \overset{\textrm{extend}}{\longrightarrow}
  \begin{tikzpicture}
    [baseline={([yshift=-0.5ex]current bounding box.center)},scale=0.9,font=\scriptsize]
    \draw[graviton] (0,0) node[above] {$1'$} -- (0,-1);
    \draw[graviton] (1,0) node[above] {$1''$} -- (1,-1);
    \draw[graviton] (2,0) node[above] {2}-- (2,-1);
    \draw (0,0) -- (-0.4,0);
    \draw[dotted] (-0.4,0) -- (-1,0);
    \draw (0,-1) -- (-0.4,-1);
    \draw[dotted] (-0.4,-1) -- (-1,-1);
    \draw (1,-1) -- (2,-1);
    \draw (0,0) -- (1,0);
    \draw[dotted] (2,0) -- (2.6, 0);
    \draw[dotted] (2,-1) -- (2.6, -1);

    \draw[dashed] (1,0) -- (2,0);
    \draw[dashed] (0,-1) -- (2,-1);
  \end{tikzpicture}\,,
\end{equation}
where WL vertex 1 was split into two vertices $1'$ and $1''$.
Following this rule, the cut WL propagator between vertices 1 and 2 must connect vertex $1''$ to 2 and not $1'$ to 2 in the extended graph.

\subsubsection{Closure bases and worldline planarity}
\label{sec:closure-bases}
For each worldline $a$, let
\begin{equation}
  X_a=\{x_{a,1},\ldots,x_{a,n_a}\}
\end{equation}
denote the WL vertices carrying independent proper-time integrations.
As discussed in Sec.~\ref{sec:proper-time-integrals}, these give $n_a$ constraints.
We want to eventually choose a unimodular basis adapted to the overall momentum transfer $q$,
\begin{equation}
  \prod_{i=1}^{n_a}
  \hat\delta(P_i\cdot u_a)
  =
  \hat\delta(q\cdot u_a)
  \prod_{\alpha=1}^{n_a-1}
  \hat\delta(\ell_\alpha\cdot u_a),
  \label{eq:closure-external-split}
\end{equation}
where $P_i$ is the net momentum flow at $x_{a,i}$.
The first factor is the ``external'' (Fourier) delta function, while the remaining $n_a-1$ factors are the delta function constraints for each internal loop integration.

\begin{figure}[h!]
  \centering
  \begin{tikzpicture}[
    font=\scriptsize,
    fourier/.style={circle, draw=black, fill=white, inner sep=0pt, minimum size=7pt},
  ]
    \begin{scope}[]
      \fill (0,0) circle (0.04);
      \fill (1,0) circle (0.04);
      \fill (2,0) circle (0.04);
      \fill (3,0) circle (0.04);

      \node[fourier] (Fa) at (1.5,1.2) {};

      \draw[dashed] (0,0) node[below] {$x_{a,1}$} -- (Fa) node[above=5pt] {$F_a$};
      \draw[dashed] (1,0) node[below] {$x_{a,2}$} -- (Fa);
      \draw[dashed] (2,0) node[below] {$x_{a,3}$} -- (Fa);
      \draw[dashed] (3,0) node[below] {$x_{a,4}$} -- (Fa);

      \node at (1.5,-0.8) {(a)};
    \end{scope}

    \begin{scope}[xshift=4.5cm]
      \fill (0,0) circle (0.04);
      \fill (1,0) circle (0.04);
      \fill (2,0) circle (0.04);
      \fill (3,0) circle (0.04);

      \node[fourier] (Fa) at (1.5,1.2) {};

      \draw[dashed,gray] (0,0) node[below,black] {$x_{a,1}$} -- (Fa) node[above=5pt,black] {$F_a$};
      \draw[dashed,gray] (1,0) node[below,black] {$x_{a,2}$} -- (Fa);
      \draw[dashed] (0,0) -- (1,0);
      \draw[dashed] (2,0) node[below] {$x_{a,3}$} -- (Fa);
      \draw[dashed] (3,0) node[below] {$x_{a,4}$} -- (Fa);

      \node at (1.5,-0.8) {(b)};
    \end{scope}

    \begin{scope}[xshift=9cm]
      \fill (0,0) circle (0.04);
      \fill (1,0) circle (0.04);
      \fill (2,0) circle (0.04);
      \fill (3,0) circle (0.04);

      \node[fourier] (Fa) at (1.5,1.2) {};

      \draw[dashed,gray] (0,0) node[below left,black] {$x_{a,1}$} -- (Fa) node[above=5pt,black] {$F_a$};
      \draw[dashed] (1,0) node[below] {$x_{a,2}$} -- (Fa);
      \draw[dashed] (0,0) .. controls (0.5,-0.5) and (1.5,-0.5) .. (2,0);
      \draw[dashed,gray] (2,0) node[below right,black] {$x_{a,3}$} -- (Fa);
      \draw[dashed] (3,0) node[below] {$x_{a,4}$} -- (Fa);

      \node at (1.5,-0.8) {(c)};
    \end{scope}

  \end{tikzpicture}

  \caption{
  (a) Tent representation of the proper-time constraints for four WL integration vertices, ending at the external Fourier vertex $F_a$.
  (b) and (c) Combination of two spokes through the relative directions $c_{12}$ and $c_{13}$.
  The gray dashed lines represent the two possible choices for the remaining connection to the Fourier node that together with the relative direction close the cycle.
  }
  \label{fig:closure-tour}
\end{figure}
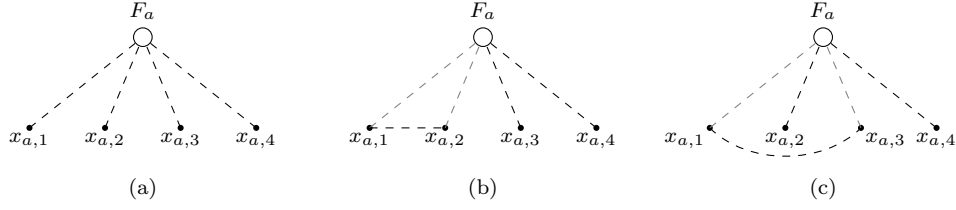

Figure~\ref{fig:closure-tour}(a) gives a graphical representation of this decomposition.
The original vertex delta function constraints are collected at an auxiliary \emph{Fourier} vertex $F_a$, producing a ``tent'' diagram.
$F_a$ is only a bookkeeping device and is not part of the original tree.
Combining two spokes eliminates their common Fourier direction, and the resulting relative direction can be represented by a direct closure edge
\begin{equation}
  c_{ij}=(x_{a,i},x_{a,j})\,.
\end{equation}
Adding $c_{ij}$ to the original tree $T$ closes this direction into a cycle.
Figure~\ref{fig:closure-tour}(b) and (c) shows the examples $c_{12}$ and $c_{13}$.

\paragraph{Equivalent closure bases.}

A complete closure basis of the type in eq.~\eqref{eq:closure-external-split} consists of $n_a-1$ independent edges, which form a spanning tree on $X_a$ \cite{Biggs1993}.\footnote{
More formally, the relative closure directions form a basis of $H_1(T,X_a;\mathbb Z) \simeq \widetilde H_0(X_a;\mathbb Z) \simeq \mathbb Z^{n_a-1}$.
The corresponding dual basis belongs to $H^1(T,X_a;\mathbb Z)$; see \cite{Hatcher2002}.
}
Different spanning trees therefore give different bases of the same $(n_a-1)$-dimensional space of relative constraints, while we enforce the external $q$-constraint to be unchanged.

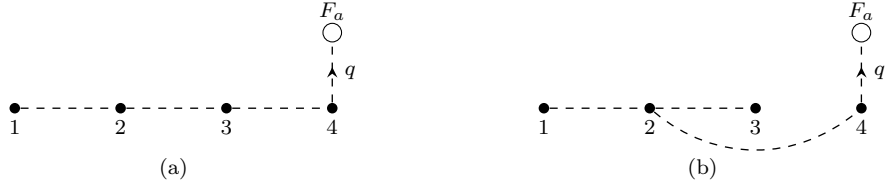
\begin{figure}[h!]
  \centering
  \begin{tikzpicture}[
    font=\scriptsize,
    augWL/.style={circle, fill=black, inner sep=1.4pt},
    fourier/.style={circle, draw=black, fill=white,
      inner sep=0pt, minimum size=7pt},
    closure/.style={line width=0.5pt, draw=black, dashed},
    external/.style={line width=0.5pt, draw=black, dashed}
  ]

    \begin{scope}[xshift=-3.5cm]
      \node[augWL] (a1) at (0,0) {};
      \node[augWL] (a2) at (1.4,0) {};
      \node[augWL] (a3) at (2.8,0) {};
      \node[augWL] (a4) at (4.2,0) {};

      \node[below=3pt] at (a1) {$1$};
      \node[below=3pt] at (a2) {$2$};
      \node[below=3pt] at (a3) {$3$};
      \node[below=3pt] at (a4) {$4$};

      \draw[closure] (a1) -- (a2);
      \draw[closure] (a2) -- (a3);
      \draw[closure] (a3) -- (a4);

      \node[fourier] (Fa) at ($(a4)+(0,1.0)$) {};
      \node[above=3pt] at (Fa) {$F_a$};
      \draw[external,quark] (a4) -- node[right=3pt] {$q$} (Fa);

      \node at (2.1,-0.8) {(a)};
    \end{scope}

    \begin{scope}[xshift=3.5cm]
      \node[augWL] (b1) at (0,0) {};
      \node[augWL] (b2) at (1.4,0) {};
      \node[augWL] (b3) at (2.8,0) {};
      \node[augWL] (b4) at (4.2,0) {};

      \node[below=3pt] at (b1) {$1$};
      \node[below=3pt] at (b2) {$2$};
      \node[below=3pt] at (b3) {$3$};
      \node[below=3pt] at (b4) {$4$};

      \draw[closure] (b1) -- (b2);
      \draw[closure] (b2) -- (b3);
      \draw[closure] (b2) to[out=-40,in=-140] (b4);

      \node[fourier] (Fb) at ($(b4)+(0,1.0)$) {};
      \node[above=3pt] at (Fb) {$F_a$};
      \draw[external,quark] (b4) -- node[right=3pt] {$q$} (Fb);

      \node at (2.1,-0.8) {(b)};
    \end{scope}

  \end{tikzpicture}

  \caption{
  Two closure bases for four (ordered) WL integration vertices: the serialized basis in (a) and an alternative spanning-tree basis in (b).
  The cut edge ending at the external Fourier vertex $F_a$ represents the same external $q$-constraint in both cases.
  }
  \label{fig:closure-realisations}
\end{figure}

For four ordered WL integration vertices, two examples are
\begin{equation}
  C_a^{\mathrm{ser}}
  =
  \{c_{12},c_{23},c_{34}\},
  \qquad
  C_a'
  =
  \{c_{12},c_{23},c_{24}\}\,.
\end{equation}
They are shown in Fig.~\ref{fig:closure-realisations}.

For the two closure bases above, the corresponding bases of relative closure directions are related by the unimodular transformation
\begin{equation}
  U=
  \begin{pmatrix}
    1&0&0\\
    0&1&0\\
    0&1&1
  \end{pmatrix}.
\end{equation}
The dual momentum variables therefore transform as
\begin{equation}
  \boldsymbol{\ell}'
  =(U^{-1})^T\boldsymbol{\ell}
  =
  \begin{pmatrix}
    1&0&0\\
    0&1&-1\\
    0&0&1
  \end{pmatrix}
  \begin{pmatrix}
    \ell_1\\
    \ell_2\\
    \ell_3
  \end{pmatrix}.
\end{equation}
More generally, changes between spanning-tree closure bases are described by
\begin{equation}
  U\in GL(n_a-1,\mathbb Z).
\end{equation}
Since $|\det U|=1$,
\begin{equation}
  \prod_{i=1}^{n_a-1}\hat\delta(z_i')
  =
  \prod_{i=1}^{n_a-1}\hat\delta(z_i),
  \qquad
  z_i=u_a\cdot\ell_i,
\end{equation}
so all spanning-tree bases impose the same relative proper-time constraints.
The serialized closure may therefore be chosen without loss of generality.
\paragraph{Closure choice and IBP reduction.}

Although equivalent at the level of the proper-time constraints, different closure bases can lead to IBP systems of different complexity for a fixed momentum assignment.

For the two choices in Fig.~\ref{fig:closure-realisations}, suppressing the common first constraint and overall factors of $2\pi$, let
\begin{equation}
  x=u_a\cdot\ell_2,
  \qquad
  y=u_a\cdot\ell_3.
\end{equation}
The remaining constraints are
\begin{equation}
  \Delta_{\mathrm{ser}}=\delta(x)\delta(y),
  \qquad
  \Delta'=\delta(x-y)\delta(y).
\end{equation}
Their individual factors have different momentum dependence:
\begin{align}
  \partial_y\Delta_{\mathrm{ser}}
  &=\delta(x)\delta'(y),\\
  \partial_y\Delta'
  &=-\delta'(x-y)\delta(y)
    +\delta(x-y)\delta'(y).
\end{align}
Thus only one cut depends on $y$ in the serialized representation, whereas both contribute in the alternative basis.
The additional terms are redundant but need not cancel manifestly in an IBP reduction, potentially enlarging the intermediate system.
The serialized closure therefore provides a convenient canonical choice for IBP reduction.
\paragraph{Multiple worldlines.}

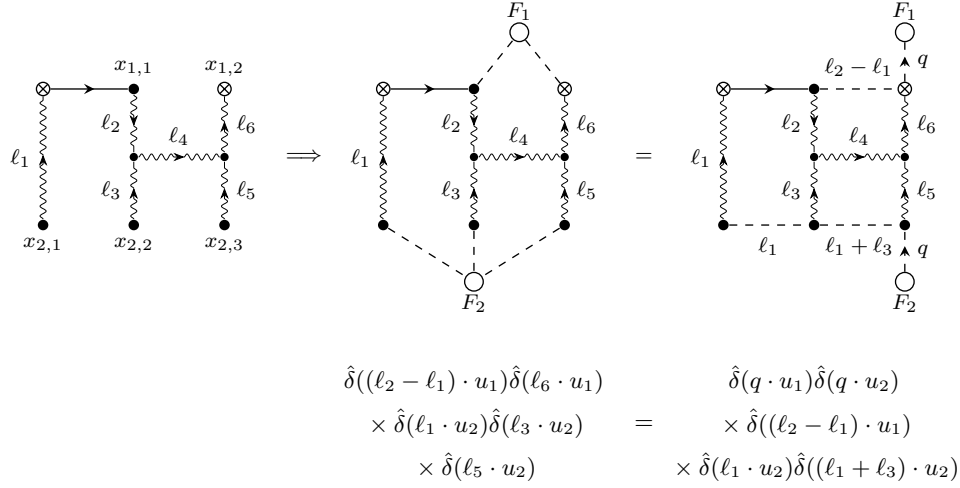
\begin{figure}[h!]
  \centering
  \begin{tikzpicture}[
    font=\scriptsize,
    line width=0.5pt,
    wl/.style={circle, fill=black, inner sep=1.4pt},
    bulk/.style={circle, fill=black, inner sep=1.1pt},
    fourier/.style={circle, draw=black, fill=white,
      inner sep=0pt, minimum size=7pt},
    cut/.style={dashed, line width=0.5pt}
  ]

    \foreach \panel/\shift in {0/-4.5cm,1/0cm,2/4.5cm} {
      \begin{scope}[xshift=\shift]
        \node[xSource, fill=white] (\panel-I) at (-1.2,0.9) {};
        \node[wl] (\panel-A) at (0,0.9) {};
        \node[xSource, fill=white] (\panel-B) at (1.2,0.9) {};

        \node[wl] (\panel-C) at (-1.2,-0.9) {};
        \node[wl] (\panel-D) at (0,-0.9) {};
        \node[wl] (\panel-E) at (1.2,-0.9) {};

        \node[bulk] (\panel-V) at (0,0) {};
        \node[bulk] (\panel-W) at (1.2,0) {};

        \draw[quark] (\panel-I) -- (\panel-A);

        \draw[directed graviton] (\panel-C) --
          node[left=3pt] {$\ell_1$} (\panel-I);

        \draw[directed graviton] (\panel-A) --
          node[left=3pt] {$\ell_2$} (\panel-V);

        \draw[directed graviton] (\panel-D) --
          node[left=3pt] {$\ell_3$} (\panel-V);

        \draw[directed graviton] (\panel-V) --
          node[above=3pt] {$\ell_4$} (\panel-W);

        \draw[directed graviton] (\panel-E) --
          node[right=3pt] {$\ell_5$} (\panel-W);

        \draw[directed graviton] (\panel-W) --
          node[right=3pt] {$\ell_6$} (\panel-B);

        \ifnum\panel=0\relax
          \node[above=3pt] at (\panel-A) {$x_{1,1}$};
          \node[above=3pt] at (\panel-B) {$x_{1,2}$};

          \node[below=3pt] at (\panel-C) {$x_{2,1}$};
          \node[below=3pt] at (\panel-D) {$x_{2,2}$};
          \node[below=3pt] at (\panel-E) {$x_{2,3}$};
        \fi

        \ifnum\panel=1\relax
          \node[fourier] (f1) at (0.6,1.65) {};
          \node[fourier] (f2) at (0,-1.65) {};
          \node[above=3pt] at (f1) {$F_1$};
          \node[below=3pt] at (f2) {$F_2$};

          \draw[cut] (\panel-A) -- (f1);
          \draw[cut] (\panel-B) -- (f1);

          \draw[cut] (\panel-C) -- (f2);
          \draw[cut] (\panel-D) -- (f2);
          \draw[cut] (\panel-E) -- (f2);

          \node[anchor=north, font=\footnotesize, inner sep=0pt] (\panel-deltas) at (0,-2.75)
          {$\begin{gathered}
            \hat\delta((\ell_2-\ell_1)\cdot u_1)
            \hat\delta(\ell_6\cdot u_1)
            \\
            {}\times
            \hat\delta(\ell_1\cdot u_2)
            \hat\delta(\ell_3\cdot u_2)
            \\
            {}\times
            \hat\delta(\ell_5\cdot u_2)
          \end{gathered}$};
        \fi

        \ifnum\panel=2\relax
          \draw[cut] (\panel-A) --
            node[above=3pt] {$\ell_2-\ell_1$}
            (\panel-B);

          \draw[cut] (\panel-C) --
            node[below=3pt] {$\ell_1$}
            (\panel-D);

          \draw[cut] (\panel-D) --
            node[below=3pt] {$\ell_1+\ell_3$}
            (\panel-E);

          \node[fourier] (f1ser) at (1.2,1.65) {};
          \node[fourier] (f2ser) at (1.2,-1.65) {};
          \node[above=3pt] at (f1ser) {$F_1$};
          \node[below=3pt] at (f2ser) {$F_2$};
          \draw[cut,quark] (\panel-B) --
            node[right=3pt] {$q$} (f1ser);
          \draw[cut,quark] (f2ser) --
            node[right=3pt] {$q$} (\panel-E);

          \node[anchor=north, font=\footnotesize, inner sep=0pt] (\panel-deltas) at (0,-2.75)
          {$\begin{gathered}
            \hat\delta(q\cdot u_1)
            \hat\delta(q\cdot u_2)
            \\
            {}\times
            \hat\delta((\ell_2-\ell_1)\cdot u_1)
            \\
            {}\times
            \hat\delta(\ell_1\cdot u_2)
            \hat\delta((\ell_1+\ell_3)\cdot u_2)
          \end{gathered}$};
        \fi
      \end{scope}
    }

    \node at (-2.25,0) {$\Longrightarrow$};
    \node at (2.25,0) {$=$};

    \node[font=\footnotesize]
      at ($(1-deltas.center)!0.5!(2-deltas.center)$) {$=$};

  \end{tikzpicture}

  \caption{
  A 4PM tree with one WL iteration (left), the direct representation of its proper-time constraints with tent-like structure(middle), and an equivalent serialized representation (right).
  The diagrams end at the external Fourier vertices $F_1$ and $F_2$.
  In the serialized graph, the cuts attached to these vertices represent $\hat\delta(q\cdot u_1)$ and $\hat\delta(q\cdot u_2)$, respectively.
  }
  \label{fig:proper-time-split}
\end{figure}

The construction applies independently to each physical worldline.
For $a=1,\ldots,r$, the serialized choice is
\begin{equation}
  S_{\mathrm{ser}}
  =
  \bigsqcup_{a=1}^{r}\mathsf P_{n_a},
\end{equation}
where $\mathsf P_{n_a}$ is the path graph on the $n_a$ WL integration vertices.
Together with the uncut WL propagators, this arranges each physical worldline into a single uninterrupted string, reminiscent of the classical limit of $2\to2$ scattering amplitudes \cite{Mogull2021}.
Figure~\ref{fig:proper-time-split} illustrates the construction: relative constraints become internal cuts, while the overall constraints are represented by cuts ending at the external Fourier vertices $F_a$.
\paragraph{Worldline planarity and graviton non-planarity.}

The serialized closure guarantees a planar realization of the worldline subgraph, but not necessarily of the complete topology.
In the canonical representation, the two WL strings are drawn as the upper and lower boundaries with their vertex orderings fixed.
We define the non-planarity level as the minimum number of graviton-edge crossings among such drawings, including both relative orientations of the WL strings.
Level zero therefore corresponds to a planar drawing compatible with the prescribed WL ordering.

\begin{figure}[h!]
  \centering

  \begin{subfigure}[t]{0.47\textwidth}
    \centering
    \begin{tikzpicture}[
      scale=0.75,
      augWL/.style={
        circle,
        fill=black,
        inner sep=1.4pt
      },
      gravV/.style={
        circle,
        fill=black,
        inner sep=1.2pt
      }
    ]

      \node[augWL] (a1) at (-2.0, 1.25) {};
      \node[gravV] (v1) at (-2.0, 0.00) {};
      \node[augWL] (b1) at (-2.0,-1.25) {};

      \node[augWL] (a2) at (-1.0, 1.25) {};
      \node[gravV] (v2) at (-1.0, 0.00) {};
      \node[augWL] (b2) at (-1.0,-1.25) {};

      \node[gravV] (v4) at ( 0.0, 0.00) {};

      \node[xSource] (a3) at ( 1.0, 1.25) {};
      \node[gravV] (v3) at ( 1.0, 0.00) {};
      \node[augWL] (b3) at ( 1.0,-1.25) {};

      \draw[graviton] (a1) -- (v1) -- (b1);
      \draw[graviton] (a2) -- (v2) -- (b2);
      \draw[graviton] (a3) -- (v3) -- (b3);

      \draw[graviton] (v1) to[out=30,in=150] (v4);
      \draw[graviton] (v2) -- (v4);
      \draw[graviton] (v3) -- (v4);

      \node[above=2pt] at (a1) {$1$};
      \node[above=2pt] at (a2) {$1$};
      \node[above=2pt] at (a3) {$1$};

      \node[below=2pt] at (b1) {$2$};
      \node[below=2pt] at (b2) {$2$};
      \node[below=2pt] at (b3) {$2$};

    \end{tikzpicture}
    \caption{A tree example with two physical worldlines before introducing the closure edges.}
    \label{fig:aligned-tree-example}
  \end{subfigure}
  \hfill
  \begin{subfigure}[t]{0.47\textwidth}
      \centering
      \begin{tikzpicture}[
      scale=0.75,
      augWL/.style={
        circle,
        fill=black,
        inner sep=1.4pt
      },
      closure/.style={
        line width=0.5pt,
        draw=black,
        dashed
      },
      gravV/.style={
        circle,
        fill=black,
        inner sep=1.2pt
      }
    ]

      \node[augWL] (a1) at (-2.0, 1.25) {};
      \node[gravV] (v1) at (-2.0, 0.00) {};
      \node[augWL] (b1) at (-2.0,-1.25) {};

      \node[augWL] (a2) at (-1.0, 1.25) {};
      \node[gravV] (v2) at (-1.0, 0.00) {};
      \node[augWL] (b2) at (-1.0,-1.25) {};

      \node[gravV] (v4) at ( 0.0, 0.00) {};

      \node[xSource] (a3) at ( 1.0, 1.25) {};
      \node[gravV] (v3) at ( 1.0, 0.00) {};
      \node[augWL] (b3) at ( 1.0,-1.25) {};

      \draw[graviton] (a1) -- (v1) -- (b1);
      \draw[graviton] (a2) -- (v2) -- (b2);
      \draw[graviton] (a3) -- (v3) -- (b3);
      \draw[closure] (a1) -- (a2);
      \draw[closure] (a2) -- (a3);

      \draw[closure] (b1) -- (b2);
      \draw[closure] (b2) -- (b3);

      \draw[graviton] (v1) to[out=30,in=150] (v4);
      \draw[graviton] (v2) -- (v4);
      \draw[graviton] (v3) -- (v4);

      \node[above=2pt] at (a1) {$1$};
      \node[above=2pt] at (a2) {$1$};
      \node[above=2pt] at (a3) {$1$};

      \node[below=2pt] at (b1) {$2$};
      \node[below=2pt] at (b2) {$2$};
      \node[below=2pt] at (b3) {$2$};

    \end{tikzpicture}
      \caption{The same graph after introducing the serialized closure edges on both worldlines.}
      \label{fig:aligned-example-closure}
    \end{subfigure}

  \caption{
  A 5PM example illustrating residual non-planarity.
  The left panel shows a tree graph with three graviton attachments on each worldline before introducing the closure edges.
  The right panel shows the same graph after choosing the serialized closure on both worldlines.
  Although the worldline subgraph is planar, the complete graph remains non-planar due to the graviton connectivity.}
  \label{fig:nonplanar-example}
\end{figure}
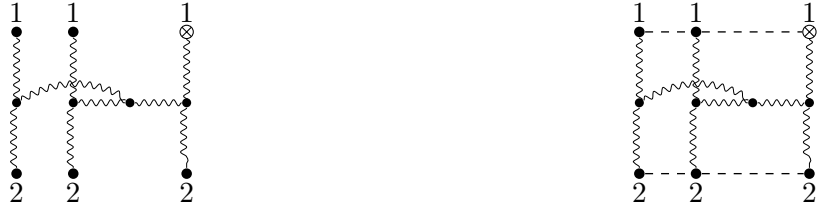

Any remaining obstruction arises from the graviton connectivity.
A 5PM example is shown in Fig.~\ref{fig:nonplanar-example}; related topologies appear in~\cite{Driesse:2026qiz}.

The obstruction is independent of the closure basis: contracting each worldline closure tree produces a $K_{3,3}$ minor, so no choice of closure basis can render the complete graph planar.

These conventions complete the canonical graphical representation of the integral topologies used in the implementation described below.
\section{\texttt{FeynRIR}: Feynman Rule Integrands in \texttt{Rust}}\label{sec:feynrir}

This publication provides an open source \texttt{Rust} library, \texttt{FeynRIR}, that implements the algorithms and integrand representations presented in this paper.
The \texttt{Git} repository is hosted under \url{https://gitlab.com/kaelingre/feynrir}.
We furthermore provide a \texttt{Git} repository with precomputed integrands for the variation of the effective action and the impulse $\Delta p_1^\mu$ under \url{https://gitlab.com/kaelingre/weft_integrands}.
We would like to encourage the community to use the \texttt{GitLab} website to submit bug reports, feature requests, merge requests, and to join discussions on the implementation.

\subsection{The \texttt{Symbolica} ecosystem}\label{sec:symbolica}

\paragraph{\texttt{Symbolica}, \texttt{Graphica}, \texttt{Numerica}.}
\texttt{FeynRIR} is built on top of the \texttt{Symbolica} ecosystem, a set of \texttt{Rust} crates designed for fast symbolic and numeric computer algebra.
The ecosystem consists of three main components:

\begin{itemize}
  \item \texttt{Symbolica} is a fast, memory-efficient computer algebra system implemented in \texttt{Rust}.
    It provides core symbolic-manipulation routines with performance competitive with, and often superior to, established systems like \texttt{FORM}, \texttt{Mathematica}, \texttt{FLINT}, or \texttt{Fermat} for the types of symbolic workloads encountered in Feynman-integrand construction.
    See e.g. \cite{Smirnov:2023yhb} for benchmark comparisons in a related high-energy physics context.
  \item \texttt{Graphica} is an open source graph library used throughout the \texttt{FeynRIR} workflow to represent Feynman diagrams and topologies.
    It provides a canonicalization algorithm for mixed-directed, colored (edges \& nodes) graphs, which is essential for WEFT graphs.
  \item \texttt{Numerica} is a community-driven high-performance numerical library that accompanies \texttt{Symbolica}.
    It provides numerical high-precision (exact and floating-point) arithmetics, finite-field capabilities, matrix operations, and adaptive Monte Carlo integration.
\end{itemize}

The choice to implement \texttt{FeynRIR} in \texttt{Rust} using \texttt{Symbolica} was motivated by the memory and performance characteristics of the integrand construction problem, in particular the challenge posed by the tensor contraction of huge algebraic expressions.
At higher PM orders the number of intermediate algebraic expressions grows to millions of terms, and the total symbolic footprint can exceed hundreds of gigabytes.
The \texttt{Rust} ecosystem provides a memory-safe, zero-overhead abstraction layer with excellent thread safety guarantees, allowing straightforward parallelization through the \texttt{rayon} library.
Together with \texttt{Symbolica}'s flat-expression representation and canonical forms, these features should help \texttt{FeynRIR} extend the range of higher-PM integrands that can be constructed efficiently.

\paragraph{Licensing.}
Starting with \texttt{Symbolica} version~3, package authors who hold a \texttt{Symbolica} license can obtain a special library license covering the use of \texttt{Symbolica} within their package.
\texttt{FeynRIR} uses this mechanism, allowing users to call any function provided by the package, including multi-threaded computations, without obtaining a separate \texttt{Symbolica} license.
Users need their own \texttt{Symbolica} license only for parallel \texttt{Symbolica} operations outside the package's licensed functionality, such as directly calling \texttt{Symbolica} functions or manipulating \texttt{Symbolica} expressions in their own code.

\subsection{Obtaining and compiling \texttt{FeynRIR}}
\texttt{FeynRIR} is available on \href{https://crates.io/crates/feynrir}{crates.io} as \texttt{feynrir} and requires \texttt{Rust} and its package manager \href{https://github.com/rust-lang/cargo}{\texttt{Cargo}}.
To add the library to an existing \texttt{Rust} project, run the following command in a terminal from that project's directory:
\begin{terminal}
$ cargo add feynrir
\end{terminal}
Alternatively, clone the \texttt{Git} repository and build the library and its example executables locally:
\begin{terminal}
$ git clone https://gitlab.com/kaelingre/feynrir.git
$ cd feynrir
$ cargo build --release
\end{terminal}
\texttt{Cargo} automatically downloads and builds the dependencies.
The \href{https://docs.rs/feynrir/latest/feynrir/}{API documentation on docs.rs} describes the library's functions and types.
It can also be built locally and opened in a browser by running
\begin{terminal}
$ cargo doc --open
\end{terminal}

\paragraph{Example programs.}
The repository includes two programs illustrating the calculations described in this paper:
\begin{itemize}
  \item \texttt{var\_eff\_action} constructs the integrands for the variation of the effective action, which determines the force on a worldline.
  \item \texttt{impulse} includes the corrections to the trajectories and constructs the integrands for the momentum change $\Delta p_1^\mu$.
    It exports both graph--numerator pairs and purely algebraic expressions.
\end{itemize}
Run either program from the cloned repository with the corresponding terminal command:
\begin{terminal}
$ cargo run --release --bin var_eff_action
$ cargo run --release --bin impulse
\end{terminal}
Both examples compute the 1--4PM orders by default and write their results below \texttt{/tmp}.
The maximum PM order and output directory can be changed near the start of the respective source file in \texttt{src/bin}.

\paragraph{\texttt{Mathematica} paclet for graph drawing.}
\texttt{FeynRIR} ships with a \texttt{Mathematica} paclet for visualizing graphs.
The current version 1.0 of the paclet can be installed from the cloud by evaluating the following input cell in \texttt{Mathematica}:
\begin{mathematicacell}
PacletInstall[CloudObject[
  "https://www.wolframcloud.com/obj/kaelingre/FeynRIR-1.0.paclet"
]];
\end{mathematicacell}
The paclet can afterward be loaded via
\begin{mathematicacell}
Needs["feynrir`"];
\end{mathematicacell}
The welcome message contains a link to the documentation with many examples.
The paclet is also included in the \texttt{Git} repository of \texttt{FeynRIR} and can be built locally by running the appropriate sections in the \texttt{mma/build.wl} file.

\subsection{Representation of integrands}\label{sec:integrand_repr}

\texttt{FeynRIR} exports integrands by default in a graph-numerator pair format.
The numerator is represented as an algebraic expression compatible with the \texttt{Mathematica} format.
The numerator contains the contracted tensor structures, while the graph specifies the propagators, integration measures and exponential factors, as described in Secs.~\ref{sec:construction} and~\ref{sec:iterations}.
Table~\ref{tab:integrand_symbols} summarizes the notation for the symbols and functions appearing in the integrands.

\begin{table}[htbp]
\centering
\small
\renewcommand{\arraystretch}{1.2}
\begin{tabular}{@{}p{0.30\textwidth}p{0.66\textwidth}@{}}
\toprule
Exported symbol & Meaning \\
\midrule
\texttt{p[i(,mu)]} & $D$-dimensional momentum vector $p_i^\mu$ associated to a graviton propagator with edge ID $i+1$ in the exported graph.
The Lorentz index is optional.\\
\texttt{l[i(,mu)]} & $D$-dimensional loop momentum vector $\ell_i^\mu$, $i=0,\dots,L$, including the momentum exchange $q$ as a superposition thereof.
The Lorentz index is optional.\\
\texttt{u[a(,mu)]} & Incoming velocity $u_a^\mu$ of body $a=1,2$, with $u_a^2=1$.
The Lorentz index is optional.\\
\texttt{g}, \texttt{d} & Relative Lorentz factor $\gamma=u_1\cdot u_2$ and spacetime dimension $D=4-2\epsilon$. \\
\texttt{mu} & The single free Lorentz index of the variation or impulse. \\
\texttt{dot[v,w]} & Minkowski scalar product $v\cdot w$. \\
\texttt{x[a,t,d(,mu)]} & Worldline symbol with $d$ proper time derivatives: $d=0$ denotes $i x_a^\mu(\tau)$, $d=1$ denotes $\dot x_a^\mu(\tau)$, and $d=2$ denotes $i\ddot x_a^\mu(\tau)$.
These conventions keep the numerator free of explicit imaginary units, as explained in Sec.~\ref{sec:iteration_insertion}.
The proper time argument $\tau$ is an integer number $i$ corresponding to the vertex index $i+1$ of the corresponding WL source or sink vertex in the corresponding Feynman diagram.
The Lorentz index is optional.\\
\texttt{flagtd} & Marker for total derivatives at the final sink: Set to 1 for the equations of motion and to 0 if integrating over the final proper time to obtain an integrated observable. \\
\bottomrule
\end{tabular}
\caption{Symbols and functions appearing in the numerators.}
\label{tab:integrand_symbols}
\end{table}

\paragraph{Interpreting the graphs.}
The graph--numerator output contains tree graphs, with the proper-time integrations still encoded in the WL nodes.
All tensor and velocity factors are already contained in the numerator.
A WL node can have several incoming graviton legs when iteration sinks are fused into it during the insertion of accelerations.
For a node $j$ on body $a$, let $h_j$ count all directly attached graviton legs and $K_j$ denote their net outgoing momentum.
A source has one outgoing graviton and $h_j-1$ incoming gravitons, whereas a sink has $h_j$ incoming gravitons and no outgoing graviton.
The corresponding graph factors are
\begin{align}\label{eq:export_wl_nodes}
 \begin{tikzpicture}
   [baseline={([yshift=-0.5ex]current bounding box.center)},font=\scriptsize]
   \coordinate (A) at (0,0);
   \fill (A) circle (0.05);
   \node[above=0.1] at (A) {$a,j$};
   \draw[directed graviton] (A) -- (-0.9,-1) node[below] {$p_0$};
   \draw[directed graviton] (0,-1) node[below] {$p_1$} -- (A);
   \draw[directed graviton] (0.9,-1) node[below] {$p_{h_j-1}$} -- (A);
   \node at (0.45,-0.85) {$\cdots$};
 \end{tikzpicture}
 &= i m_a\kappa^{h_j}\int_{-\infty}^{\infty}\dd\tau_j\,
 e^{iK_j\cdot x_a(\tau_j)},
 \\[1ex]
 \begin{tikzpicture}
   [baseline={([yshift=-0.5ex]current bounding box.center)},font=\scriptsize]
   \node[xSource] (A) at (0,0) {};
   \node[above=0.1] at (A) {$a,j$};
   \draw[directed graviton] (-0.9,-1) node[below] {$p_1$} -- (A);
   \draw[directed graviton] (0.9,-1) node[below] {$p_{h_j}$} -- (A);
   \node at (0,-0.85) {$\cdots$};
 \end{tikzpicture}
 &= i m_a\kappa^{h_j}\,e^{iK_j\cdot x_a(\tau_j)}.
 \label{eq:export_wl_sink}
\end{align}
Thus $K_j=p_0-\sum_{s=1}^{h_j-1}p_s$ for the source and $K_j=-\sum_{s=1}^{h_j}p_s$ for the sink, consistent with the general definition
\begin{equation}\label{eq:export_node_momentum}
 K_j=\sum_{\substack{e\text{ graviton}\\e\text{ outgoing at }j}}p_e
     -\sum_{\substack{e\text{ graviton}\\e\text{ incoming at }j}}p_e.
\end{equation}
Each attached graviton supplies one power of $\kappa$; $h_j=1$ recovers the ordinary one-graviton rules, with an empty incoming sum for the source.
The final sink remains at fixed proper time $\tau_j$ for the variation of the effective action.
To obtain the impulse, one also integrates the final sink over $\int_{-\infty}^{\infty}\dd\tau_j$.
The graviton propagator includes its momentum integration, and a graviton vertex of valence $r$ imposes momentum conservation:
\begin{align}
 \begin{tikzpicture}
   [baseline={([yshift=-0.5ex]current bounding box.center)},font=\scriptsize]
   \draw[directed graviton] (0,0) -- node[above] {$p$} (1,0);
 \end{tikzpicture}
 &=\int_p iR(p)
 \equiv\int\frac{\dd^D p}{(2\pi)^D}\,
 \frac{i}{(p^0+i0)^2-\mathbf p^2},
 \label{eq:export_fourier}
 \\[1ex]
 \begin{tikzpicture}
   [baseline={([yshift=-0.5ex]current bounding box.center)},font=\scriptsize]
   \coordinate (A) at (0,0);
   \fill (A) circle (0.05);
   \draw[directed graviton] (A) -- (0:0.7) node[right] {$p_{e_1}$};
   \draw[directed graviton] (45:0.7) node[above right] {$p_{e_2}$} -- (A);
   \draw[directed graviton] (90:0.7) node[above] {$p_{e_3}$} -- (A);
   \draw[directed graviton] (135:0.7) node[above left] {$p_{e_4}$} -- (A);
   \fill (180:0.5) circle (0.02);
   \fill (220:0.5) circle (0.02);
   \fill (260:0.5) circle (0.02);
   \fill (300:0.5) circle (0.02);
 \end{tikzpicture}
 &=i\kappa^{r-2}(2\pi)^D\delta^{(D)}\!\left(p_{e_1}-\sum_{s=2}^{r}p_{e_s}\right).
 \label{eq:export_graviton_vertex}
\end{align}
Here $e_1$ labels the outgoing graviton edge and $e_2,\ldots,e_r$ the incoming edges.
Together with the numerator and the inverse symmetry factor, these give minus the variation in eq.~\eqref{eq:varSeff}.

After inserting iterations, the numerator depends only on momenta and incoming velocities.
The phases in eqs.~\eqref{eq:export_wl_nodes} and \eqref{eq:export_wl_sink} are then evaluated on the straight worldlines $x_a^{(0)}(\tau)=b_a+u_a\tau$.
A WL propagator of power $r$ on body $a$ contributes
\begin{equation}\label{eq:export_wl_momentum}
 \begin{tikzpicture}
   [baseline={([yshift=-0.5ex]current bounding box.center)},font=\scriptsize]
   \draw[quark] (0,0) -- node[above] {$k\cdot u_a$} node[below] {$r$} (1,0);
 \end{tikzpicture}
 =\frac{i}{m_a(k\cdot u_a+i0)^r},
\end{equation}
where $k$ is the total momentum entering its initial sink through graviton propagators and earlier worldline iterations.

Collecting the phases of all WL nodes sharing a proper time $\tau_j$ gives a net outgoing momentum $P_j$, equal to the sum of their $K_j$ contributions.
Unlike $K_j$, which only contains the directly attached graviton momenta, $P_j$ also includes the momenta inherited through incoming WL iterations.
Each source integration, and the final sink integration for the impulse, therefore yields
\begin{equation}\label{eq:export_proper_time}
 \int_{-\infty}^{\infty}\dd\tau_j\,
 e^{iP_j\cdot(b_a+u_a\tau_j)}
 =e^{iP_j\cdot b_a}\hat\delta(P_j\cdot u_a),
 \qquad \hat\delta(z)=2\pi\delta(z).
\end{equation}
The resulting delta functions can be represented by cut WL propagators, closing the exported trees into loop graphs.

The external Fourier momentum $q$ is defined by writing the combined impact-parameter dependence as $e^{-iq\cdot b}$, with $b=b_1-b_2$.
If $V_a$ denotes all WL source and sink vertices belonging to body $a$, momentum conservation at the graviton vertices gives
\begin{equation}\label{eq:export_transfer_momentum}
 \begin{aligned}
 q&=-\sum_{j\in V_1}K_j=\sum_{j\in V_2}K_j,
 \\
 e^{ib_1\cdot\sum_{j\in V_1}K_j+ib_2\cdot\sum_{j\in V_2}K_j}
 &=e^{-iq\cdot b}.
 \end{aligned}
\end{equation}
Thus $q$ is the net graviton momentum flowing into all WL1 vertices, equivalently out of all WL2 vertices.

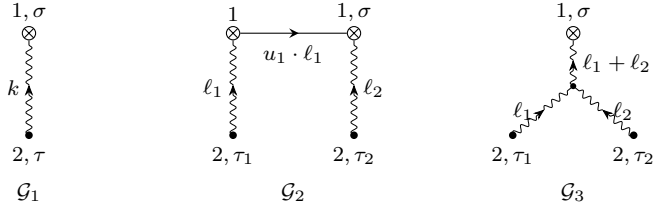
\begin{figure}[htbp]
\centering
\begin{tikzpicture}[font=\scriptsize]
  \begin{scope}
    \coordinate (S) at (0,-1.35);
    \node[xSource] (T) at (0,0) {};
    \draw[directed graviton] (S) -- node[left=2pt] {$k$} (T);
    \fill (S) circle (0.05);
    \node[below=3pt] at (S) {$2,\tau$};
    \node[above=3pt] at (T) {$1,\sigma$};
    \node at (0,-2.1) {$\mathcal G_1$};
  \end{scope}
  \begin{scope}[shift={(2.7,0)}]
    \coordinate (S1) at (0,-1.35);
    \coordinate (S2) at (1.6,-1.35);
    \node[xSource] (T1) at (0,0) {};
    \node[xSource] (T2) at (1.6,0) {};
    \draw[directed graviton] (S1) -- node[left=2pt] {$\ell_1$} (T1);
    \draw[directed graviton] (S2) -- node[right=2pt] {$\ell_2$} (T2);
    \draw[quark] (T1) -- node[below=3pt] {$u_1\cdot\ell_1$} (T2);
    \fill (S1) circle (0.05);
    \fill (S2) circle (0.05);
    \node[below=3pt] at (S1) {$2,\tau_1$};
    \node[below=3pt] at (S2) {$2,\tau_2$};
    \node[above=3pt] at (T1) {$1$};
    \node[above=3pt] at (T2) {$1,\sigma$};
    \node at (0.8,-2.1) {$\mathcal G_2$};
  \end{scope}
  \begin{scope}[shift={(7.2,0)}]
    \coordinate (S1) at (-0.8,-1.35);
    \coordinate (S2) at (0.8,-1.35);
    \coordinate (B) at (0,-0.70);
    \node[xSource] (T) at (0,0) {};
    \draw[directed graviton] (S1) -- node[left=2pt] {$\ell_1$} (B);
    \draw[directed graviton] (S2) -- node[right=2pt] {$\ell_2$} (B);
    \draw[directed graviton] (B) -- node[right=2pt] {$\ell_1+\ell_2$} (T);
    \fill (S1) circle (0.05);
    \fill (S2) circle (0.05);
    \fill (B) circle (0.04);
    \node[below=3pt] at (S1) {$2,\tau_1$};
    \node[below=3pt] at (S2) {$2,\tau_2$};
    \node[above=3pt] at (T) {$1,\sigma$};
    \node at (0,-2.1) {$\mathcal G_3$};
  \end{scope}
\end{tikzpicture}
\caption{Examples of graph--numerator contributions: a 1PM exchange, a 2PM worldline iteration and a 2PM graviton self-interaction.
Labels at WL vertices indicate the body and, where shown, the proper time.
Crossed circles denote WL sink vertices.
The internal sink of $\mathcal G_2$ inherits the proper time $\sigma$ of the final sink.}
\label{fig:algebraic_examples}
\end{figure}

\begin{samepage}
\paragraph{Examples.}
For the graphs $\mathcal G_1$, $\mathcal G_2$ and $\mathcal G_3$ in Fig.~\ref{fig:algebraic_examples}, let $N_1^\mu$, $N_2^\mu$ and $N_3^\mu$ denote generic impulse numerators, including the respective inverse symmetry factors.
After performing the proper-time integrals, their contributions are
\begin{equation}\label{eq:export_impulse}
 \Delta p_{1,\mathcal G_1}^\mu
 =-i\kappa^2m_1m_2
 \int_k e^{-ik\cdot b}\hat\delta(k\cdot u_1)\hat\delta(k\cdot u_2)R(k)N_1^\mu,
\end{equation}
\end{samepage}
\begin{equation}
\begin{aligned}
 \Delta p_{1,\mathcal G_2}^\mu={}&-i\kappa^4m_1m_2^2
 \int_{\ell_1}\int_{\ell_2}e^{-i(\ell_1+\ell_2)\cdot b}
 \hat\delta(\ell_1\cdot u_2)\hat\delta(\ell_2\cdot u_2)
 \\
 &\times\hat\delta((\ell_1+\ell_2)\cdot u_1)
 R(\ell_1)R(\ell_2)\frac{N_2^\mu}{\ell_1\cdot u_1+i0},
\end{aligned}
\end{equation}
\begin{equation}
\begin{aligned}
 \Delta p_{1,\mathcal G_3}^\mu={}&-i\kappa^4m_1m_2^2
 \int_{\ell_1}\int_{\ell_2}e^{-i(\ell_1+\ell_2)\cdot b}
 \hat\delta(\ell_1\cdot u_2)\hat\delta(\ell_2\cdot u_2)
 \\
 &\times\hat\delta((\ell_1+\ell_2)\cdot u_1)
 R(\ell_1)R(\ell_2)R(\ell_1+\ell_2)N_3^\mu.
\end{aligned}
\end{equation}
The WL propagator of power 1 in $\mathcal G_2$ supplies the linear denominator, while momentum conservation at the graviton vertex in $\mathcal G_3$ fixes the outgoing graviton momentum to $\ell_1+\ell_2$.
The transfer momentum is $q=k$ in $\mathcal G_1$ and $q=\ell_1+\ell_2$ in both $\mathcal G_2$ and $\mathcal G_3$; in $\mathcal G_2$, no individual graviton edge carries $q$.

\subsection{Benchmarks}\label{sec:benchmarks}

We benchmark the two example programs through 4PM on an AMD Ryzen 9 9950X processor with 16 physical cores, 32 hardware threads and 64\,MiB of L3 cache.
The machine provides 123.5\,GiB of usable RAM.
For these 1--4PM runs, both programs use 32 \texttt{rayon} worker threads and are compiled with \texttt{Rust} 1.98.1 in the default release profile, using \texttt{Symbolica} 3.0.0, on \texttt{Linux}.

As a validation of the construction, we have integrated the 1--4PM radiation-reacted impulse integrands and reproduced the collected results, region-by-region, in~\cite{Dlapa:2023hsl}.

Table~\ref{tab:benchmarks} gives the construction time at each PM order, excluding initialization and file export.
The variation of the effective action is constructed using the cached results from lower orders.
Since this calculation is also a step in computing the impulse, the impulse column reports only the additional time to insert the already computed trajectory corrections and simplify the result, given the effective-action contributions and lower-order iterations.
The time to prepare these inputs is excluded from the impulse column.
Each timing is taken from the corresponding phase of a single run.
The memory figures give the maximum RAM used during each run, including file export except for the marked 5PM entries.

\begin{table}[htbp]
\centering
\small
\renewcommand{\arraystretch}{1.2}
\begin{tabular}{@{}crrrrr@{}}
\toprule
 & \multicolumn{2}{c}{Variation of the effective action}
 & \multicolumn{3}{c}{Impulse from the given action} \\
\cmidrule(lr){2-3}\cmidrule(l){4-6}
PM order
 & Time & Peak memory
 & Time & Peak memory & Total file size \\
\midrule
1 & 0.375\,ms & 0.110\,GiB & 0.171\,ms & 0.111\,GiB & 95\,B \\
2 & 4.03\,ms & 0.128\,GiB & 1.30\,ms & 0.130\,GiB & 5.19\,KiB \\
3 & 165\,ms & 0.375\,GiB & 73.9\,ms & 0.406\,GiB & 526.7\,KiB \\
4 & 17.4\,s & 27.7\,GiB & 10.2\,s & 34.0\,GiB & 67.34\,MiB \\
\midrule
5 & 62.82\,min & 2.57\,TiB$^\ast$ & 36.58\,min & 2.57\,TiB$^\ast$ & 9.20\,GiB \\
\bottomrule
\end{tabular}
\caption{Construction times, peak memory usage, and total impulse file sizes at each PM order with 32 worker threads.
The 1--4PM results use the Ryzen 9 9950X; the 5PM results use the EPYC 7J13 cluster node.
The impulse time covers insertion of prepared iterations and simplification, excluding construction of the effective action.
File sizes sum all uncompressed graph--numerator files across the self-force sectors at each PM order in the integrand repository's \texttt{impulse/} directory.
$^\ast$The 5PM peak memory values of 2.57\,TiB exclude export; the full 5PM run, including export, peaks at 3.46\,TiB.}
\label{tab:benchmarks}
\end{table}

\paragraph{5PM.}
For the 5PM benchmarks we use an AMD EPYC 7J13 cluster node with 4\,TiB of RAM, running with 32 worker threads.
Constructing the 5PM variation of the effective action takes 62.82\,min, with a peak memory usage of 2.57\,TiB before export.
Obtaining the impulse from the prepared action and trajectory corrections takes a further 36.58\,min.
The impulse construction also peaks at 2.57\,TiB before export; including export, the full 5PM run reaches a peak memory usage of 3.46\,TiB.
The exported 5PM impulse files in graph-numerator format have a total size of 9.20\,GiB.
The high memory usage reflects the aggressive caching in \texttt{FeynRIR}, which minimizes recomputation by retaining intermediate results.
For integrands at higher loop orders, reducing the caching lowers memory usage at the cost of longer computing times.

\section{Conclusions}\label{sec:conclusions}

This manuscript presented an efficient graph-centered construction and representation for WEFT integrands based on graph-numerator pairs.
The key insight is a clean separation between the topological structure of a Feynman graph --- propagators, vertices, momentum routing, and $i0$ prescriptions --- and the minimal algebraic tensor content carried by the numerator.
This separation has important consequences that are immediately relevant for the subsequent analytic integration.
Beyond its direct impact on the analytic integration, this separation also makes the construction itself efficient.
The main consequences are as follows:

\begin{itemize}
  \item \textbf{Efficient construction and canonicalization.}
    The recursive multi-set construction schedules each unique contraction exactly once through a dependency DAG, and caches intermediate results across PM orders and across different observables (Sec.~\ref{sec:construction}).
    Serialization and selfie identities reorganize the graph component, requiring only scalar coefficients and momentum relabeling in the numerator.
    Deferring the proper-time integrals until after the family has been matched keeps the topologies as tree graphs so that isomorphism and symmetry detection remain cheap (Sec.~\ref{sec:ibp_proper_time}).
    In practice the construction phases take less than a second through 3PM and 17.4\,s for the 4PM variation of the effective action on a desktop machine, with the 5PM variation constructed in 62.82\,min on a cluster node (Sec.~\ref{sec:benchmarks}).
    
  \item \textbf{A single integral family per SF sector.}
    By \emph{mass polynomiality}, the impulse naturally splits into independent SF sectors, so the full reduction decomposes into a set of smaller, independent problems.
    Each SF sector at fixed PM order embeds into a single generalized integral family (Sec.~\ref{sec:ibp_proper_time}).
    This family's propagator set is overcomplete only globally, but linearly independent within each individual sector.
    Because the propagator labeling is carried by the graph itself, there is no single master loop-momentum routing and no need to introduce a fixed set of purely formal irreducible scalar products.

  \item \textbf{Trivialized symmetries and minimal rule sets.}
    The overcompleteness, together with the graphical representation of sectors/topologies, makes symmetries among them trivial to identify and renders the sector-to-subsector maps algorithmically accessible.
    Symbolic IBP reduction rules therefore need only be derived for one representative of each symmetry equivalence class; e.g.~at 3PM in the 0SF/0$\overline{\textrm{SF}}$ sector, only 5 of the 40 non-zero sectors require such rules (Sec.~\ref{sec:topo_example}).
    \texttt{FeynRIR} implements part of this strategy in an experimental \emph{topology library}.

  \item \textbf{Per-sector freedom of labeling.}
    Each sector may choose its own loop-momentum routing, its own minimal set of independent scalar products, and its integral ordering, and optimize them independently for syzygy generation and for (symbolic) IBP rule discovery.
    Mapping to a subsector then amounts to a simple linear transformation of the numerator ISPs.
\end{itemize}

Taken together these properties open new avenues for the (symbolic) IBP reduction.
The graph-numerator representation supplies a canonical, representation-independent starting point for the reduction, with all necessary sector-to-subsector maps and symmetry relations algorithmically accessible.

The algorithms presented here are directly generalizable to other observables, such as the change of angular momentum~$\Delta J$ or position~$\Delta x$, or the scattering waveform.
In fact, an integrand for $\Delta x$ is readily obtained from the impulse integrand by a simple re-interpretation of the proper time integration associated to the final sink vertex:
\begin{equation}
  \int_{-\infty}^\infty\dd\sigma \cF_a^\mu[x](\sigma) \rightarrow \int_{-\infty}^\infty\dd\sigma' \int_{-\infty}^{\sigma'}\dd\sigma \cF_a^\mu[x](\sigma)\,.
\end{equation}
We plan to implement more observables in future versions of \texttt{FeynRIR}.

\section*{Acknowledgments}

We thank Stefano De Angelis, Christoph Dlapa, Gustav Uhre Jakobsen, Zhengwen Liu, Jan Plefka, Rafael A. Porto, and Radu Roiban for insightful discussions on the ideas presented in this work.
We also thank Zhengwen Liu for providing access to his computing cluster for benchmarking the generation of the 5PM integrands.
G.K. is especially grateful to Jakob Neef for their collaboration on \texttt{WoLF}, the unpublished \texttt{C++} predecessor of \texttt{FeynRIR}.
G.K. also thanks Ben Ruijl for extensive support with \texttt{Symbolica} and for implementing new features that benefited this project.
G.K. acknowledges OpenAI for providing access to \texttt{ChatGPT} and \texttt{Codex} through the \texttt{ChatGPT} for Academic Researchers program.
J.L is supported by the Excellence Cluster ORIGINS funded by the Deutsche Forschungsgemeinschaft (DFG, German Research Foundation) under Germany’s Excellence Strategy – EXC-2094-390783311.

Finally, we note that AI tools were used for limited, auxiliary purposes: \texttt{Codex}/\texttt{ChatGPT}, as well as a local \texttt{vLLM} deployment of \texttt{Qwen3.8}, were used to identify bugs and logical errors in the source code (\texttt{Rust} and \texttt{Mathematica}), to improve the language and formatting of the manuscript, and to generate the documentation for the \texttt{Mathematica} graph drawing paclet.
The authors take full responsibility for all content of this paper.

\appendix

\section{Algebraically minimal WEFT Feynman rules}\label{app:feynman_rules}

In order to simplify the notation we discuss the construction of algebraically minimal Feynman rules in pure Yang-Mills theory first.
The gravitational case works in the same way and is briefly discussed below.

\subsection{A generic gauge-fixed YM Lagrangian}

We follow the discussion in \cite{Donoghue:2017pgk,Rafie-Zinedine:2018izq} for this construction, while working out many of the details.

The YM Lagrangian is invariant under the gauge transformation
\begin{equation}
  A_\mu^a(x) \rightarrow (A^{(\alpha)})_\mu^a(x) = A_\mu^a(x) + \frac{1}{g}D_\mu \alpha^a(x) = A_\mu^a(x) +\frac{1}{g}\partial_\mu \alpha^a(x)+f^{abc}A^b_\mu(x)\alpha^c(x)\,.
\end{equation}
When performing the path integral over $A$ (say in a partition function), we, hence, integrate over the same configuration infinitely many times.
Optimally, we'd like to only integrate over a single representative $[A]$ of each equivalence class under this transformation.
A neat way to achieve this goes back to Faddeev-Popov, which we will discuss now.

Equivalently to picking a representative for each equivalence class, we can demand a gauge fix condition at each point $x$ in spacetime:
\begin{equation}
  G^a[A](x) = F^a(x)\,,
\end{equation}
where for each color index $G^a$ is a scalar functional of $A$.
We use the following trick and insert the following identity into the path integral
\begin{equation}
  1 = \int \cD\alpha \delta(G^a[A^{(\alpha)}]-F^a) \det\left(\frac{\delta G^a[A^{(\alpha)}](x)}{\delta \alpha^b(y)}\right)\,,
\end{equation}
where the determinant is the continuum generalization of
\begin{equation}
  \begin{aligned}
    1 &= \int \dd^n \ba \delta^{(n)}(\bg(\ba))\det\left(\frac{\partial g_i(\ba)}{\partial a_j}\right)\\
    &= \int \dd^n \bg(\ba) \delta^{(n)}(\bg(\ba))\,,
  \end{aligned}
\end{equation}
I.e. it is a determinant over the continuum ``matrix indices'' $(x,y)$ and the color indices $(a,b)$
For the example of the partition function, we'd then find
\begin{equation}
  Z = Z[0] \int \cD\alpha \cD A \,\delta(G^a[A^{(\alpha)}] - F^a) \det\left(\frac{\delta G^a[A^{(\alpha)}](x)}{\delta \alpha^b(y)}\right) e^{i S_\textrm{YM}[A]}\,.
\end{equation}
Next, we insert yet another identity
\begin{equation}
  1 = N(\xi) \int \cD F\, e^{-\frac{i}{2\xi}\int \dd^Dx\,F(x)^2}\,,
\end{equation}
which is a functional Gaussian integral with a normalization constant $N(\xi)$, where $\xi$ is an arbitrary parameter.
We are allowed to insert this identity, since the result for the partition function does not depend on $F$.
Note that we use the abbreviation $F(x)^2=F^a(x) F^a(x)$ that includes a summation/contraction of the color indices.
One finds the form of the partition function
\begin{equation}
  Z = Z[0] N(\xi) \int \cD\alpha\cD F\cD A\, \delta(G^a[A^{(\alpha)}] - F^a) \det\left(\frac{\delta G^a[A^{(\alpha)}](x)}{\delta \alpha^b(y)}\right) e^{i S_\textrm{YM}[A] -\frac{i}{2\xi}\int \dd^Dx\,F(x)^2 }\,.
\end{equation}
After performing the path integral over $F$ we immediately find
\begin{equation}
  Z = Z[0] N(\xi) \int \cD\alpha\cD A\, \det\left(\frac{\delta G^a[A^{(\alpha)}](x)}{\delta \alpha^b(y)}\right) e^{i S_\textrm{YM}[A] -\frac{i}{2\xi}\int \dd^Dx\,G[A^{(\alpha)}](x)^2 }\,.
\end{equation}
We use the gauge invariance of the YM action $S_\textrm{YM}[A] = S_\textrm{YM}[A^{(\alpha)}]$, followed by a substitution of the integration $A_\mu\rightarrow A_\mu-\frac{1}{g}D_\mu \alpha$ (which has a trivial transformation of the integration measure).
This allows us then to factorize out the path integral over $\alpha$:
\begin{equation}
  \begin{aligned}
    Z &= Z[0] N(\xi) \int \cD \alpha \int \cD A \det\left(\frac{\delta G^a[A](x)}{\delta \alpha^b(y)}\right) e^{i S_\textrm{YM}[A] -\frac{i}{2\xi}\int \dd^Dx\,G[A](x)^2 }\\
    &= N'(\xi) \int \cD A \det\left(\frac{\delta G^a[A](x)}{\delta \alpha^b(y)}\right) e^{i S_\textrm{YM}[A] -\frac{i}{2\xi}\int \dd^Dx\,G[A](x)^2 }\,,
  \end{aligned}
\end{equation}
with a formally infinite normalization constant $N'(\xi)$, which stems from the infinite equivalent configurations that we previously integrated over.
This normalization constant drops out in correlation functions/amplitudes due to the cancellation in the LSZ procedure between numerator and denominator.
We have also assumed that the determinant is independent of $\alpha$, which the lecture notes cited above claim is generally true.

In a final step we use the continuum version of a common Grassmann identity
\begin{equation}
  \int \cD\bar{c}\cD c\, e^{-\int \dd^Dx \dd^Dy\, \bar{c}(x)M(x,y)c(y)} = \det(M(x,y))\,,
\end{equation}
where $M$ is a functional continuum matrix.
Note that both, the Grassmann fields $\bar{c}$, $c$ and the matrix $M$, need to be dressed with color in the case of YM theory.
This gives us the final version of our gauge-fixed partition function
\begin{equation}
  Z = N'(\xi) \int \cD A \cD\bar\chi \cD\chi e^{i S_\textrm{YM}[A] -\frac{i}{2\xi}\int \dd^Dx\,G[A](x)^2 - i \int\dd^Dx\dd^Dy \bar\chi(x) \frac{\delta G[A](x)}{\delta\alpha(y)} \chi(y) }\,,
\end{equation}

In summary, we have now a generic version of the gauge fixed YM action, that we carefully write with all color indices as
\begin{equation}
  S_\textrm{tot}[A,\bar\chi,\chi] = S_\textrm{YM}[A] -\frac{1}{2\xi}\int \dd^Dx\,G^a[A](x)G^a[A](x) - \int\dd^Dx\dd^Dy \bar\chi^a(x) \frac{\delta G^a[A](x)}{\delta\alpha^b(y)} \chi^b(y)\,,
\end{equation}
where $G^a[A](x)$ must be non-singular (w.r.t. the $\alpha$-variation of the argument $A^{(\alpha)}$) functionals in the gauge field $A$.

\subsection{A generic gauge-fixed  GR Lagrangian}

Let us repeat the discussion of the previous section to GR.

For $g_{\mu\nu}=\eta_{\mu\nu}+\kappa h_{\mu\nu}$, the EH Lagrangian is invariant under the infinitesimal gauge transformation
\begin{equation}
  \begin{aligned}
    h_{\mu\nu}(x) \rightarrow h_{\mu\nu}^{(\alpha)}(x) &= h_{\mu\nu}(x) + \partial_\mu \alpha_\nu(x)+ \partial_\nu \alpha_\mu(x)\\
    &\quad+\kappa\Bigl[h_{\mu\sigma}(x)\partial_\nu \alpha^\sigma(x) + h_{\nu\sigma}(x)\partial_\mu \alpha^\sigma(x)+\alpha^\sigma(x)\partial_\sigma h_{\mu\nu}(x)\Bigr]\,.
  \end{aligned}
\end{equation}
The gauge fix condition now generalizes to $D$ conditions
\begin{equation}
  G_\mu[h](x) = F_\mu(x)\,,
\end{equation}
The Faddeev-Popov trick becomes
\begin{equation}
  1 = \int \cD^D\balpha \delta(G_\mu[h^{(\alpha)}]-F_\mu) \det\left(\frac{\delta G_\mu[h^{(\alpha)}](x)}{\delta \alpha_\nu(y)}\right)\,,
\end{equation}
where the determinant is over the continuum spacetime variables $(x,y)$ and the Lorentz indices $(\mu,\nu)$.

The rest of the steps is equivalent to the YM theory procedure and we find in the end
\begin{equation}
  S_\textrm{tot}[h,\bar\chi^\mu,\chi^\nu] = S_\textrm{EH}[h] -\frac{1}{2\xi}\int \dd^Dx\,G_\mu[h](x)G^\mu[h](x) - \int\dd^Dx\dd^Dy \bar\chi^\mu(x) \frac{\delta G_\mu[h](x)}{\delta\alpha_\nu(y)} \chi_\nu(y)\,.
\end{equation}
The gauge condition through the order needed for four-point vertices can be parameterized as
\begin{equation}\label{eq:generic_gr_gauge}
  \begin{aligned}
    G_\mu[h] &= \kappa\left(c_1\partial^\nu h_{\mu\nu}+c_2\partial_\mu h\right)\\
    &\quad+\kappa^2\Bigl[h\left(c_3\partial_\mu h+c_5\partial^\nu h_{\mu\nu}\right)
      +h^{\nu\rho}\left(c_4\partial_\mu h_{\nu\rho}+c_6\partial_\nu h_{\mu\rho}\right)\\
    &\qquad\quad+h_\mu{}^\nu\left(c_7\partial^\rho h_{\nu\rho}+c_8\partial_\nu h\right)\Bigr]\\
    &\quad+\kappa^3\Bigl[h_\mu{}^\nu h_\nu{}^\rho\left(c_9\partial_\rho h+c_{10}\partial^\sigma h_{\rho\sigma}\right)
      +h h_\mu{}^\nu\left(c_{11}\partial_\nu h+c_{12}\partial^\rho h_{\nu\rho}\right)\\
    &\qquad\quad+h_\mu{}^\nu h^{\rho\sigma}\left(c_{13}\partial_\nu h_{\rho\sigma}+c_{14}\partial_\rho h_{\nu\sigma}\right)
      +h^2\left(c_{15}\partial_\mu h+c_{16}\partial^\nu h_{\mu\nu}\right)\\
    &\qquad\quad+h h^{\nu\rho}\left(c_{17}\partial_\mu h_{\nu\rho}+c_{18}\partial_\nu h_{\mu\rho}\right)
      +h_{\nu\rho}h^{\nu\rho}\left(c_{19}\partial_\mu h+c_{20}\partial^\sigma h_{\mu\sigma}\right)\\
    &\qquad\quad+h^\nu{}_{\sigma}h^{\rho\sigma}\left(c_{21}\partial_\mu h_{\nu\rho}+c_{22}\partial_\nu h_{\mu\rho}\right)\Bigr]
      +\cO(\kappa^4)\,.
  \end{aligned}
\end{equation}
Here $h={h_\mu}^\mu$, indices are raised with $\eta_{\mu\nu}$, and each derivative acts only on the field immediately following it.
For completeness, the common De Donder Gauge condition is
\begin{equation}
  G_\mu[h] = \kappa\left(\partial^\nu h_{\mu\nu}-1/2 \partial_\mu h\right)\,.
\end{equation}

\subsection{A generic total derivative Lagrangian for GR}

We are allowed to add any total derivative to a Lagrangian without changing the (classical) physics.
Through four fields, we write these terms as
\begingroup
\allowdisplaybreaks
\begin{align}
  \kappa^2\cL_\textrm{TD}[h]
  &=\partial^\mu\Bigl\{\kappa\left(d_1\partial_\mu h+d_2\partial^\nu h_{\mu\nu}\right)\notag\\
  &\quad+\kappa^2\Bigl[h\left(d_3\partial_\mu h+d_5\partial^\nu h_{\mu\nu}\right)
    +h^{\nu\rho}\left(d_4\partial_\mu h_{\nu\rho}+d_6\partial_\nu h_{\mu\rho}\right)\notag\\
  &\qquad\quad+h_\mu{}^\nu\left(d_7\partial^\rho h_{\nu\rho}+d_8\partial_\nu h\right)\Bigr]\notag\\
  &\quad+\kappa^3\Bigl[h_\mu{}^\nu h_\nu{}^\rho\left(d_9\partial_\rho h+d_{10}\partial^\sigma h_{\rho\sigma}\right)
    +h h_\mu{}^\nu\left(d_{11}\partial_\nu h+d_{12}\partial^\rho h_{\nu\rho}\right)\notag\\
  &\qquad\quad+h_\mu{}^\nu h^{\rho\sigma}\left(d_{13}\partial_\nu h_{\rho\sigma}+d_{14}\partial_\rho h_{\nu\sigma}\right)
    +h^2\left(d_{15}\partial_\mu h+d_{16}\partial^\nu h_{\mu\nu}\right)\notag\\
  &\qquad\quad+h h^{\nu\rho}\left(d_{17}\partial_\mu h_{\nu\rho}+d_{18}\partial_\nu h_{\mu\rho}\right)
    +h_{\nu\rho}h^{\nu\rho}\left(d_{19}\partial_\mu h+d_{20}\partial^\sigma h_{\mu\sigma}\right)\notag\\
  &\qquad\quad+h^\nu{}_{\sigma}h^{\rho\sigma}\left(d_{21}\partial_\mu h_{\nu\rho}+d_{22}\partial_\nu h_{\mu\rho}\right)\Bigr]\notag\\
  &\quad+\kappa^4\Bigl[h_\mu{}^\nu h_\nu{}^\rho h_\rho{}^\sigma\left(d_{23}\partial_\sigma h+d_{24}\partial^\lambda h_{\sigma\lambda}\right)
    +h h_\mu{}^\nu h_\nu{}^\rho\left(d_{25}\partial_\rho h+d_{26}\partial^\sigma h_{\rho\sigma}\right)\notag\\
  &\qquad\quad+h_\mu{}^\nu h_\nu{}^\rho h^{\sigma\lambda}\left(d_{27}\partial_\rho h_{\sigma\lambda}+d_{28}\partial_\sigma h_{\rho\lambda}\right)
    +h^2h_\mu{}^\nu\left(d_{29}\partial_\nu h+d_{30}\partial^\rho h_{\nu\rho}\right)\notag\\
  &\qquad\quad+h h_\mu{}^\nu h^{\rho\sigma}\left(d_{31}\partial_\nu h_{\rho\sigma}+d_{32}\partial_\rho h_{\nu\sigma}\right)
    +h_\mu{}^\nu h_{\rho\sigma}h^{\rho\sigma}\left(d_{33}\partial_\nu h+d_{34}\partial^\lambda h_{\nu\lambda}\right)\notag\\
  &\qquad\quad+h_\mu{}^\nu h^{\rho\sigma}h_\rho{}^\lambda\left(d_{35}\partial_\nu h_{\sigma\lambda}+d_{36}\partial_\sigma h_{\nu\lambda}\right)
    +h^3\left(d_{37}\partial_\mu h+d_{38}\partial^\nu h_{\mu\nu}\right)\notag\\
  &\qquad\quad+h^2h^{\nu\rho}\left(d_{39}\partial_\mu h_{\nu\rho}+d_{40}\partial_\nu h_{\mu\rho}\right)
    +h h_{\nu\rho}h^{\nu\rho}\left(d_{41}\partial_\mu h+d_{42}\partial^\sigma h_{\mu\sigma}\right)\notag\\
  &\qquad\quad+h h^\nu{}_{\sigma}h^{\rho\sigma}\left(d_{43}\partial_\mu h_{\nu\rho}+d_{44}\partial_\nu h_{\mu\rho}\right)
    +h_{\nu\rho}h^{\nu\rho}h^{\sigma\lambda}\left(d_{45}\partial_\mu h_{\sigma\lambda}+d_{46}\partial_\sigma h_{\mu\lambda}\right)\notag\\
  &\qquad\quad+h_\nu{}^\rho h_\rho{}^\sigma h_\sigma{}^\nu\left(d_{47}\partial_\mu h+d_{48}\partial^\lambda h_{\mu\lambda}\right)\notag\\
  &\qquad\quad+h^{\nu\rho}h_\nu{}^\sigma h_\rho{}^\lambda\left(d_{49}\partial_\mu h_{\sigma\lambda}+d_{50}\partial_\sigma h_{\mu\lambda}\right)\Bigr]\Bigr\}
    +\cO(\kappa^5)\,.\label{eq:generic_gr_total_derivative}
\end{align}
\endgroup
The coefficients $d_i$ are independent of the gauge parameters $c_i$.

\subsection{Assembling the full action}

The full WEFT action is assembled as
\begin{equation}
  \cS_\textrm{WEFT}[h,x_a] = \cS_\textrm{EH}[h] + \cS_\textrm{GF}[h] + \cS_\textrm{TD}[h] + \cS_\textrm{WL}[h,x_a]\,,
\end{equation}
where the Einstein-Hilbert action is
\begin{equation}
  \cS_\textrm{EH}[h] = -\frac{2}{\kappa^2} \int\dd^D x \sqrt{-\det g} R\,,
\end{equation}
and the worldline action in Brink--Di Vecchia--Howe form~\cite{Brink:1976sc,Deser:1976rb} is
\begin{equation}
  \cS_\textrm{WL}[h,x_a] = -\sum_a \frac{m_a}{2} \int \dd\tau g_{\mu\nu}(x_a(\tau))\dot{x}_a^\mu(\tau)\dot{x}_a^\nu(\tau)\,.
\end{equation}
\texttt{FeynRIR} uses the following expression for the Riemann scalar in terms of the metric
\begin{equation}
  \begin{aligned}
    R &= {R^a}_{bcd} {\delta_a}^c g^{bd}\\
    &= -1/4 g^{ab}g^{cd}g^{ef}(g_{cd,a}g_{ef,b} + 2 g_{df,b}g_{ca,e} - 3 g_{ca,e}g_{db,f} - 4 g_{cd,a}g_{be,f} + 4 g_{ca,d}g_{be,f})\\
    &\quad - g^{ab}g^{cd}(g_{cd,a,b} - g_{bd,a,c})\,.
  \end{aligned}
\end{equation}
Using (closed-form) combinatorial expressions for the expansion of the metric and the square root of the determinant it is a simple task to expand the Einstein-Hilbert action.

\subsection{An algebraically minimal choice}\label{app:optimal_gr_choice}

For $g_{\mu\nu}=\eta_{\mu\nu}+\kappa h_{\mu\nu}$, $\kappa^2=32\pi G$ and $\xi=-\kappa^2/2$, Table~\ref{tab:optimal_gr_parameters} gives an algebraically minimal choice through four points.
The coefficient ordering is defined explicitly in eqs.~\eqref{eq:generic_gr_gauge} and~\eqref{eq:generic_gr_total_derivative}.

\begin{table}[htbp]
  \centering\small
  \renewcommand{\arraystretch}{1.35}
  \begin{tabular}{@{}cll@{}}
    \toprule
    Order & Coefficients & Values \\
    \midrule
    $\kappa$ & $(c_{1},c_{2})$ & $(-1,\tfrac{1}{2})$ \\
     & $(d_{1},d_{2})$ & $(-2,2)$ \\
    \midrule
    $\kappa^2$ & $(c_{3},\ldots,c_{8})$ & $(\tfrac{1}{8},-\tfrac{1}{2},-\tfrac{1}{4},\tfrac{1}{2},\tfrac{1}{2},-\tfrac{1}{2})$ \\
     & $(d_{3},\ldots,d_{8})$ & $(-1,2,1,-1,-3,2)$ \\
    \midrule
    $\kappa^3$ & $(c_{9},\ldots,c_{22})$ & $(\tfrac{1}{4},-\tfrac{1}{2},-\tfrac{1}{8},\tfrac{1}{8},\tfrac{1}{2},-\tfrac{1}{2},\tfrac{3}{64},\tfrac{1}{32},-\tfrac{1}{8},\tfrac{1}{8},-\tfrac{1}{16},\tfrac{1}{8},1,-\tfrac{1}{2})$ \\
     & $(d_{9},\ldots,d_{22})$ & $(-2,3,1,-\tfrac{3}{2},-2,2,-\tfrac{1}{4},\tfrac{1}{4},1,-\tfrac{1}{2},\tfrac{1}{2},-\tfrac{1}{2},-2,1)$ \\
    \midrule
    $\kappa^4$ & $(d_{23},\ldots,d_{36})$ & $(2,-\tfrac{13}{4},-1,\tfrac{3}{2},2,-\tfrac{9}{4},\tfrac{1}{4},-\tfrac{3}{8},-1,1,-\tfrac{1}{2},\tfrac{3}{4},2,-\tfrac{7}{4})$ \\
     & $(d_{37},\ldots,d_{50})$ & $(-\tfrac{1}{24},\tfrac{1}{24},\tfrac{1}{4},-\tfrac{1}{8},\tfrac{1}{4},-\tfrac{1}{4},-1,\tfrac{1}{2},-\tfrac{1}{2},\tfrac{1}{4},-\tfrac{1}{3},\tfrac{1}{3},2,-\tfrac{3}{4})$ \\
    \bottomrule
  \end{tabular}
  \caption{An algebraically minimal choice through four points, grouped by powers of $\kappa$ in $G_\mu$ and $\kappa^2\cL_\textrm{TD}$, with $\kappa^2\propto G$.}
  \label{tab:optimal_gr_parameters}
\end{table}

The resulting bulk Lagrangian is given below, with indices raised by $\eta_{\mu\nu}$ and $h={h_\mu}^\mu$.
Its cubic and quartic parts contain $6$ and $18$ monomials, respectively, with no second derivatives acting on a single field.
\begingroup\small
\allowdisplaybreaks[1]
\begin{align}
&\cL_\textrm{EH}+\cL_\textrm{GF}+\cL_\textrm{TD}=\frac12 h_{bc,a} h^{bc,a}-\frac14 h{}_{,a} h{}^{,a}\notag\\
&\quad+\kappa\Biggl\{\frac h4 h_{bc,a} h^{bc,a}-\frac h8 h{}_{,a} h{}^{,a}+h^{ab}\Bigl[h_{cd,a} h_{b}{}^{d,c}-\frac12 h_{cd,a} h^{cd}{}_{,b}-h_{ad,c} h_{b}{}^{d,c}+\frac12 h_{ab,c} h{}^{,c}\Bigr]\Biggr\}\notag\\
&\quad+\kappa^2\Biggl\{\frac{h^2-2h_{\mu\nu}h^{\mu\nu}}{16} h_{bc,a} h^{bc,a}+\frac{h_{\mu\nu}h^{\mu\nu}}{16} h{}_{,a} h{}^{,a}-\frac{h^2}{8} h_{ab}{}^{,a} h_{c}{}^{b,c}\notag\\
&\qquad+\frac h2 h^{ab}\Bigl[h_{cd,a} h_{b}{}^{d,c}-\frac12 h_{cd,a} h^{cd}{}_{,b}-h_{ad,c} h_{b}{}^{d,c}+\frac12 h_{ab,c} h{}^{,c}\Bigr]\notag\\
&\qquad+h^{ab}h_{a}{}^{c}\Bigl[\frac12 h_{de,b} h^{de}{}_{,c}-h_{de,b} h_{c}{}^{e,d}+h_{be,d} h_{c}{}^{e,d}+\frac14 h_{be,d} h_{c}{}^{d,e}-h_{bc,d} h_{e}{}^{d,e}\Bigr]\notag\\
&\qquad+h^{ab}h^{cd}\Bigl[h_{ce,a} h_{d}{}^{e}{}_{,b}-\frac34 h_{ce,a} h_{b}{}^{e}{}_{,d}-\frac12 h_{ce,a} h_{bd}{}^{,e}-\frac14 h_{ab,e} h_{cd}{}^{,e}+\frac12 h_{ac,e} h_{bd}{}^{,e}\Bigr]\Biggr\}+\cO(\kappa^3).\label{eq:optimal_gr_lagrangian}
\end{align}
\endgroup

\bibliographystyle{JHEP}
\bibliography{refs}

\end{document}